\documentclass[manuscript]{aastex}

\usepackage{txfonts}
\usepackage{natbib}
\bibpunct{(}{)}{;}{a}{}{,} % to follow the A&A style

\newcommand{\hi}{{H\sc{i}~}}

\newcommand{\kms}{km\,s$^{-1}$}
\newcommand{\kmss}{km\,s$^{-1}~$}

\shorttitle{Cold Milky Way \hi filaments}
\shortauthors{Kalberla et al.}
\begin{document}

\title{Cold Milky Way \hi gas in filaments  }

\author{P.\ M.\ W.\ Kalberla\altaffilmark{1}, 
%\email{pkalberla@astro.uni-bonn.de}
        J. Kerp\altaffilmark{1},
        U. Haud\altaffilmark{2}, 
        B. Winkel\altaffilmark{3},
        N. Ben Bekhti\altaffilmark{1},
        L. Fl\"oer\altaffilmark{1},
        D. Lenz\altaffilmark{1},
}

\affil{Argelander-Institut f\"ur Astronomie, Universit\"at Bonn, Auf dem H\"ugel 71, 53121 Bonn, Germany}
\and
\affil{Tartu Observatory, 61602 T\~oravere, Tartumaa, Estonia,}
\and
\affil{Max--Planck--Institut f\"ur Radioastronomie, Auf dem H\"ugel 69, 53121 Bonn, Germany}

\begin{abstract} 

  We investigate data from the Galactic Effelsberg--Bonn \hi Survey
  (EBHIS), supplemented with data from the third release of the Galactic
  All Sky Survey (GASS III) observed at Parkes.  We explore the all sky
  distribution of the local Galactic \hi gas with $|v_{\rm LSR}| < 25 $
  \kms on angular scales of 11\arcmin~ to 16\arcmin.  Unsharp masking
  (USM) is applied to extract small scale features. We find cold
  filaments that are aligned with polarized dust emission and conclude
  that the cold neutral medium (CNM) is mostly organized in sheets that
  are, because of projection effects, observed as filaments. These
  filaments are associated with dust ridges, aligned with the magnetic
  field measured on the structures by {\it Planck} at 353 GHz. The CNM
  above latitudes $|b|>20^\circ$ is described by a log-normal
  distribution, with a median Doppler temperature $T_{\rm D} = 223$\,K,
  derived from observed line widths that include turbulent
  contributions. The median neutral hydrogen (HI) column density is
  $N_{\rm HI} \simeq 10^{19.1}\,{\rm cm^{-2}}$. These CNM structures are
  embedded within a warm neutral medium (WNM) with $N_{\rm HI} \simeq
  10^{20}\,{\rm cm^{-2}}$. Assuming an average distance of 100 pc, we
  derive for the CNM sheets a thickness of $\la 0.3$ pc. Adopting a
  magnetic field strength of $B_{\rm tot} = (6.0 \pm 1.8)\mu$G, proposed
  by Heiles \& Troland 2005, and assuming that the CNM filaments are
  confined by magnetic pressure, we estimate a thickness of 0.09
  pc. Correspondingly the median volume density is in the range $ 14 \la
  n \la 47\,{\rm cm^{-3}}$.

\end{abstract}

% -------------------------------
% -------------------------------
  \maketitle
%
%________________________________________________________________

\section{Introduction}
\label{intro}

The  Leiden/Argentine/Bonn (LAB) survey \citep{Kalberla2005} is
  currently the prime resource for Galactic \hi survey data. LAB
  comprises data from the Leiden Dwingeloo survey (LDS),
  \citet{Atlas1997}, (beam width at full width at half maximum (FWHM) $
  \sim 0.6 \degr$) and from the survey of the Instituto Argentino de
  Radioastronomía (IAR), \citet{Bajaja2005} (FWHM $ \sim 0.5
  \degr$). This survey provides information for a broad range of
  applications, ranging from cosmology to Milky Way research. Its
  success is based on the high reliability of the LAB data. LAB is the
  first HI full-sky survey comprising high quality flux calibration,
  radio--frequency interference (RFI) mitigation and most importantly, a
  correction of a major instrumental problem of single dish radio
  telescopes, the stray radiation \citep{Kalberla1980}. Though, the LAB
  survey has some shortcomings. The angular resolution is only moderate
  and even worse: the HI sky is not fully Nyquist sampled
  \citep{Shannon1949}.
   
Within the last few years, major efforts have been made to use two of
the world's largest single dish radio telescopes for a new Galactic all sky
\hi survey; on the southern hemisphere the Parkes Telescope and the on
the northern hemisphere the Effelsberg telescope. For declinations
$\delta < 1 \degr$, the Galactic All Sky Survey (GASS) was first
published by \cite{Naomi2009}, the second data release (GASS II)
includes corrections for stray radiation \citep{Kalberla2010}. Recently
it was superseded by the third data release (GASS III) with improved
performance after removal of residual instrumental problems
\citep{Kalberla2015}.

The first data release of the Galactic portion of the Effelsberg Bonn
\hi Survey (EBHIS) just became available \citep{Winkel2016}.  GASS III
and EBHIS were prepared in parallel in Bonn aiming to be merged to a
common all sky \hi survey.  It is then mandatory to evaluate the
performance of the {\it common} data product. As a first test we have
chosen a topic that is sensitive to residual instrumental problems. At
the same time we have selected a theme that is interesting on its own,
turbulent structures and filaments in the interstellar medium
(ISM). Here we use the term ``filament'' to describe ``a single thread
or a thin flexible threadlike object'' (Merriam-Webster). The observed
filaments are projections of 3-D objects onto the plane of the sky. We
describe these features ``as seen'' without any constrains on the
detailed geometry.

 Already in 1994, when the northern part of the LAB, the LDS, became
available \citep{Atlas1997}, Henk van de Hulst pointed out (private
communication) that such features of the \hi distribution are probably
best appropriate for a large scale characterization of turbulent
structures in the ISM.
 
Unsharp masking (USM) is a technique to enhance the contrast of small
scale features while suppressing large scale features. A spatially
smoothed representation of the area of interest is calculated and
subtracted from the original data product. Digital-imaging software
packages like Adobe Photoshop and GIMP use this method to enhance partly
contrast and sharpness of an image.

Here we use the USM technique to demonstrate what is gained by using the
world's largest radio telescopes for a Galactic \hi survey in comparison
to the well established LAB survey. Using the USM techniques allows to
investigate the spatial distribution and to perform a parameterization
of \hi filaments across the whole sky.

 USM techniques with partly stunning results were applied first in
  astronomy by \citet{Malin1978} as an analogous photographic masking
  technique to enhance pictures of faint nebulosities. Using digital
  data processing, \citet{Sofue1979} were the first to apply the USM
  technique in radio astronomy. They successfully applied a Gaussian
  smoothing to determine scanning effects. The width of the smoothing
  kernel was twice the telescope beam. USM techniques can only be
  applied to remove additive effects, e.g. emission of the atmosphere,
  and not multiplicative ones, e.g. caused by gain variations or
  atmospheric damping. 

  Since then, USM found a number of applications in astronomy, either to
  identify instrumental problems or to detect faint sources. Most recently,
  \citet{Planck2016} applied USM techniques for the
  characterization of ridges and calculation of the associated excess
  column densities of polarized dust emission at
  353\,GHz. For the GASS, \citet{Kalberla2011,Kalberla2015} used among
  other methods USM to identify residual instrumental problems. It
  turned out that USM data show numerous genuine \hi filaments that may be
  mistakenly identified as instrumental artifacts when using median
  filtering or other automatic RFI excision methods. Here we apply USM
  methods to demonstrate that GASS III and EBHIS discloses a wealth of
  real \hi filaments without suffering from residual instrumental
  effects that may mimic these structures. 

Our investigations are timely. A tight correlation between gas and dust
is expected \citep{Boulanger1996}, but the question arises on what
spatial scale this correlation is observed. Moreover, does the \hi gas
correlate with the magnetized interstellar medium? Very recently, {\it
  Planck\/} all-sky maps of the linearly polarized dust emission
at 353\,GHz became available. This data reveals that dust-ridges in the
total intensity maps have their counterparts in the corresponding Stokes
Q and/or U maps, suggesting that the dusty interstellar medium is
physically related to the magnetized medium
\citep{Planck2016}. While the diffuse ISM appears to be aligned
to the magnetic field lines of forces, towards the cold molecular gas
phase a perpendicular orientation is observed. The polarization angle
displays a coherent well ordered pattern across extended areas of
several square degrees, intercepted by individual filamentary structures
\citep{PlanckXIXa}. To investigate the question, how far filaments are
affected by interstellar turbulence, polarized thermal emission from
Galactic dust was compared with magneto hydrodynamic (MHD)
simulations \citep{PlanckXX2015b}.

Previous radio astronomical investigations concerning a relation between
\hi gas and magnetic fields are based primarily on Zeeman splitting
observations in interstellar \hi clouds observed first by
\citet{Verschuur1969} in absorption against Cas A and Tau
A. Observations of the Zeeman splitting are challenging because of the
complex antenna characteristics of radio telescopes. The Arecibo
Millennium survey took nearly 1000 hours of telescope time but yielded
only 22 detections with a 2.5 $\sigma$ signal out of 69 measured sources
\citep{Heiles2005}. \citet{HeilesCrutcher2005} comprise in their review
the most recent results.  Their conclusions are based mainly on
statistical evidence that the large scale magnetic field appears to be
in equipartition with turbulent motions in the cold neutral medium
(CNM). The median total magnetic field strength is determined to $B_{\rm
  tot} \sim (6.0 \pm 1.8)\mu$G.

Further evidence for a close association between gas and magnetic fields
came very recently from the Galactic Arecibo L-Band Feed Array \hi
(GALFA-{H\sc{i}}) survey. \citet{Clark2014,Clark2015} detected slender, linear
\hi features, denoted as ``fibers'', that extend across many degrees
at the high Galactic latitude sky. These fibers are oriented parallel to the
magnetic field lines. \citet{Clark2014} demonstrated that these fibers
trace dust polarization angles.

\section{Data Sets }
\label{data_sets}
 
For the southern sky we use the GASS III data release
\citep{Kalberla2015} which was proven to be affected at most at a very
low level by correlator problems and RFI that might mimic filamentary
features. For the northern sky we use data from the first EBHIS data
release \citep{Winkel2016}. 

For both surveys independent databases on a common HEALPix grid
\citep{Gorski2005} were generated by gridding the spectra that were
observed on-the-fly to a common nside = 1024 data structure. On such a
grid the formal HEALPix angular resolution is 3\farcm44 per pixel, on
the equator the pixel separation is 5\farcm32. Compared to the full
width half maximum (FWHM) beam-size of 10\farcm8 for the EBHIS and
16\farcm2 for GASS III this database provides an adequate sampling
\citep{Shannon1949}. The rms uncertainties of the brightness
temperatures per channel are 90\,mK for EBHIS at a FWHM width of $\delta
v = 1.44$ \kmss and 57\,mK at $\delta v = 1.0$ \kmss for GASS. When
analyzing filaments extracted by USM we are eventually going to use only
data with brightness temperatures in excess of 1 K (partly 0.3 K or even
lower), corresponding to a brightness temperature threshold of $\geq 10
\sigma$ (or $\geq 3 \sigma$). In any case the derived structures can
safely be considered to be unaffected by residual instrumental problems
\citep{Kalberla2015,Winkel2016}.

For both surveys we generate independently a smooth database by applying
a Gaussian convolution resulting in an effective FWHM beam size of
30\arcmin.  The EBHIS covers declination $\delta > -4\degr $, while for
GASS the limit is $\delta < 1\degr $. After smoothing we merge both data
sets, using EBHIS data for $\delta > -2\degr $ and GASS III data for
$\delta < -2\degr $.

When merging the original unsmoothed brightness temperature databases we
are faced with the problem that both surveys have a different spatial
resolution. Using a hard transition as before at $\delta = -2\degr $ leads
to unavoidable discontinuities. We therefore apply a linear
interpolation in angular resolution between $-3\degr < \delta < -1\degr $
from 16\farcm2 (GASS) to 10\farcm8 (EBHIS).

EBHIS and GASS are observed on a different velocity grid. To combine the
data we need to define a common velocity grid. Here we have chosen for
convenience the LAB velocities with a channel spacing of 1.03 \kms.  To
resample the data we used a cubic spline interpolation.

%-----------------------------------------------------------------------
\begin{figure*}[tbp]
\epsscale{1.}
   \centerline{
   \plotone{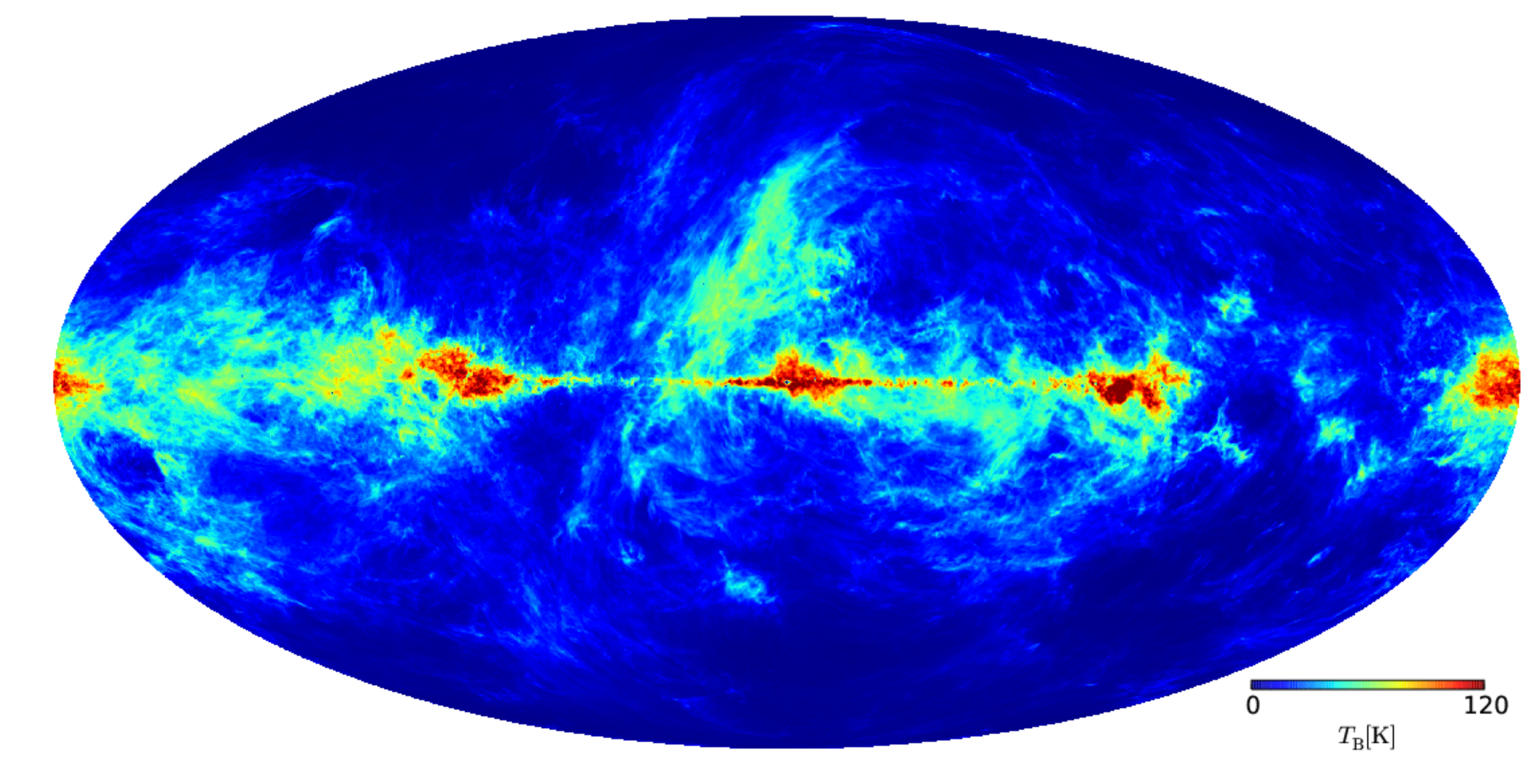}
}
   \centerline{
   \plotone{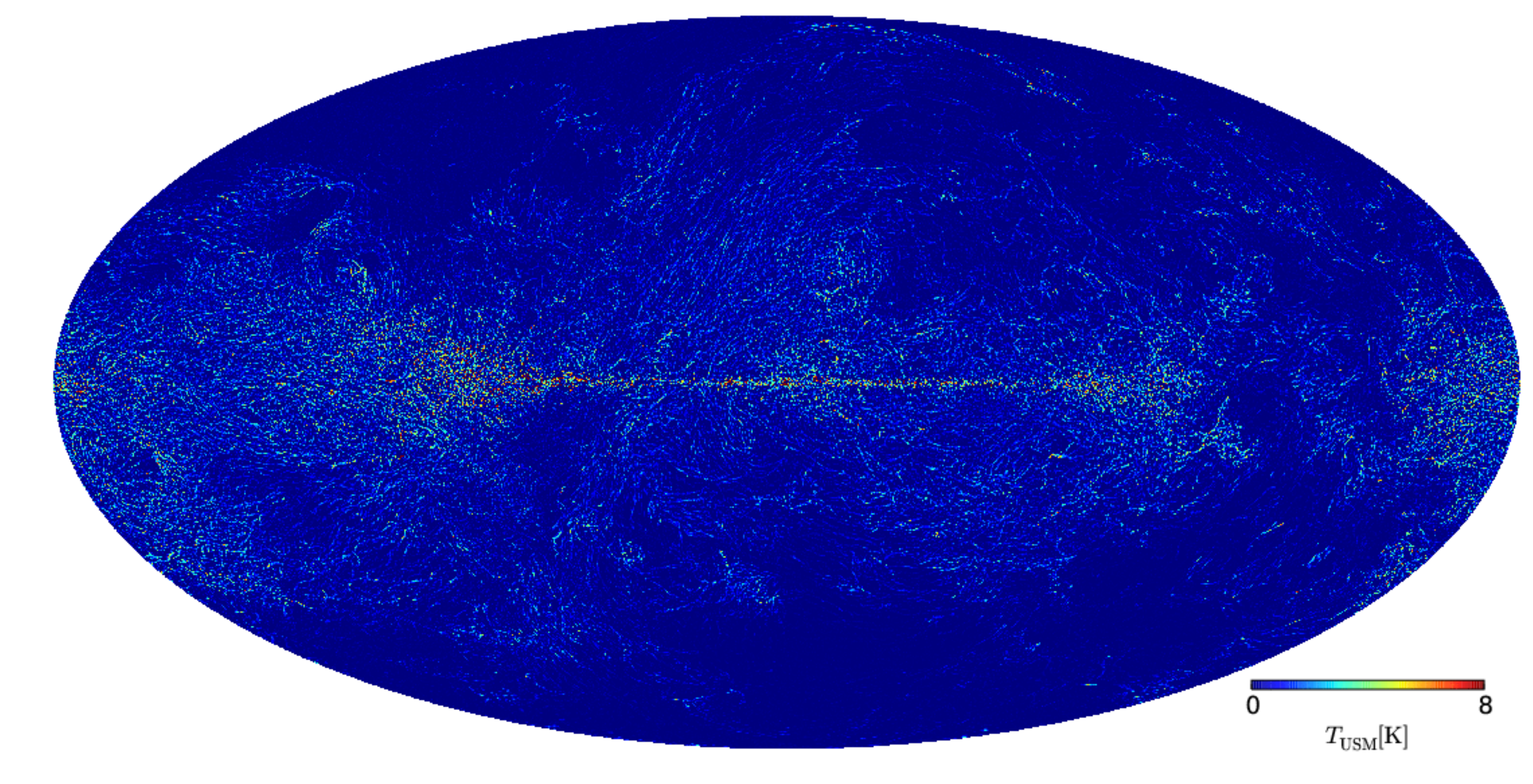}
}
\caption{All sky Mollweide display of the observed \hi brightness
  temperature distributions at a velocity of $v=0$ \kmss {\it (top)} and
  filamentary structures, derived by unsharp masking, at the same
  velocity (bottom). Plots for two more channels ($ v = \pm 8$ \kms) are
  given in the Appendix. }
        \label{Fig_Filaments1}
\end{figure*}
%-----------------------------------------------------------------------

\section{Data analysis }
\label{data_analysis}

Subsequent to the calculation of a common HEALPix databases for both
telescopes we generate an USM database by subtracting previously
smoothed data from the unsmoothed ones. The USM maps display only the
high spatial frequencies (small-scale structure) of the objects of
interest; spatial frequencies lower than the smoothing kernel of
$30\arcmin$ (large-scale structures) are suppressed.

We are interested to study small scale \hi structures on angular scales
between 11\arcmin~ (EBHIS) to 16\arcmin~ (GASS), the highest
angular resolution that is provided by all-sky surveys using two of the
world's largest single dish radio telescopes. We suppressed HI
structures spatially larger than 30\arcmin. A Gaussian smoothing kernel
was chosen since it is non-negative and non-oscillating, hence it causes
no overshoot or ringing.  In fact, we blur the data and drop all the
information on extended \hi structures that LAB provides. These so far
only marginally explored HI filamentary features at high spatial
frequencies as presented in detail below.  We aim here to derive
parameters for the global all sky distribution of local gas.

Figures \ref{Fig_Filaments1}, \ref{Fig_Filaments1b}, and
\ref{Fig_Filaments1c} display three examples. All sky brightness
temperature channel maps are displayed in the top panel in Mollweide
projection for the radial velocities of $v = 0$ \kms~
(Fig. \ref{Fig_Filaments1}), $v = -8$ \kms~ (Fig. \ref{Fig_Filaments1b})
and +8 \kms~ ( Fig. \ref{Fig_Filaments1c}). USM data that emphasize
small scale \hi fluctuations are given in the lower panels.

Visual inspection of these channel maps discloses that the USM data in
general display only filamentary structures that change significantly in
both, position and brightness temperature from channel to channel,
implying that the line width of the \hi gas under investigation is
comparable to the velocity resolution (channel spacing) of the HI survey
data. Such narrow line widths correspond to low kinetic gas
temperatures $T_{\rm kin} \leq 200$\,K. Thus, the filaments are
physically associated with \hi gas belonging to the CNM 
\citep{ARAA2009}.

Contrarily, inspecting the observed brightness temperature maps channel
by channel shows only minor changes. These maps are dominated by the
spatially more extended warm neutral medium (WNM). We interpret the USM
features as an indication that the cold filamentary structures are in
brightness temperature superimposed to the WNM, perhaps also embedded in
it. Whenever we inspect the location of a cold filamentary structures in
the USM maps we find corresponding features in the original brightness
temperature maps. The opposite search fails in the majority of the
cases, even for features that appear visually as unique HI filaments.

Inspecting individual regions at various velocities we often find
multiple filaments towards the same area of interest. These filaments
reveal for different radial velocities different orientations at the
sky. Obviously we are faced with a wealth of cold filaments showing up
in the USM maps.

%-----------------------------------------------------------------------
\begin{figure*}[tbp]
%\epsscale{0.75}
\epsscale{0.68}
   \centerline{
   \plotone{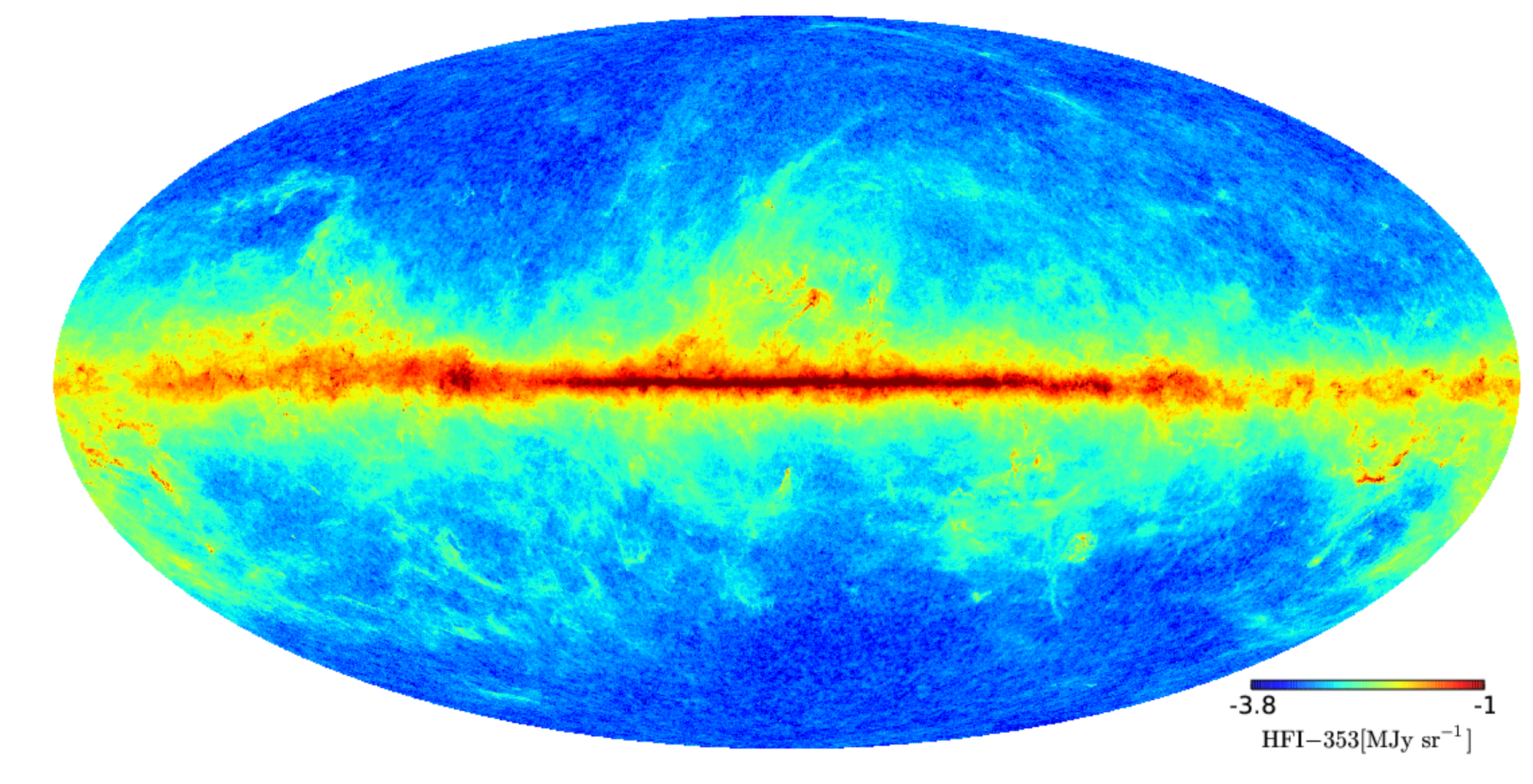}
}
   \centerline{
   \plotone{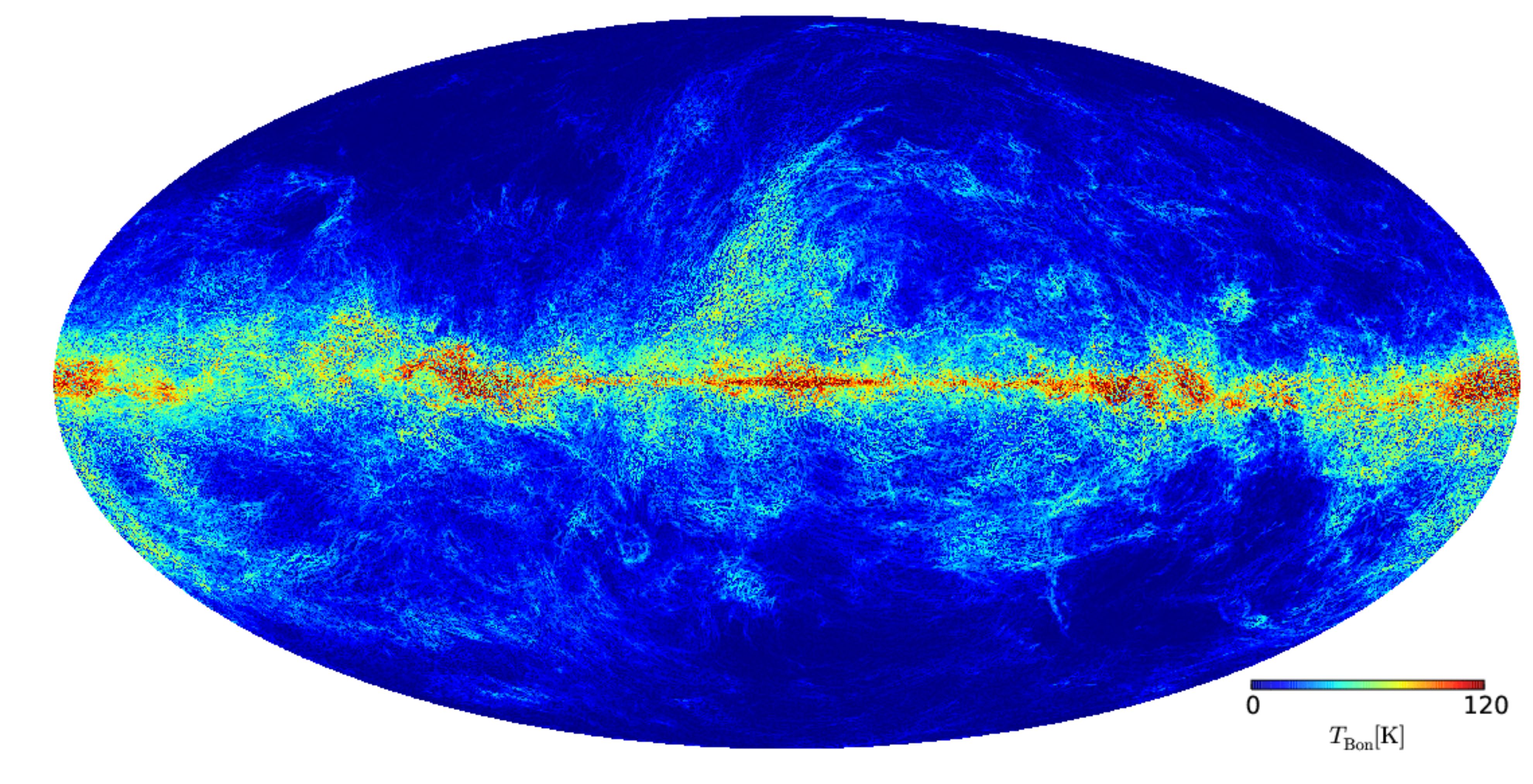}
}
   \centerline{
   \plotone{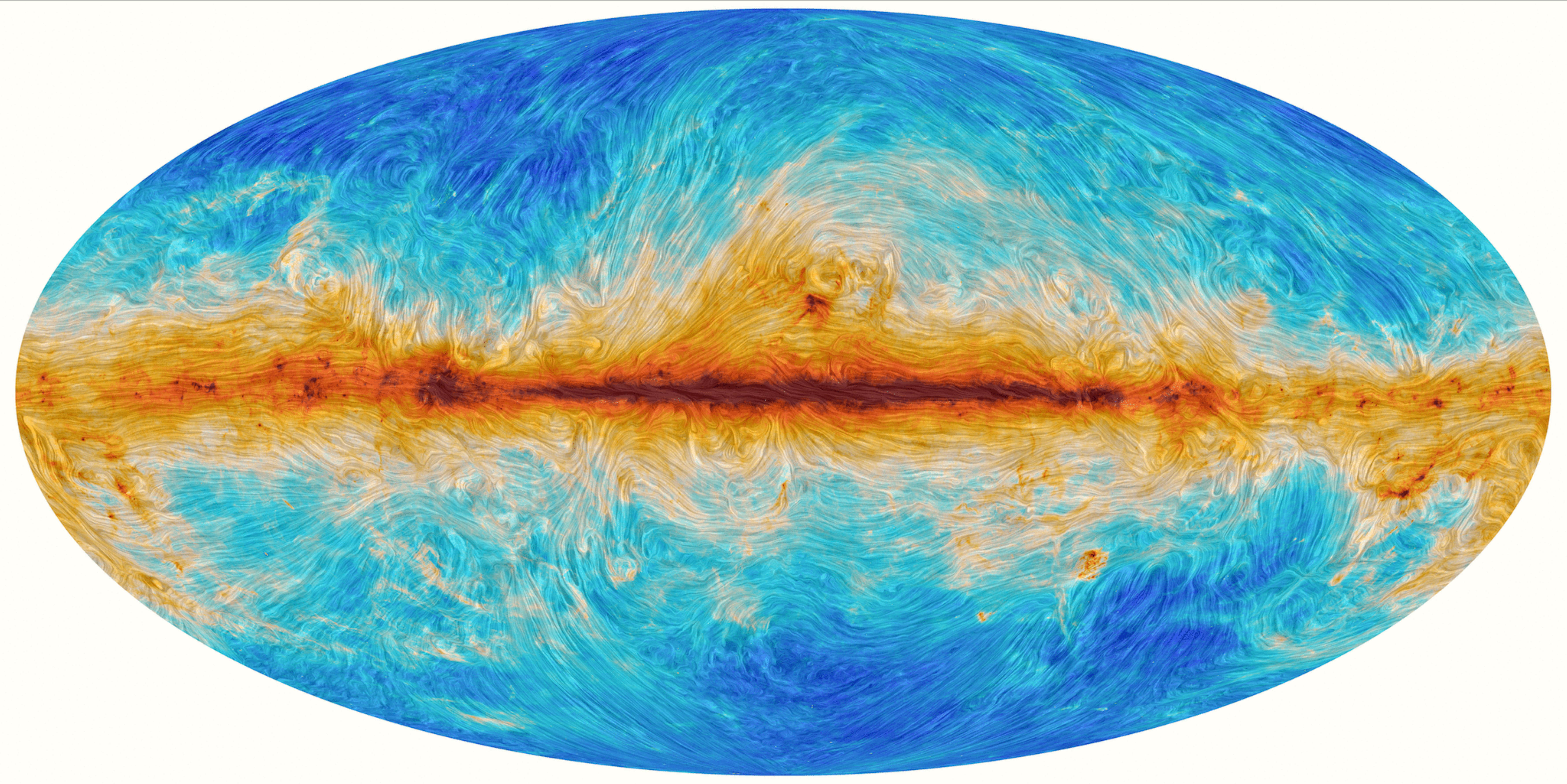}
}
   \caption{ {\it Top:} {\it Planck\/} HFI Sky Map at 353 GHz (MJy sr$^{-1}$). To
     cover the high dynamic range we use a logarithmic look-up table.
     {\it Middle:} all sky brightness temperature map $T_{\rm Bon}$ of major
     local \hi filaments (K) observed by us. {\it Bottom:} All-sky view
     of the magnetic field and total intensity of dust emission measured
     by {\it Planck\/} as published by ESA 05/02/2015. The colors represent
     intensity. The texture is based on measurements of the direction of
     the polarized light emitted by the dust, which in turn indicates
     the orientation of the magnetic field.}
   \label{Fig_Filaments2}
\end{figure*}
%-----------------------------------------------------------------------
%http://www.esa.int/spaceinimages/Images/2015/02/Polarised_emission_from_Milky_Way_dust

\subsection{Deriving global properties of the filamentary HI structures}
\label{deriving}

Given the wealth of filamentary features visible in
Fig. \ref{Fig_Filaments1} (and Figs. \ref{Fig_Filaments1b} and
\ref{Fig_Filaments1c} in the Appendix) we obviously need to restrict our
analysis to the most prominent ones. We therefore applied the following
algorithm to search for major filaments: for each position we check at
all velocities $|v_{\rm LSR}| < 25 $ \kmss for a positive peak $T_{\rm
  Uon}$ in the USM database. In addition to the strength of the USM peak
we record its velocity $v_{\rm Uon}$ and the corresponding unsmoothed
brightness temperature $T_{\rm Bon}$ at that velocity. We use the
subscript ``on'' to denote that this parameter is ``on source'' in
position $l,b$ and velocity $v$; $T_{\rm Bon}$ is at the same ($l,b,v$)
position as $T_{\rm Uon}$.  We trace for each position the USM features
in velocity as long as the intensities remain positive in the USM data
base. This allows us to determine the column density $N_{\rm HUon}$ and
the FWHM velocity width $\delta v_{\rm Uon}$.

From $\delta v_{\rm Uon}$ we derive Doppler temperatures $T_{\rm D} =
21.86~ \delta v_{\rm Uon}^2$ \citep[][Eq. 8]{Payne1980}. This
temperature is derived from the observed FWHM line width, corrected for
instrumental broadening, and is a measure for an upper limit of the
kinetic temperature $T_{\rm kin}$ \citep{Field1959}. Line broadening is
caused by a superposition of turbulent motions with $T_{\rm turb} =
21.86~ \delta v_{\rm turb}^2$ \citep{Liszt2001} and bulk motions along
the line of sight.  Only in case of a negligible contribution of the
turbulent component both temperatures would approach each other, $T_{\rm
  D} \sim T_{\rm kin} $. In Sect. \ref{environment} we discuss line of
sight effects that cause a broadening of observed WNM emission
lines. Turbulent line broadening, affecting CNM components, is discussed
in Sect. \ref{width}.

For \hi gas in equilibrium, the kinetic temperature is related to the
spin temperature, which is just the excitation temperature of the
hyperfine levels evaluated according to the Boltzmann equation
\citep{Field1959}.  Due to its long life time, the 21-cm transition is
usually collisionally excited, and the spin temperature of the gas is a
measure for the kinetic temperature. According to \citet{Field1958} the
term spin temperature is defined for temperatures derived from
absorption lines while kinetic temperatures are determined from the
width of \hi lines in thermal motion.

%-----------------------------------------------------------------------
\begin{figure*}[tbp]
\epsscale{0.48}
   \centerline{
   \plotone{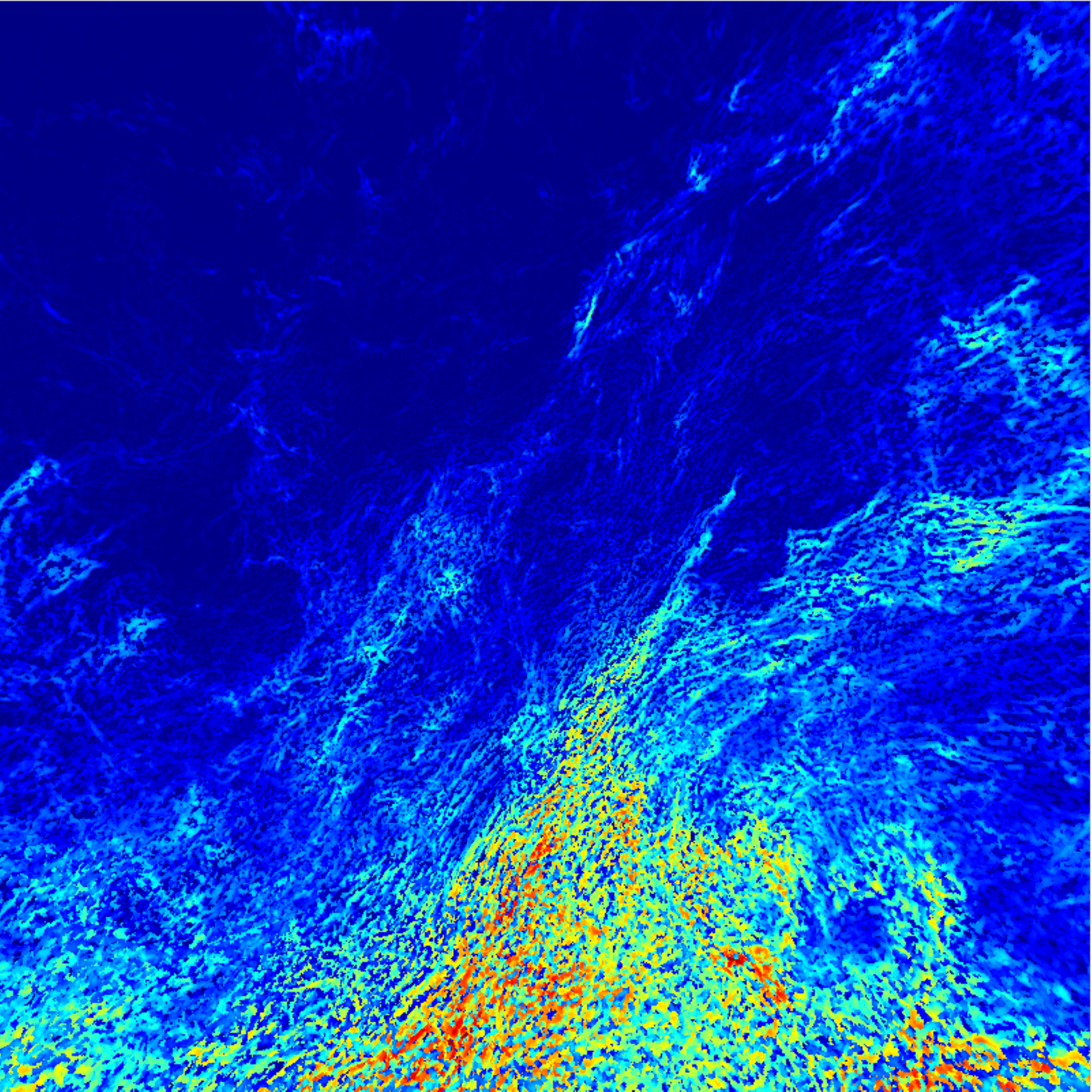}
   \plotone{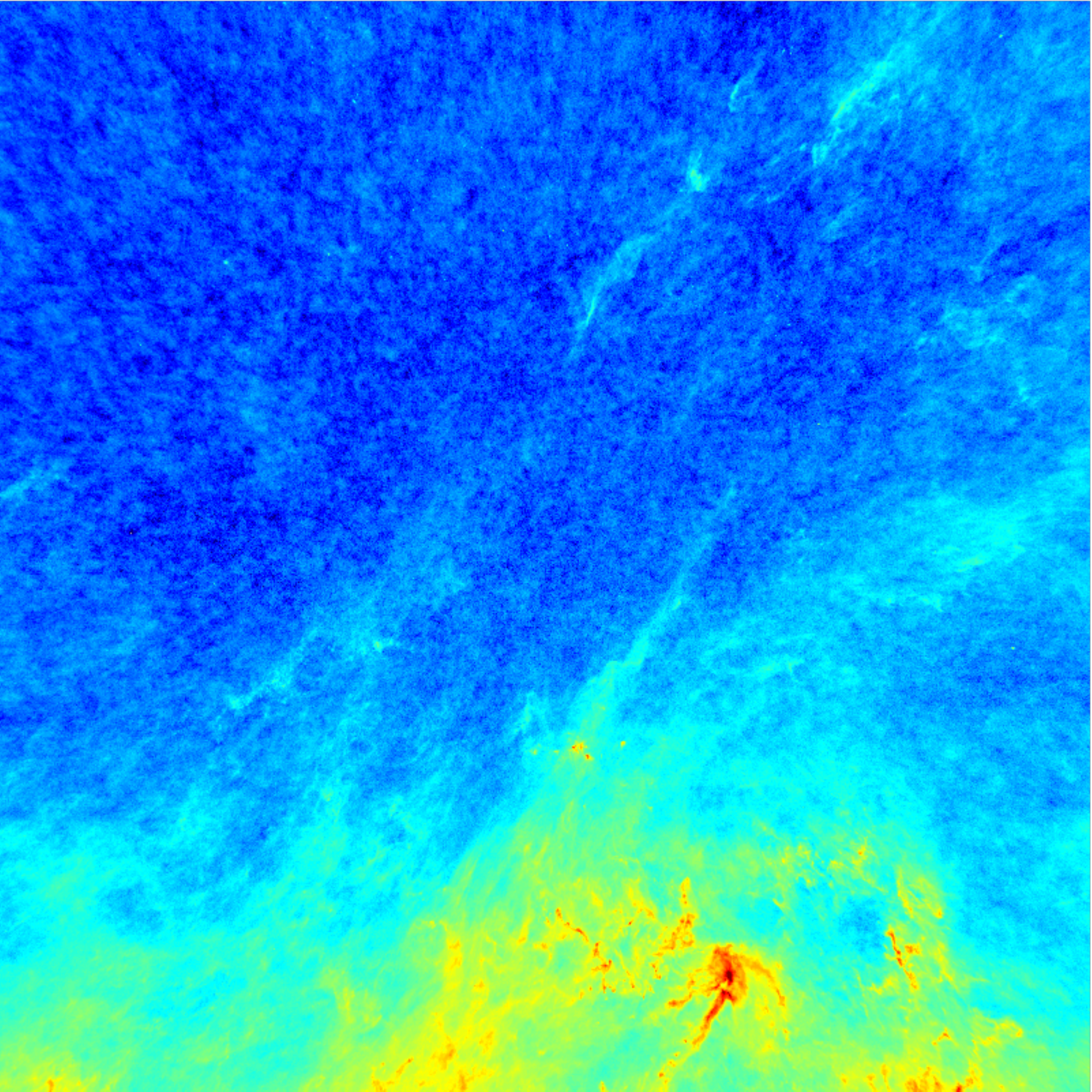}
}
\caption{Zoom in only for a visual comparison between major \hi
  filaments (left) and the {\it Planck\/} HFI dust emission at 353
  GHz. Both maps are centered at $l = 10 \degr$, $b = 60 \degr$ and
  display about $100\degr \times 100\degr$ in gnomonic projection.  }
   \label{Fig_Detail1}
\end{figure*}
%-----------------------------------------------------------------------

%-----------------------------------------------------------------------
\begin{figure*}[tbp]
\epsscale{0.48}
   \centerline{
   \plotone{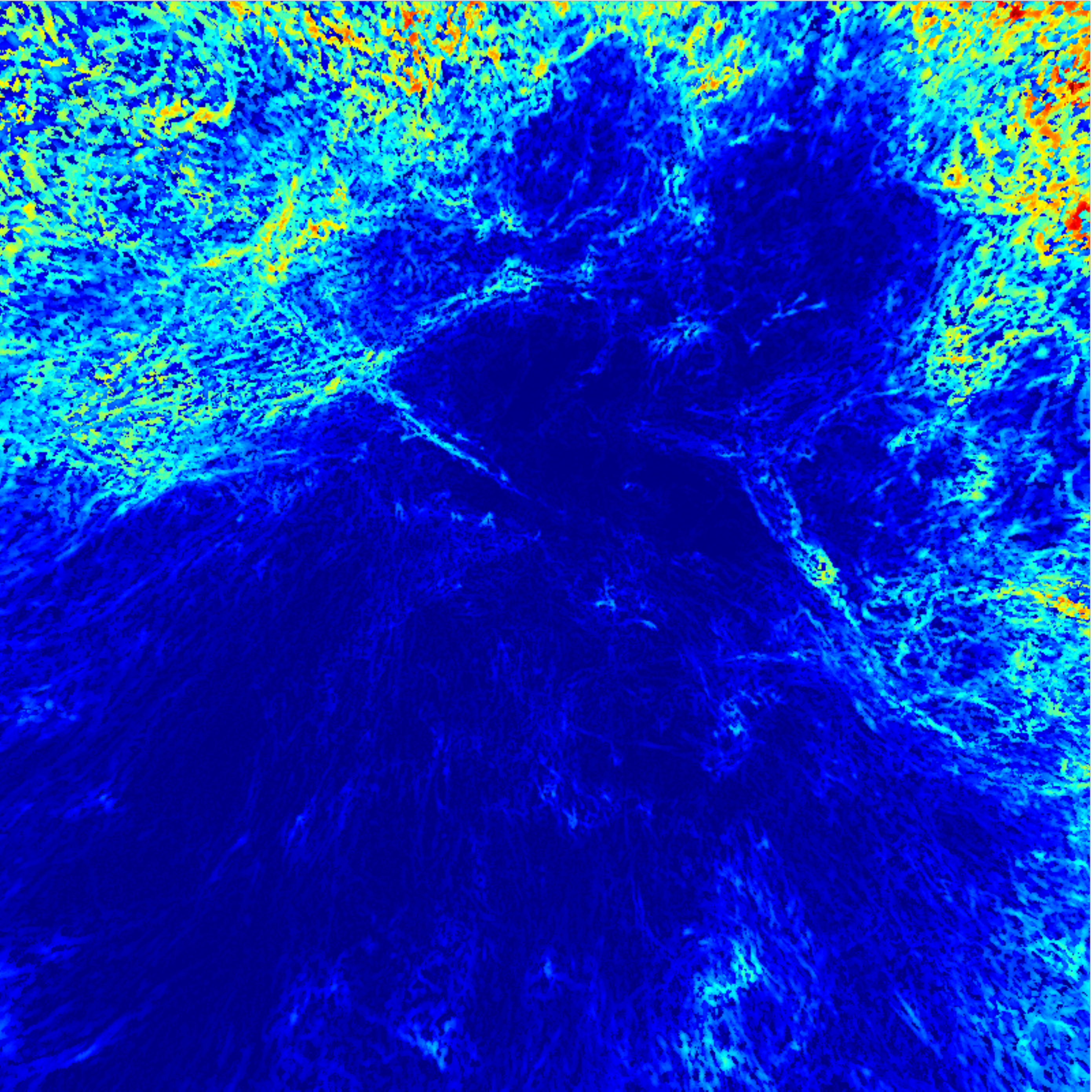}
   \plotone{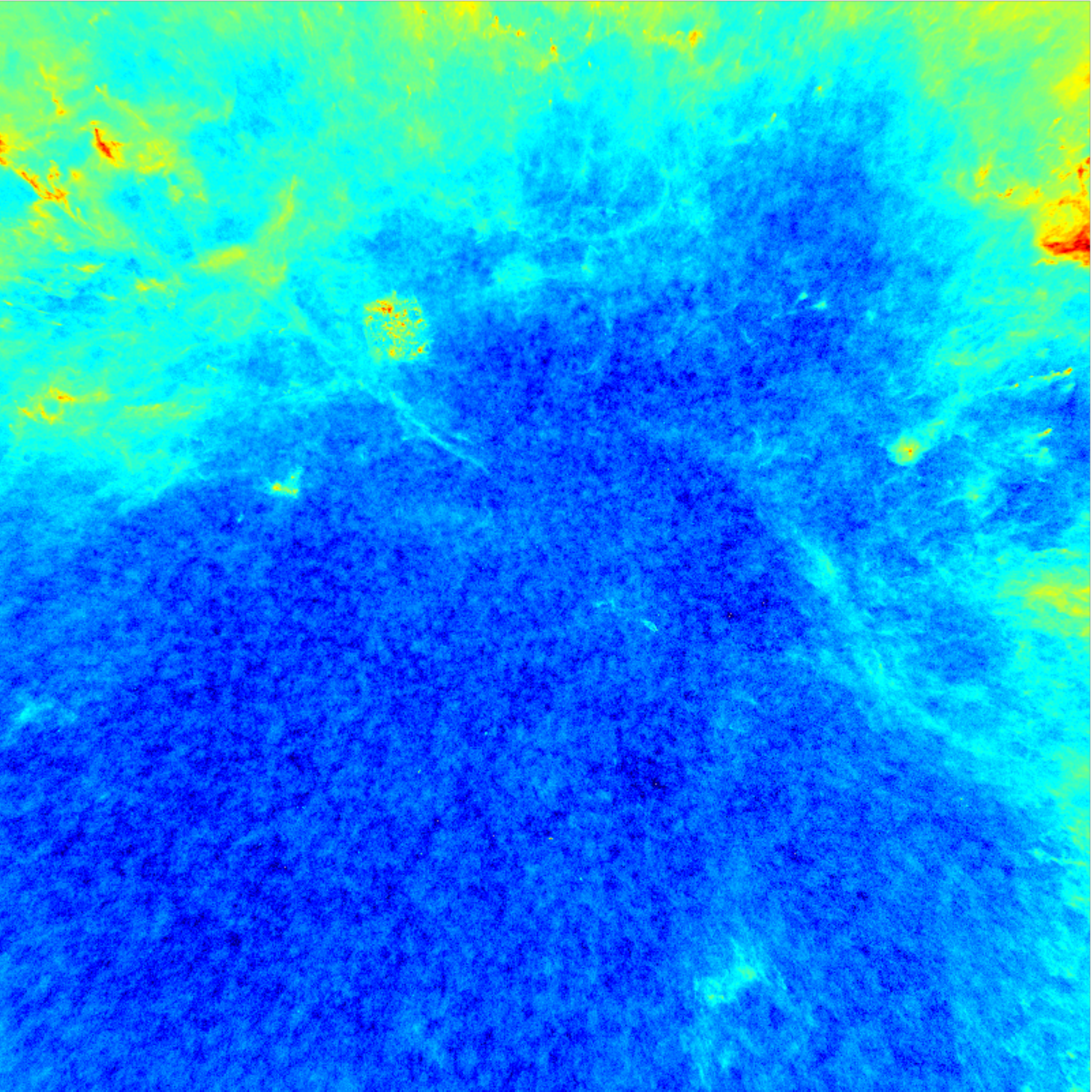}
}
\caption{Zoom in only for a visual comparison between major \hi
  filaments (left) and the {\it Planck\/} HFI dust emission at 353
  GHz. Both maps are centered at $l = 260 \degr$, $b = -60 \degr$ and
  display about $100 \degr \times 100\degr$ in gnomonic projection. Note
  that the HFI map is affected by emission from the Magellanic System
  while only local gas for $|v_{\rm LSR}| < 25 $ \kmss is included in
  the \hi data.  }
   \label{Fig_Detail2}
\end{figure*}
%-----------------------------------------------------------------------

\subsection{Selected area studies with the Arecibo telescope}
\label{Arecibo}

\citet{Clark2014,Clark2015} have used the (GALFA-{H\sc{i}}) Survey and
the first GASS data release \citep{Naomi2009} to search for filamentary
\hi features. In both cases the \hi data are not corrected for stray
radiation. We note also that GASS data prior to the third data release
\citep{Kalberla2015} may suffer from residual correlator problems. These
can cause artificial linear structures oriented in scanning direction;
for details see \citet{Kalberla2015}.

\citet{Clark2014,Clark2015} used the Rolling Hough Transform (RHT), a
machine vision method for parameterizing the coherent linearity of
structures in the image plane. The RHT initially was designed for
detection of complex patterns in bubble chamber photographs.

The first step in this case is also to apply USM to the image. The image
is convolved with a two-dimensional tophat smoothing kernel of a
user-defined diameter, DK = 10\arcmin~ or 15\arcmin~ for Arecibo, and DK =
53\arcmin~ for Parkes data. Different in our analysis is that we use a
Gaussian smoothing kernel to convolve Effelsberg and Parkes data to an
effective resolution of 30\arcmin~ for both surveys. 
 
After subtracting the smoothed data from the original observations,
\citet{Clark2014,Clark2015} threshold the data at a 70\% level to
obtain a bit mask.  The derived features, called fibers, with a smooth
coherent structure and curvature, extend typically across angular scales
of a few degrees. In the next step of the analysis the position angle is
determined to allow a detailed comparison with the polarization angles
of the reflected stellar light and magnetic field directions. Several
fibers may exist at individual positions, crossing each other
\citep[Figs. 3 and 4 of][]{Clark2014}. To derive a projected RHT field
direction, \citet{Clark2015} average Stokes parameters derived at
different velocities.

Our focus here is different, we are going to identify the all-sky
distribution of the major \hi filaments without modifying the filaments
brightness temperature characteristics traced by its curvature or
restricting their angular length. For each position we consider only the
brightest \hi filament towards each individual line of sight. We do not
aim to determine position angles here.

\section{All-sky map of \hi filaments }
\label{allsky}

Figure \ref{Fig_Filaments2} displays the results of our analysis in
comparison to dust emission and polarization measured by {\it Planck\/}.
The top panel displays the dust emission from the HFI SkyMap at 353 GHz
(I\_STOKES from the public available
HFI\_SkyMap\_353\_2048\_R2.00\_full.fits in units of MJy sr$^{-1}$).  In
the middle panel the brightness temperatures $T_{\rm Bon}$ of the major
\hi filaments are displayed. This is a presentation of the \hi
emission from the gas that hosts the major filaments that we find in the
USM analysis. The lower panel of Fig. \ref{Fig_Filaments2} shows for
comparison a composite map of the dust emission as observed by {\it
  Planck\/} at 353 GHz. Superimposed is a texture derived from the 
measurements of the orientation of the polarized light emitted by the
dust, which in turn indicates the orientation of the magnetic field. The
map is from http://www.esa.int/spaceinimages/Images/2015/02/.

Visually the upper two maps show a high degree of correlation between
$T_{\rm B}$ and dust emission except for \hi filaments close to the
Galactic plane and in regions with very luminous \hi emission. The lower
map displays the global distribution of magnetic fields, in most cases
visually aligned by \hi and dust filaments. Many of the fainter \hi
filaments show even in details a high degree of correlation between
magnetic field lines an the \hi brightness temperature (see
Sect.\,\ref{details} in particular Figs.\,\ref{Fig_Detail1} and
\ref{Fig_Detail2}). The general impression is that \hi ridges are in
most cases better defined with more detailed substructures than the
emission from dust (top).  Because of \hi substructures, \hi filaments
appear to disclose a higher degree of correlation with the magnetic
field lines than the {\it Planck\/} map at the top.

\subsection{A detailed comparison between \hi and dust maps}
\label{details}

For a more detailed comparison between \hi and dust maps we inspect two
interesting regions with prominent filamentary
structures. Figs. \ref{Fig_Detail1} and \ref{Fig_Detail2} give detailed
presentations at $l = 10 \degr$, $b = 60 \degr$ and $l = 260 \degr$, $b
= -60 \degr$. It is obvious that the USM \hi data disclose more details
at a lower sensitivity limit. Despite the lower angular resolution of
the \hi data the \hi reveals much finer details. Next to the major
ridges we frequently find striations. The dust intensity map displays a
more diffuse distribution. Here, towards the faintest portions of the
{\it Planck\/} 353\,GHz map we identify intensity fluctuations caused by
the cosmic microwave background \citep{Planck2014}. Towards the bright
far infrared emission the apparent lack of small--scale structure of the
{\it Planck\/} data is most likely that the far--infrared emission
correlates quantitatively with both, the WNM and CNM. Using USM
techniques allows us to determine the velocity where the CNM filament is
most prominent. This defines $T_{\rm Bon}$ (see Sect. \ref{deriving})
and allows to trace the filament, even in case of velocity
gradients. Alternatively, an unbound integration over the line of sight
would add contributions from unrelated \hi gas.  So, the {\it Planck\/}
FIR continuum data is indeed limited by the angular resolution of the
satellite dishes while the radial velocity information of the single
dish \hi surveys allows to differentiate between the WNM and CNM
structures.

%-----------------------------------------------------------------------
\begin{figure*}[tbp]
\epsscale{1.}
   \centerline{
   \plotone{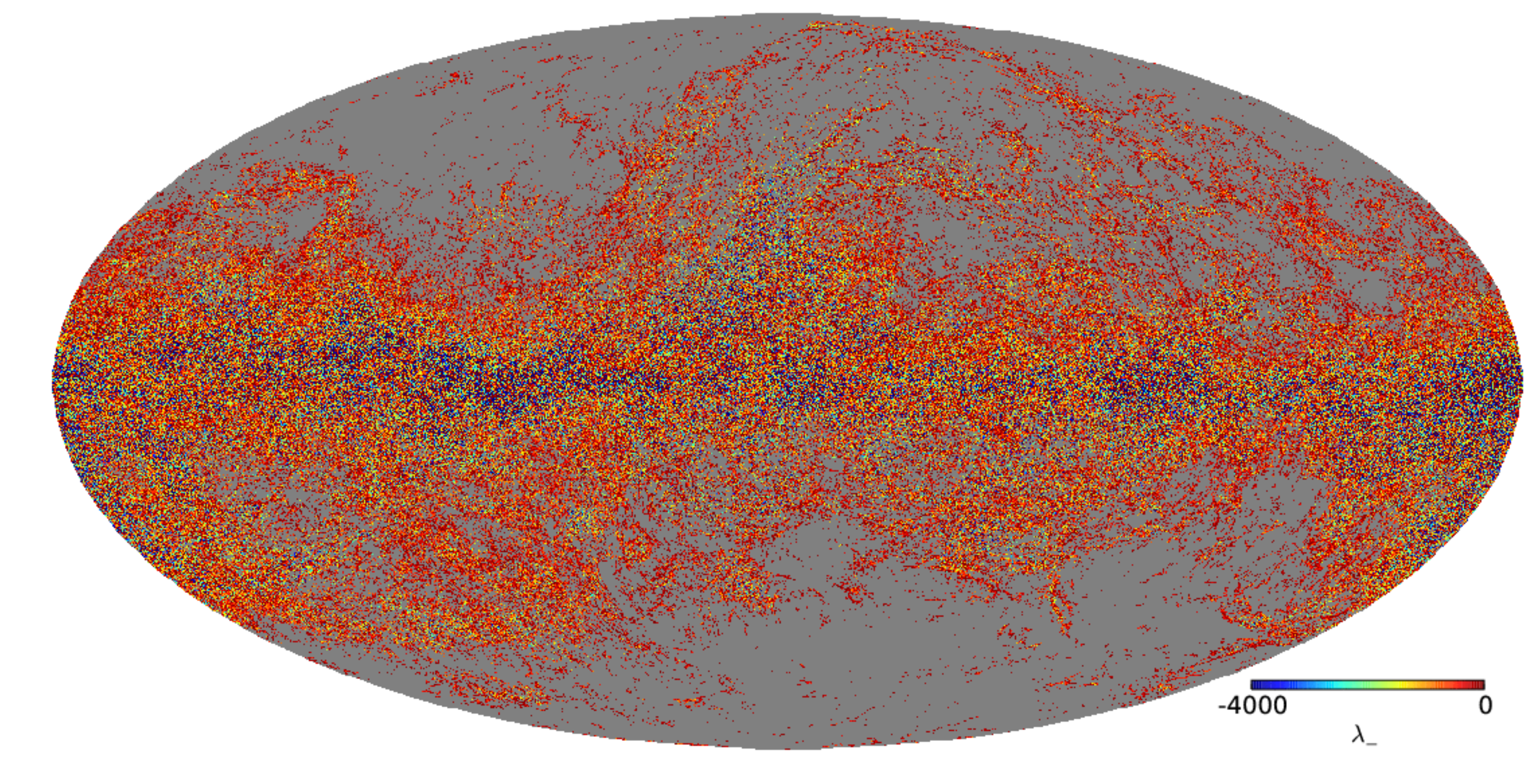}
}
   \centerline{
   \plotone{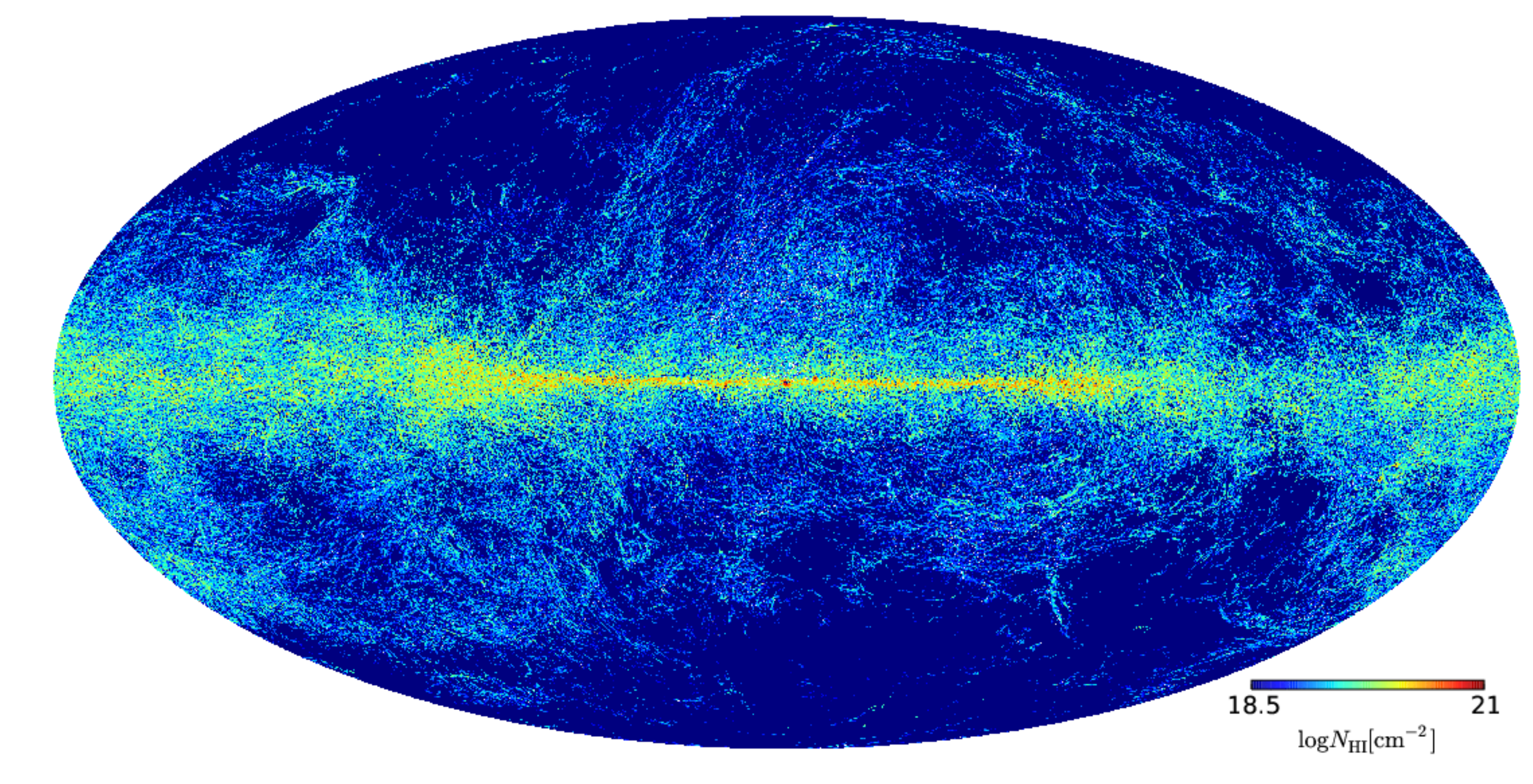}
}
   \caption{Derived properties for the major \hi filaments. {\it Top:}
     principal directions of least curvature, as determined from Hessian
     analysis, below column densities (lg$(N_{\rm H})$). }
   \label{Fig_Filaments3}
\end{figure*}

\begin{figure*}[tbp]
   \centerline{
   \plotone{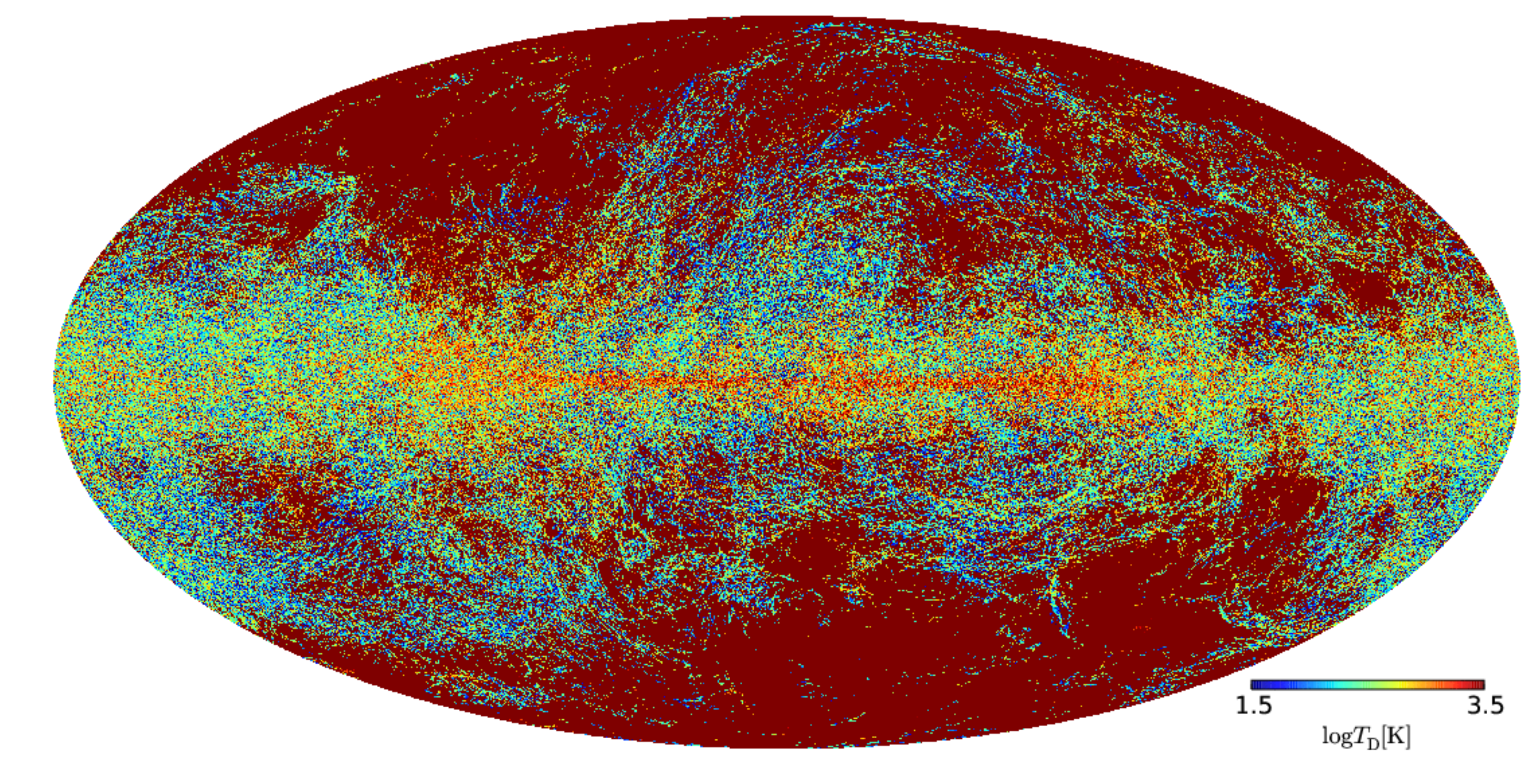}
}
   \centerline{
   \plotone{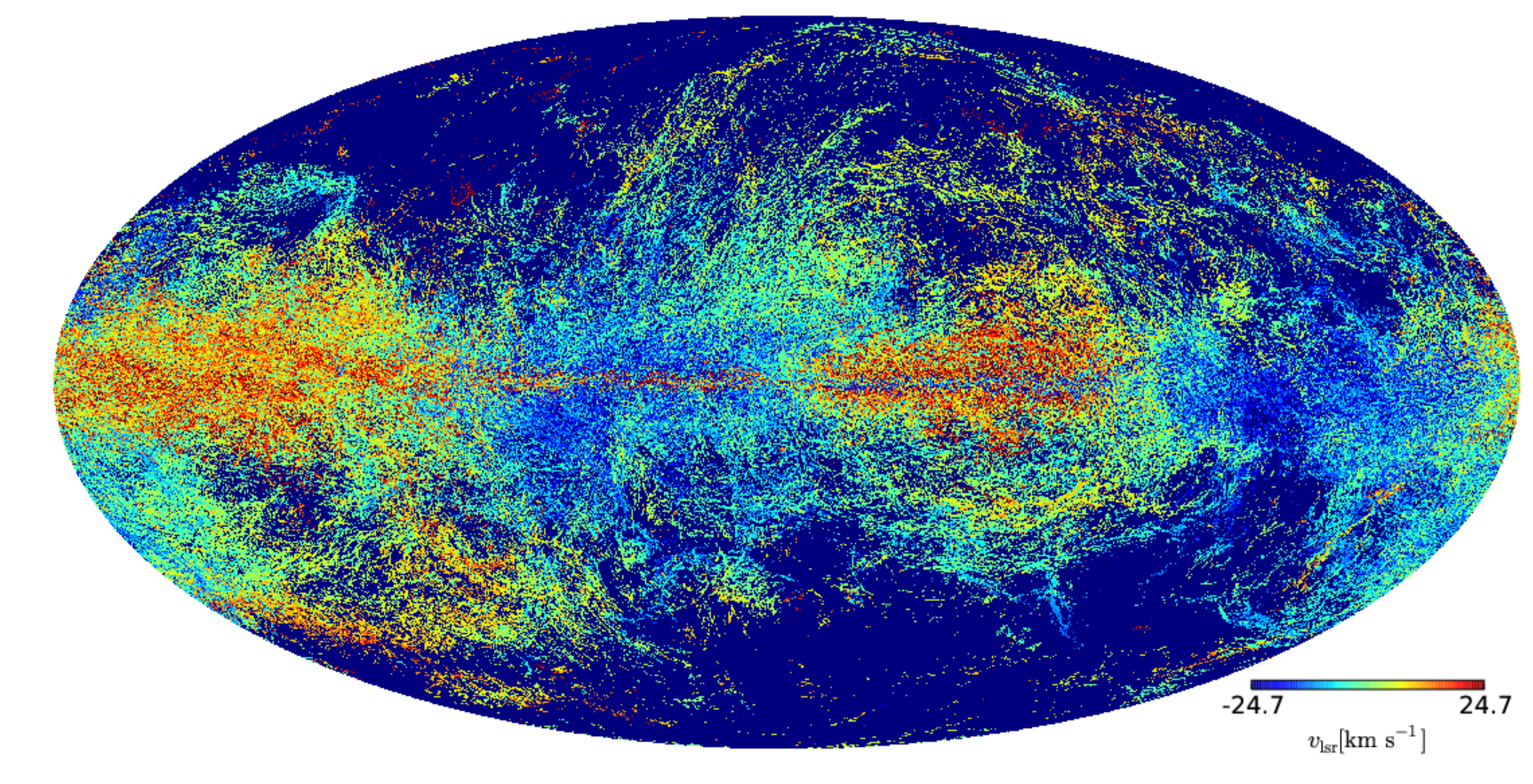}
}
\caption{Derived properties for the major \hi filaments, continued. {\it
    Top:} Doppler temperatures (lg$(T_{\rm D})$) and at the bottom the
  velocity field of the filaments (\kms). See text for a detailed
  description. }
   \label{Fig_Filaments3b}
\end{figure*}
%-----------------------------------------------------------------------

\section{Derived parameters }
\label{parameters}

In the following we will parameterize the properties of filamentary
structures. Figure \ref{Fig_Filaments3} and \ref{Fig_Filaments3b} gives
an all sky overview and displays in the top panel the principal
curvature image of the $T_{\rm B}$ distribution derived by us. Below we
plot column densities $N_{\rm H}$ for $T_{\rm Uon}$, next the derived
Doppler temperatures on logarithmic scale, lg$(T_{\rm D})$ (see
Sect. \ref{deriving} for definition of $T_{\rm D}$), and at the bottom
the velocity field of the filaments. In all cases we find well defined
structures. The purpose of this Fig. is only to highlight global
features. All details will be discussed in turn.

\subsection{Hessian analysis }
\label{Hesse}

Using the USM algorithm, we have derived a map of filamentary
structures. Now we discuss whether these structures may also be called
filaments in the sense of image analysis.

To classify the curvature of a feature within an intensity map along any
direction, the Hessian operator is a useful tool. \citet{Schisano2014}
and \citet{Planck2016} applied this method recently to
astronomical targets. For compatibility we use their notation.

The Hessian matrix is defined as 
\begin{equation}
     \label{eq:hessI} 
        H(x,y)\, \equiv \, \left ( \begin{array}{cc} H_{xx} & H_{xy }\\
            H_{yx} & H_{yy} \end{array} \right ),
\end{equation}  
here {\it x} and {\it y} refer to true angles in latitude $x=b$ and
longitude $y=l\,\cos{b}$. The second-order partial derivatives are
$H_{xx}=\partial^2 T_{\rm Bon} / \partial x^2$, $H_{xy}=\partial^2 T_{\rm Bon}
/ \partial x \partial y$, $H_{yx}=\partial^2 T_{\rm Bon} / \partial
y \partial x$, $H_{yy}=\partial^2 T_{\rm Bon} / \partial y^2$.

The eigenvalues of H, 
\begin{equation}
\label{eq:lambda}
\lambda_{\pm}=\frac{(H_{xx}+H_{yy}) \pm \sqrt{(H_{xx}-H_{yy})^2+4H_{xy}H_{yx}}}{2},
\end{equation}
are often denoted as principal directions and describe the local
curvature of the features; $\lambda_-$ is in direction of least
curvature.

We determine the derivatives for $T_{\rm Bon}$
(Fig. \ref{Fig_Filaments2}, middle panel) from symmetric differential
quotients. Due to the interleaved structure of the HEALPix database it
is necessary to interpolate the gridded data in Galactic longitudes, we
use a cubic spline interpolation.%, similar to the interpolation
%of the velocity vector as discussed in Sect. \ref{data_sets}.

Figure \ref{Fig_Filaments3} (top) displays the principal curvature
image, derived from $\lambda_-$. Images of this kind are often generated
for ridge detection in the fields of computer vision and image
analysis. Essentially this image reproduces structures seen in the $N_{\rm H}$
map below or in the middle panel of Fig. \ref{Fig_Filaments2} and may be
compared with Fig. 3 of \citet{Planck2016}. 

\begin{figure*}[tbp]
   \includegraphics[scale=0.6, angle=270]{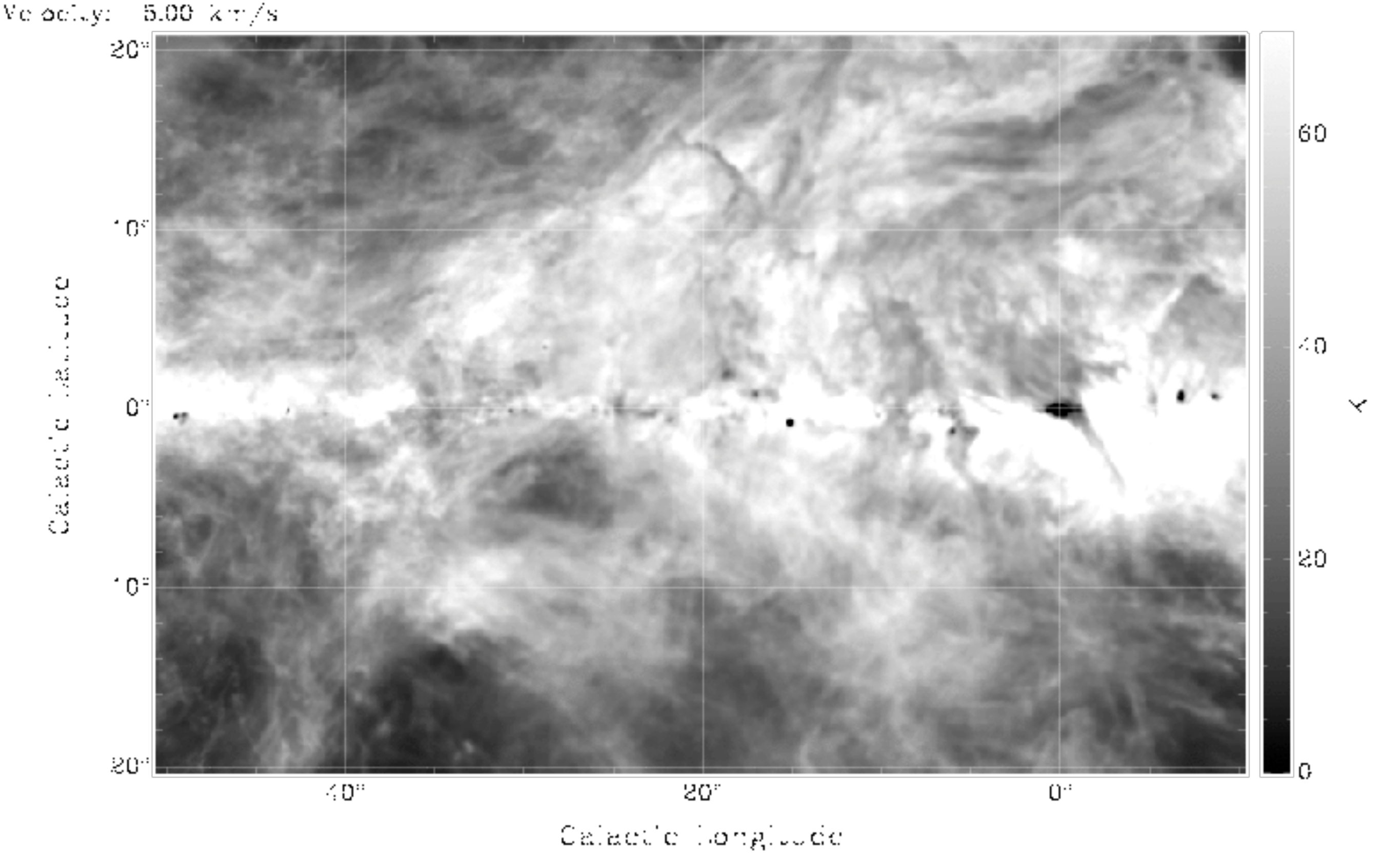}
   \includegraphics[scale=0.6, angle=270]{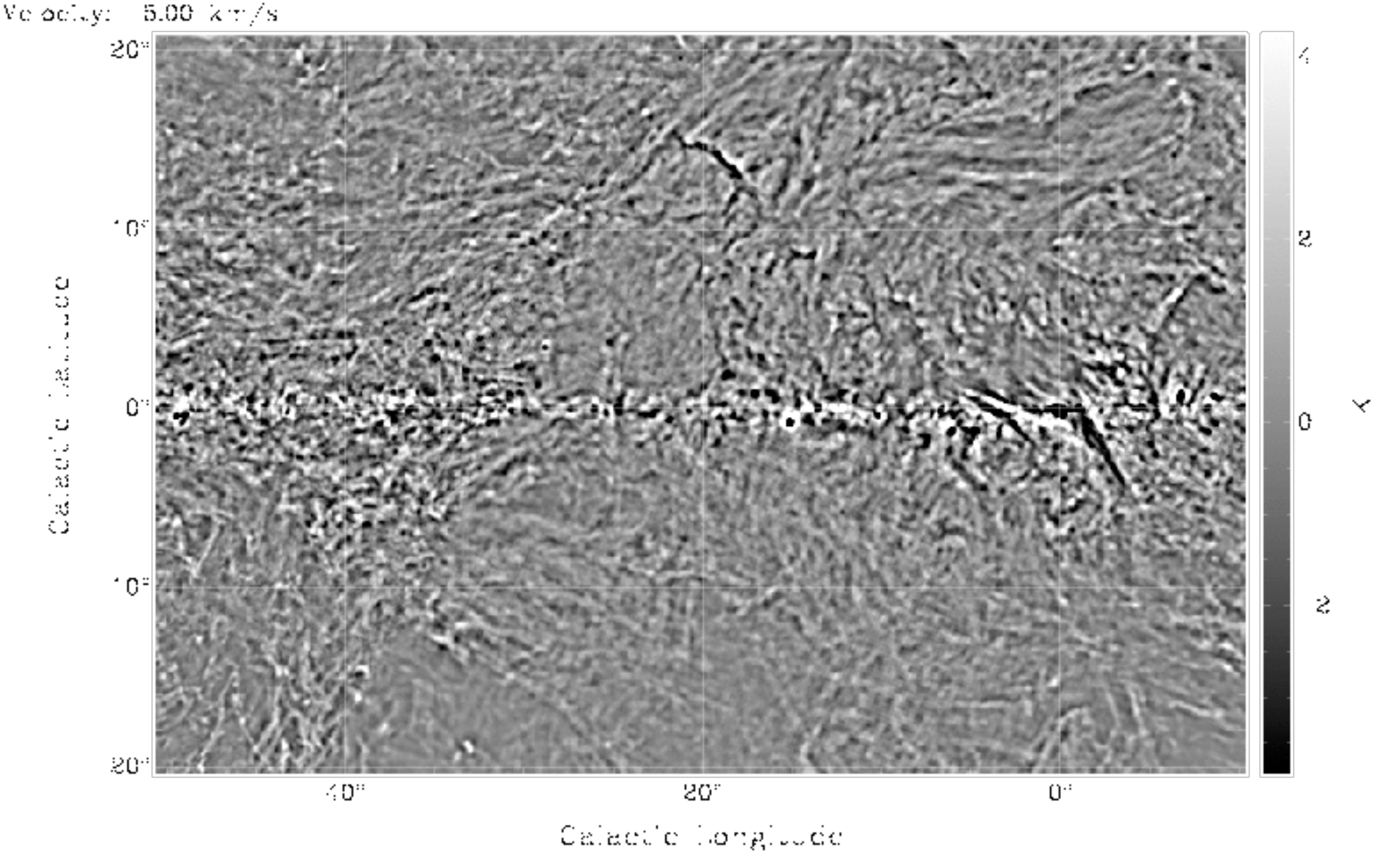}
   \caption{Brightness temperature (top) and USM filaments (bottom) 
     for single channels at a velocity of 5 \kms.  }
   \label{plane}
\end{figure*}

%-----------------------------------------------------------------------
%\begin{figure*}[btp]
\begin{figure*}[bht]
\epsscale{0.48}
   \centerline{
   \plotone{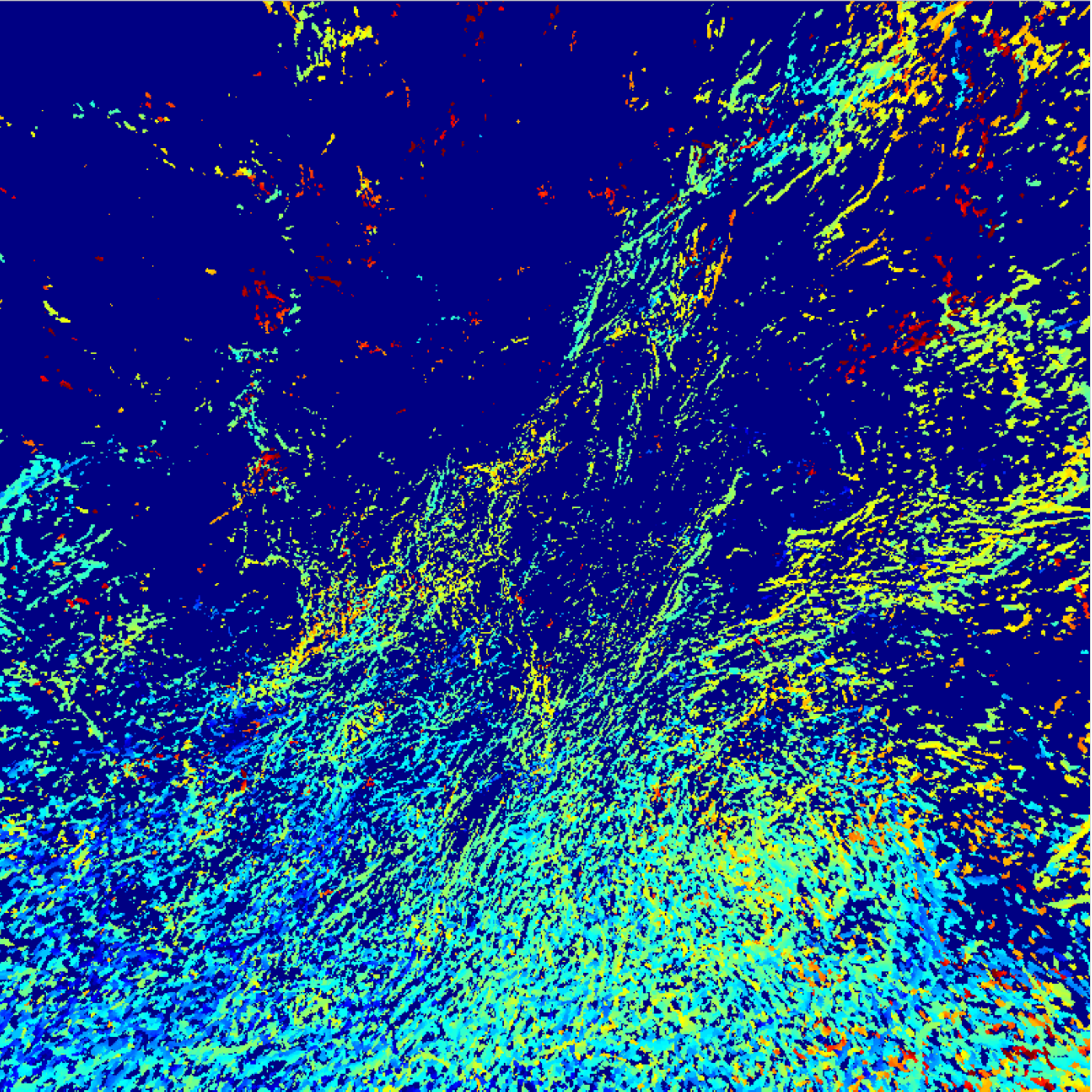}
   \plotone{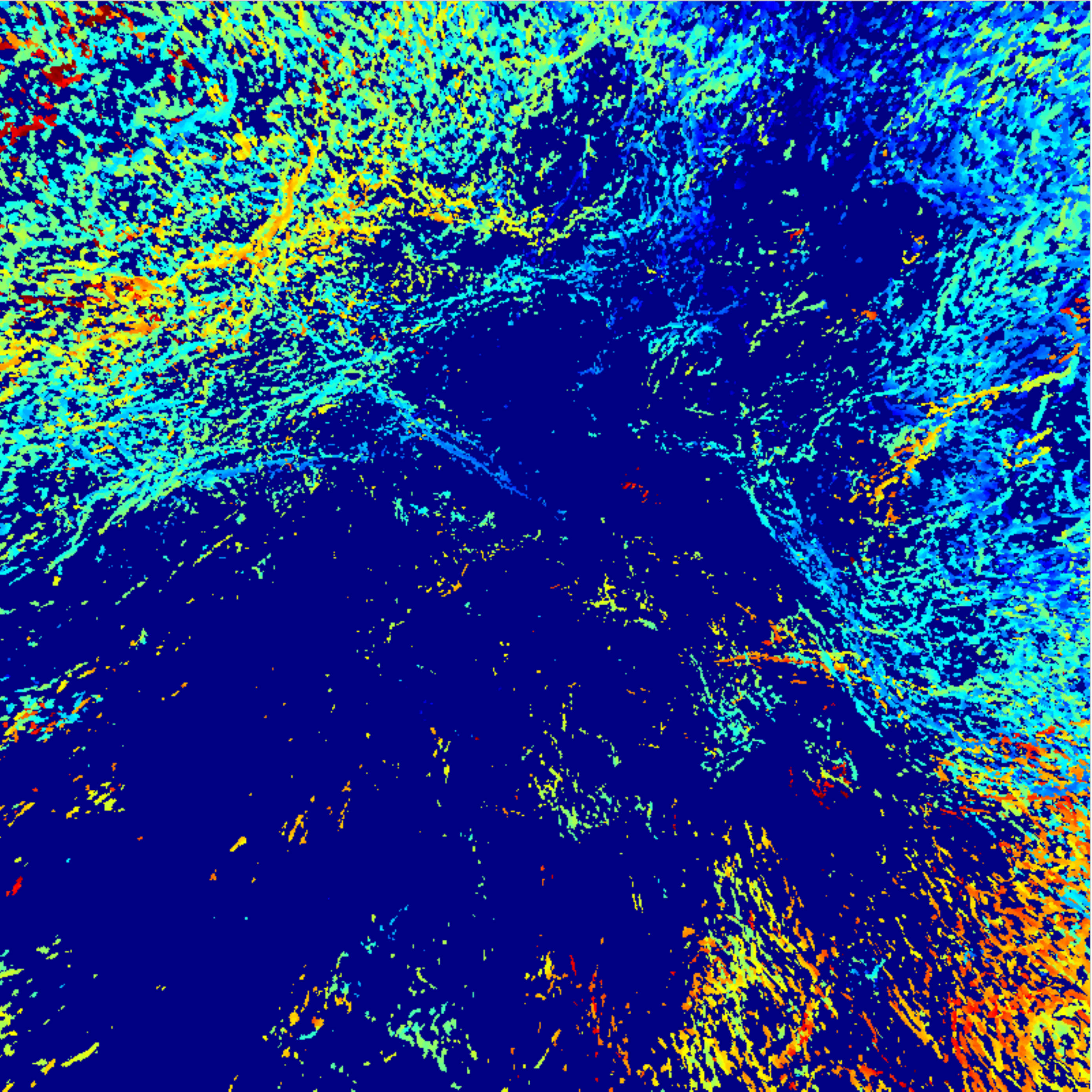}
}
\caption{To demonstrate the dominance of coherent velocity structures
  the radial velocity distribution of the major \hi filaments for the
  regions selected in Figs. \ref{Fig_Detail1} and \ref{Fig_Detail2} are
  displayed. Both plots display about $100 \degr \times 100\degr$ in
  gnomonic projection.  {\it Left:} centered at $l = 10 \degr$, $b = 60
  \degr$ , {\it Right:} centered at $l = 260 \degr$, $b = -60
  \degr$. Displayed is the full radial velocity ranged analyzed between
  $|v_{\rm LSR}|\leq\,25~ {\rm km\,s^{-1}}\,$. The color coding
  is the same as in Fig.\,\ref{Fig_Filaments3b} bottom panel.}
   \label{Fig_Detail3}
\end{figure*}
%\end{figure*}
%-----------------------------------------------------------------------

\subsection{The Galactic plane}
\label{sect_plane}
 
The Hessian analysis is sensitive to point-to-point fluctuations. The
noisy blue dotted regions close to the Galactic plane indicate that our
basic concept, to define major filaments that are dominating the \hi
distribution, approaches its limits towards low Galactic latitudes. This
does not imply that there are no well defined filaments in the 
  Galactic plane. The opposite is true, there are too many filaments
that overlay each other, causing confusion.

Figure \ref{plane} shows as an example an extended region centered at
longitude $l = 20\degr$ at a velocity of 5 \kms. To the left we plot the
observed brightness temperature, to the right the USM image. This region
shows numerous filaments crossing each other. Even within a single
spectral channel the confusion is severe. It is worth to note several
prominent diagonal features that are due to \hi self-absorption. Close
to the Galactic center position is the Riegel-Crutcher Cloud
\citep{Riegel1972,Clark2014}, another structure, almost parallel to the
the Riegel-Crutcher Cloud, is at $l \sim 6\degr$ and at $l \sim 18\degr,
b\sim 13\degr$. These structures are dark because the \hi is optically
thick, leading to self-absorption.

We conclude that the confusion in the principal curvature image is
caused by confusion from crowded \hi filaments that are superposing each
other. The affected regions in Fig. \ref{Fig_Filaments3} (top) appear
mostly to be spatially correlated to regions with a high polarization
angle dispersion function S \citet[see][Fig.\,12]{PlanckXIXa}.

To avoid biases due to confusion by superposition, we focus in the
following our analysis to latitudes $|b| > 20\degr$. The scale height of
this gas is limited, hence the observed features are relatively simple
with a low probability of confusion. However, for comparison we give
also values for the whole sky, to demonstrate that the local
interstellar medium at $|b| > 20\degr$ represents well the full sky \hi
gas distribution.

\subsection{Filament velocities on large scales}

The all sky velocity field of the USM filaments is shown in Fig.
\ref{Fig_Filaments3b} (bottom). In Fig. \ref{Fig_Detail3} some more
details on the velocities for the USM filaments displayed in
Figs. \ref{Fig_Detail1} and \ref{Fig_Detail2} are shown. All together we
find that filaments share mostly similar velocities but occasionally
there are significant gradients in radial velocity perpendicular to or
along the filaments. A more detailed discussion on velocity gradients is
given in Sect. \ref{orientation}. Detailed USM channel maps, showing
velocity gradients, are given in the Appendix.

\subsection{Filament shapes and orientations}
\label{orientation}

\begin{figure*}[tbp]
   \centerline{
   \includegraphics[scale=0.33, angle=270]{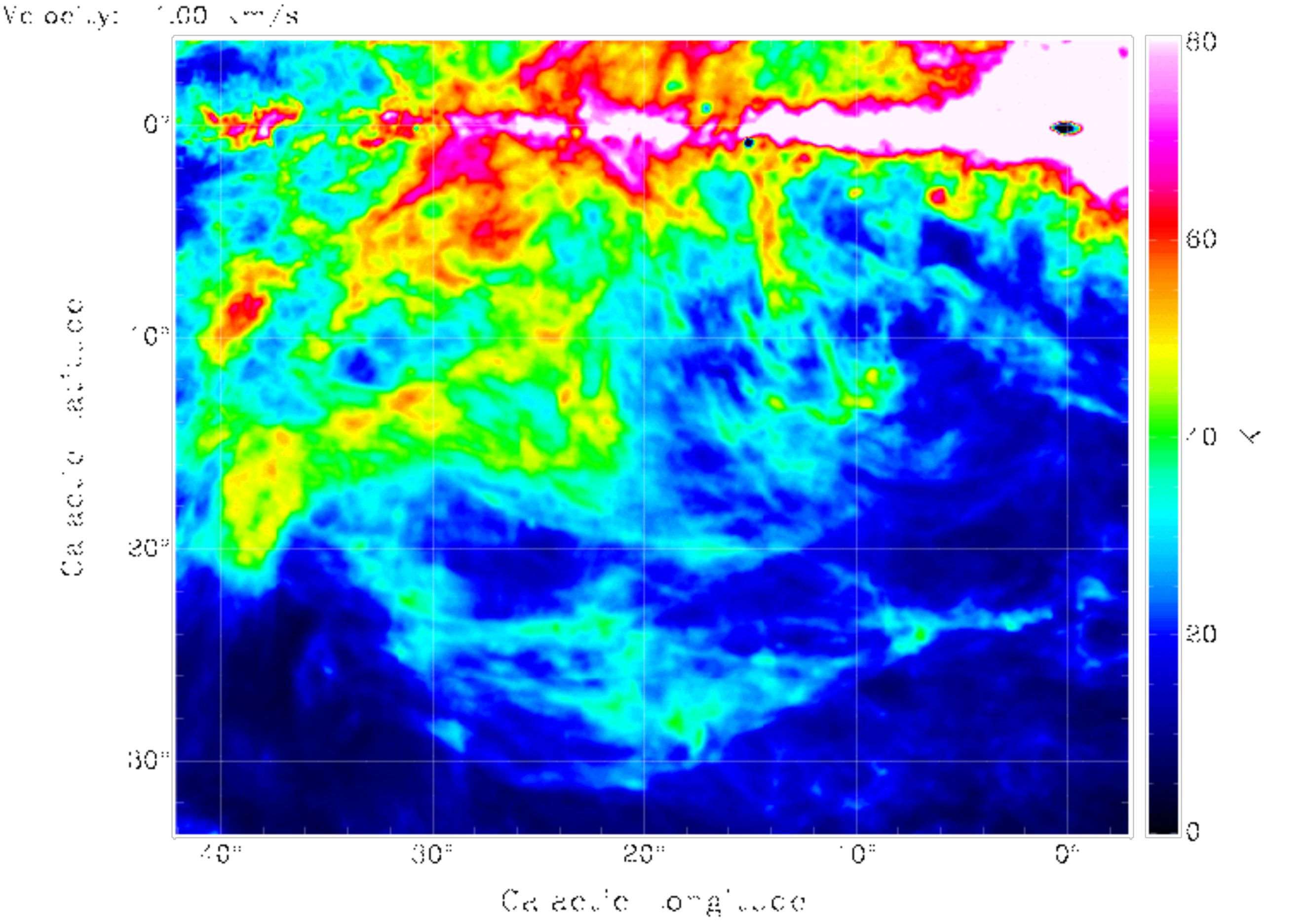}
   \includegraphics[scale=0.33, angle=270]{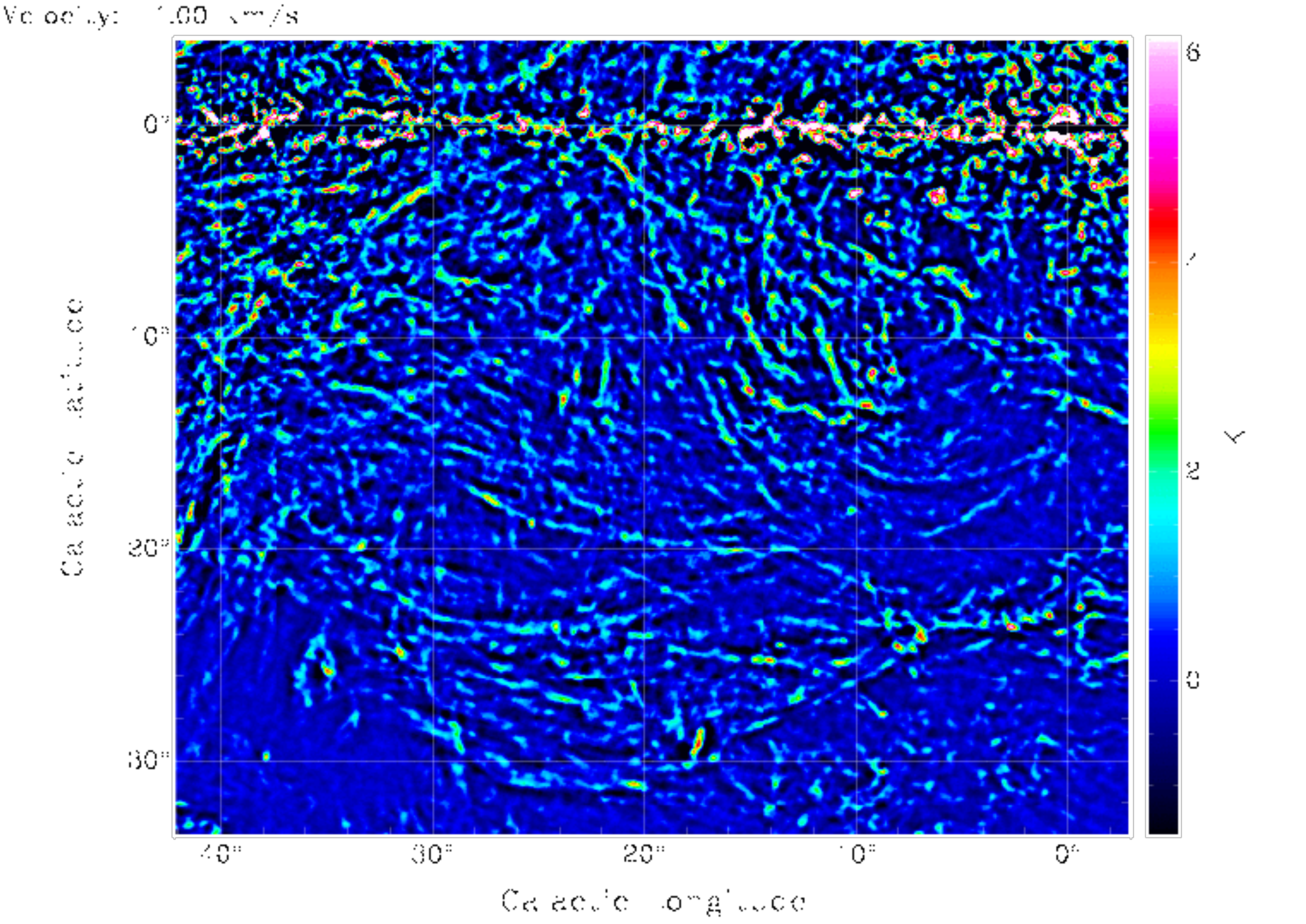}
}
   \caption{Brightness temperature (left) and USM filaments (right) 
     for a single channel at a velocity of 1 \kms.  }
   \label{FIG_orientation}
\end{figure*}

We use Fig. \ref{FIG_orientation} to discuss the observed orientation of
filaments with respect to the Galactic plane. Close to the plane it is
not easy to identify regions of low confusion levels. This example was
chosen to demonstrate that most of the filaments have thin arc-like
structures but may run in different directions.  Close to the Galactic
plane we find filaments without preferred directions. Off the plane,
here roughly at $b \sim -20\degr$, we find filaments that are bent
convex, suggestive for an activity of blast waves originating from
supernovae in the plane.

\subsubsection{Fibers or sheets?}
\label{sheet}

\begin{figure*}[tbp]
   \centerline{
   \includegraphics[scale=0.3, angle=270]{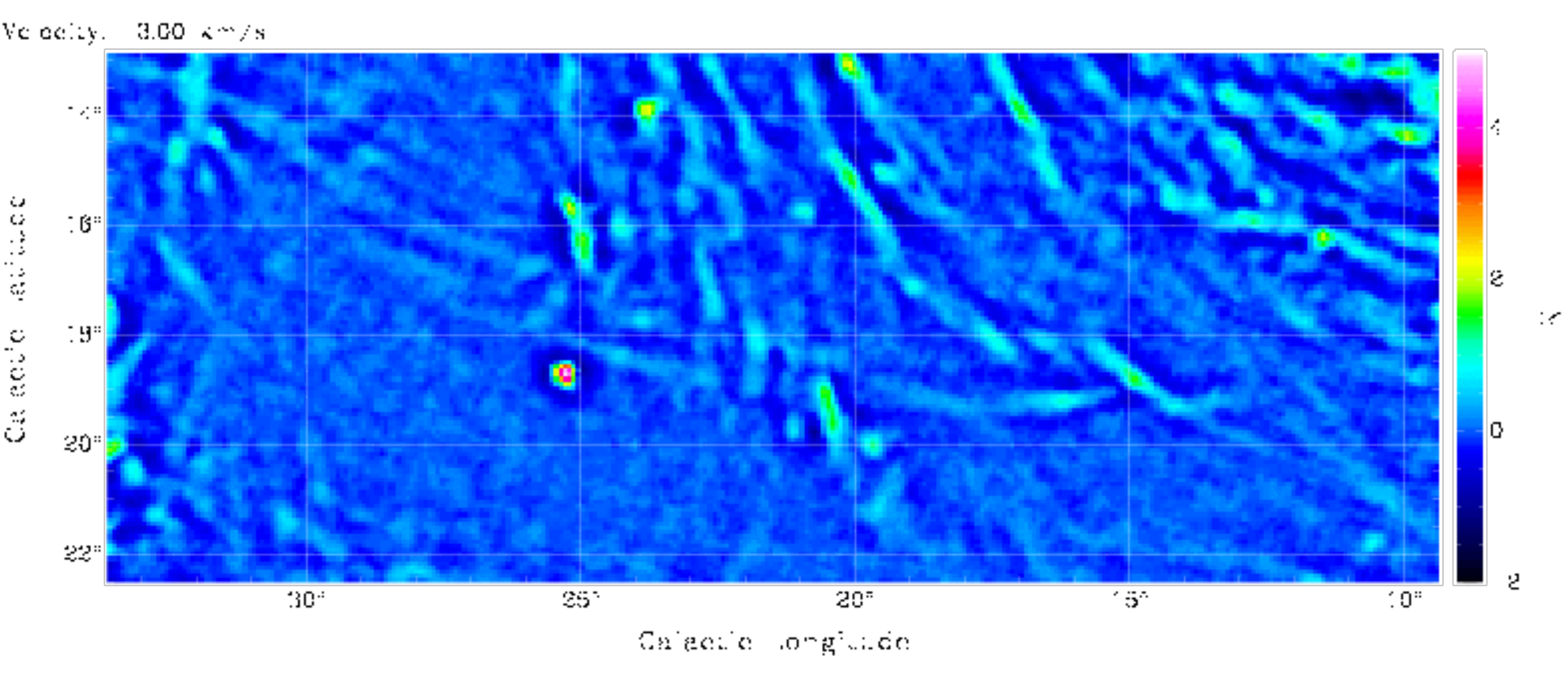}
   \includegraphics[scale=0.3, angle=270]{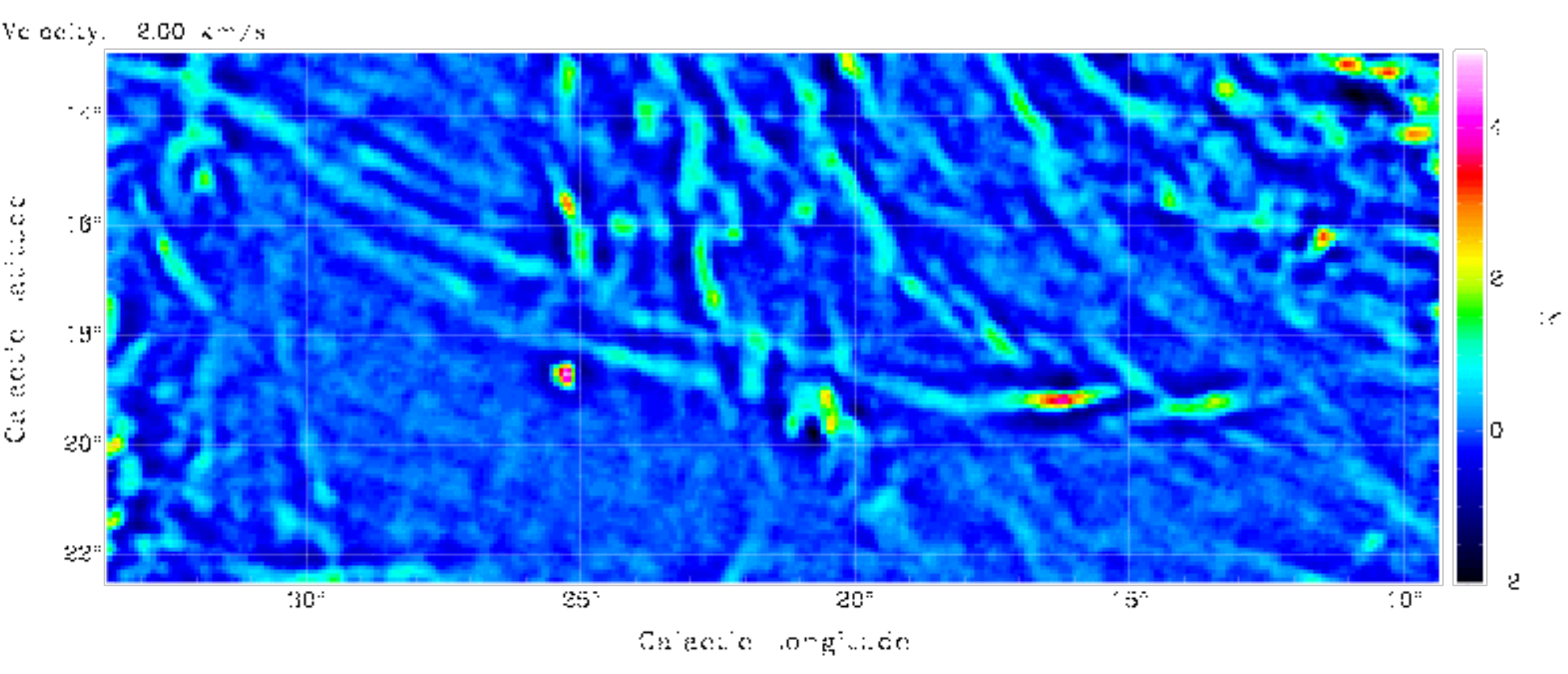}
}
   \centerline{
   \includegraphics[scale=0.3, angle=270]{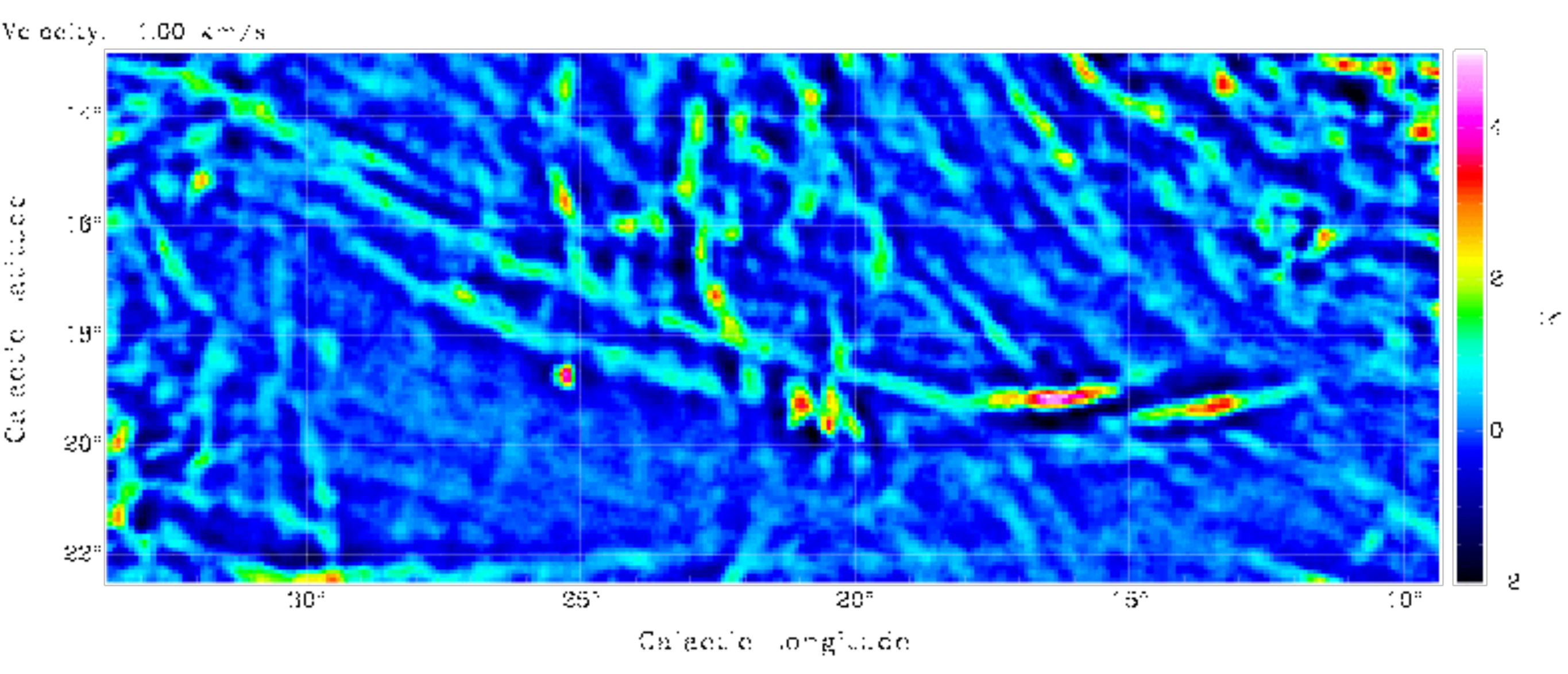}
   \includegraphics[scale=0.3, angle=270]{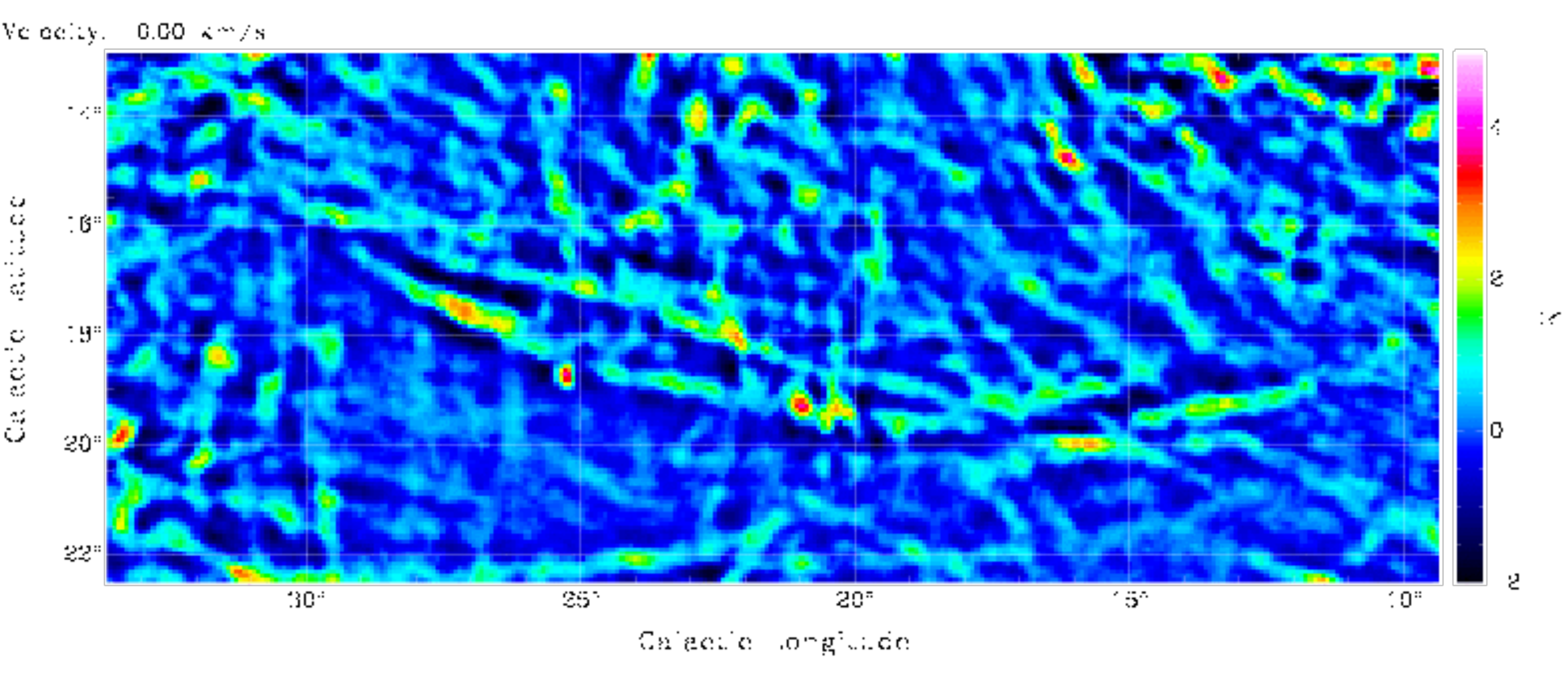}
}
   \centerline{
   \includegraphics[scale=0.3, angle=270]{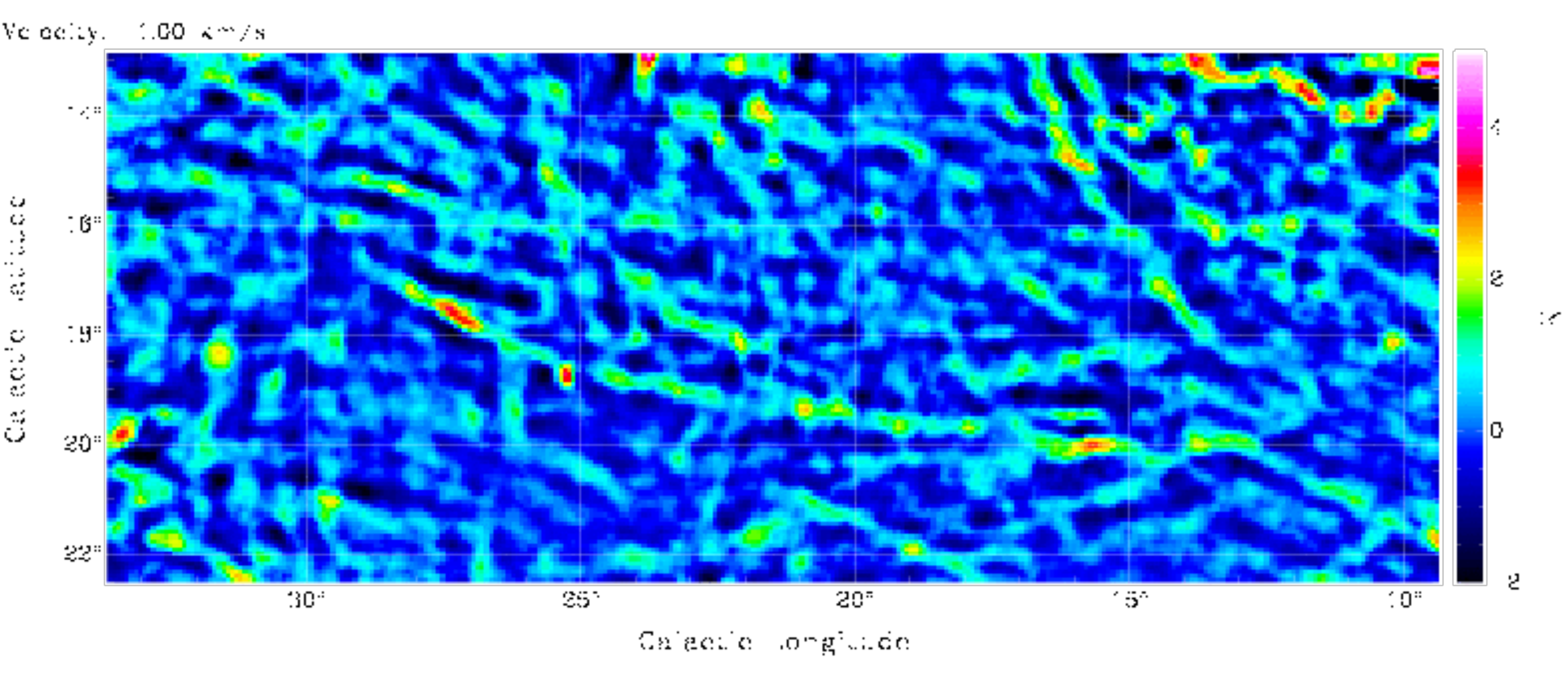}
   \includegraphics[scale=0.3, angle=270]{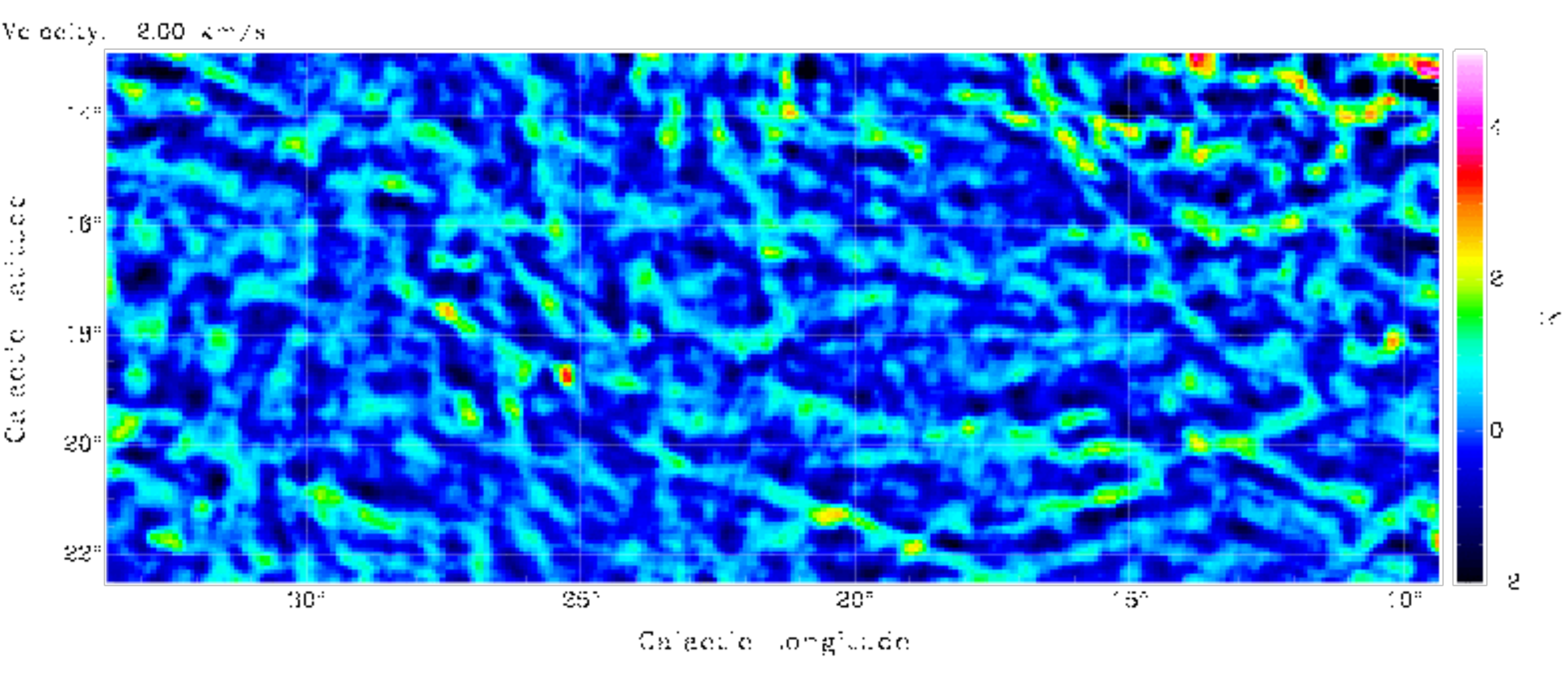}
}
   \caption{USM filaments for a single channel at velocities between -3
     and +2 \kms.  }
   \label{FIG_filament}
\end{figure*}

Figure \ref{FIG_filament} displays details of a pair of filaments
located within Fig. \ref{FIG_orientation}; both filaments having a
length of approximately $15\degr$. Tracing these features shows that
they develop subsequently in position-velocity space. The ridges appear
to move from channel to channel somewhat less then a beam width, in
total 0\fdg5 over 5 \kms~ before they disintegrate.

Apparent position shifts of \hi structures with channel velocity can be
caused by systematic velocity shifts of the gas within a more extended
distribution, e.g. as for rotation curves of Galaxies.  Significant
position changes from one channel to the next, however, can then only be
observed for a sufficiently low velocity dispersion of the gas along the
line of sight. For a channel spacing around 1.3 \kms~ this implies that
the Doppler temperature within the gas layer has to be $T_{\rm D} \la
200 $ K. The corresponding FWHM for turbulent motions in velocity is
$\delta v_{\rm Uon} \la 2 $ \kms, only a factor two to five lower than
the total correlation width $ 5 \la \Delta v_{\rm Uon} \la 10 $ \kms~
that limits the observability of the filaments (see more examples in the
appendix). Outside this velocity range the left-over fragments barely
resemble a filamentary origin.

Filamentary features of this kind have been described as
``morphologically obvious structures'' by \citet{HeilesCrutcher2005};
meaning fibers (strings) or edge-on sheets. But how to distinguish
between fibers and sheets?

Fibers in the ISM may be caused by sub-Alfvenic anisotropic turbulence
\citep{Stone1998}. In simulations such filaments are found to be aligned
along the magnetic field lines. That is what we observe but it is unclear
how such a model could explain an apparent {\it continuous} position
shift with velocity.

\citet{Hennebelle2013} studied non-self-gravitating filaments in the
ISM. He used MHD simulations to study the
formation of clumps in various conditions. It was found that filaments
are in general preferentially aligned with the strain due to the stretch
induced by turbulence. These filaments survive longer in case of
magnetized flows. A strain must affect bulk velocities within filaments.
In this case the velocity gradient is along the filament. 

According to \citet{HeilesCrutcher2005} edge-on sheets should be edge-on
seen shocks in which the field is parallel to the sheet. Such sheets are
observable as filaments when the line of sight becomes nearly parallel
to the sheet. Figures \ref{FIG_orientation} and \ref{FIG_filament}
suggest that the filaments are caused by blast waves originating from
the Galactic plane. The observed bent convex shapes imply that the
filaments are part of a shell. We are observing the shell in almost
tangential direction. It is then plausible that we have also some bending
in position along the line of sight with an associated continuous change
in projected radial velocities, mimicking a position shift perpendicular
to the observed filaments. In such a scenario edge-on sheets should be
edge-on shocks in which the field is parallel to the sheet, explaining
low observed Doppler temperatures.

We use henceforth the notation of \citet{HeilesCrutcher2005,Heiles2005}
and assume that the total column density perpendicular to the sheet is
$N_{\perp}$. If the normal vector to the sheet is oriented at angle
$\Theta$ with respect to the line of sight, we observe
 
\begin{equation}
\label{eq:Nfilament}
N_{\rm obs} = N_{\perp} / {\rm cos}(\Theta)
\end{equation}

In a similar way, if the motion of the sheet is also seen in perpendicular
direction, with a velocity $v_{\perp}$, we observe
\begin{equation}
\label{eq:vfilament1}
v_{\rm obs} = v_{\perp}  {\rm cos}(\Theta).
\end{equation}
The observed turbulent velocity dispersion along the line of sight
due to projection is given as
\begin{equation}
\label{eq:vfilament2b}
\sigma(v_{\rm turb, los}) = \sigma(v_{\rm turb}) {\rm sin}(\Theta)
\end{equation}
\citep[][Eq. 12]{Heiles2003b}.

For an angle of $\Theta \sim 90\degr$, tangential to the sheet or shell,
we obtain favorable conditions to observe high column densities $N_{\rm
  obs}$. Considering line velocities, we observe velocity
crowding \citep{Burton1971} under the same condition. This additional
condition is particularly important in case of low Doppler temperatures.
\begin{equation}
\label{eq:vfilament3}
N_{\rm obs}(v_{\rm obs}) = N_{\perp}(v_{\perp}  {\rm cos}(\Theta)) / {\rm cos}(\Theta)
\end{equation}
The observed column densities are strongly velocity dependent and the
radial velocities themselves, as bulk velocities within the sheets,
depend on projection effects. For nearly tangential viewing, velocity gradients
are favorable for an alignment of $N_{\rm obs}(v_{\rm obs})$ in filaments.

\subsection{Surface filling factor,  mass fraction }
\label{surface}

For a determination of surface filling factor and mass fraction of the USM
filaments we use major filaments as defined in Sect. \ref{data_analysis}. 

For the best defined USM filaments at $|b| > 20\degr$ with $T_{\rm Uon}
> 1$K, corresponding to a $10 \sigma$ threshold, we find that 30\% of
the sky is covered by filaments. Lowering the threshold to 0.3 K, the
coverage increases to 60\%. Repeating the analysis with a 0.1 K
constraint and including in this case also the Galactic plane we
estimate that 95\% of the sky is occupied. Filaments are ubiquitous.

\begin{figure}[tbp]
   \centerline{
\epsscale{0.6}
   \plotone{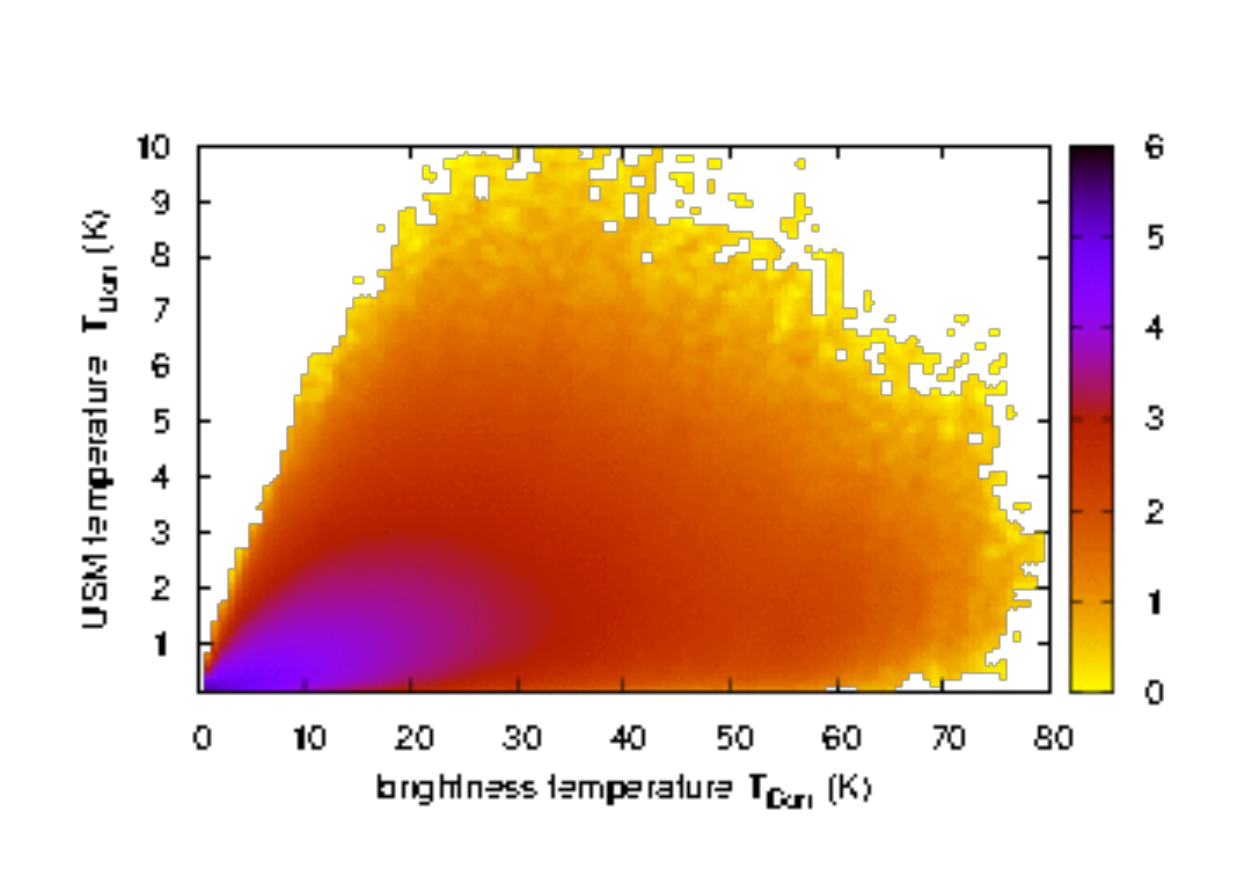}
}
   \caption{2-D density distribution for the temperature of USM
     filaments in relation to brightness temperatures of the \hi gas at
     the same velocity. The look-up table is logarithmic. }
   \label{TB_USM}
\end{figure}

The relation between $T_{\rm Uon}$ and $T_{\rm Bon}$ for all positions
$|b| > 20\degr$ is displayed in Fig. \ref{TB_USM}. We find a large scatter,
most of the filaments are found in regions with low brightness
temperatures $T_{\rm Bon}$, correspondingly the $T_{\rm Uon}$ values are low.

We estimate the mass fraction of the \hi gas in major filaments from $
F_m = \sum T_{\rm Uon} \Delta v / \sum T_{\rm Bon} \Delta v $,
integrating over all positions with $|b| > 20\degr$. For a thresholded
$T_{\rm Uon} > 1$K we find $ F_m = 0.1$ and for threshold of 0.3 K we
get $ F_m = 0.08$. Our derivation of the mass fraction assumes that the
gas in filaments as well as the surrounding warm medium is optically
thin, for detailed discussion see Sect. \ref{depth}.  The filling
  factor $F_m$, derived so far, considers only major filaments. There
  may be more than a single filament along the line of sight, we need
  accordingly to determine $F_{\rm tot} = \sum T_{\rm USM} \Delta v / \sum
  T_{\rm B} \Delta v $, integrating unconstrained across all significant
  filaments along the line of sight. This way we obtain a total fraction
  $F_{\rm tot} = 0.20 \pm .01$. 

\begin{figure}[hbt]
\epsscale{0.7}
   \centerline{
   \plotone{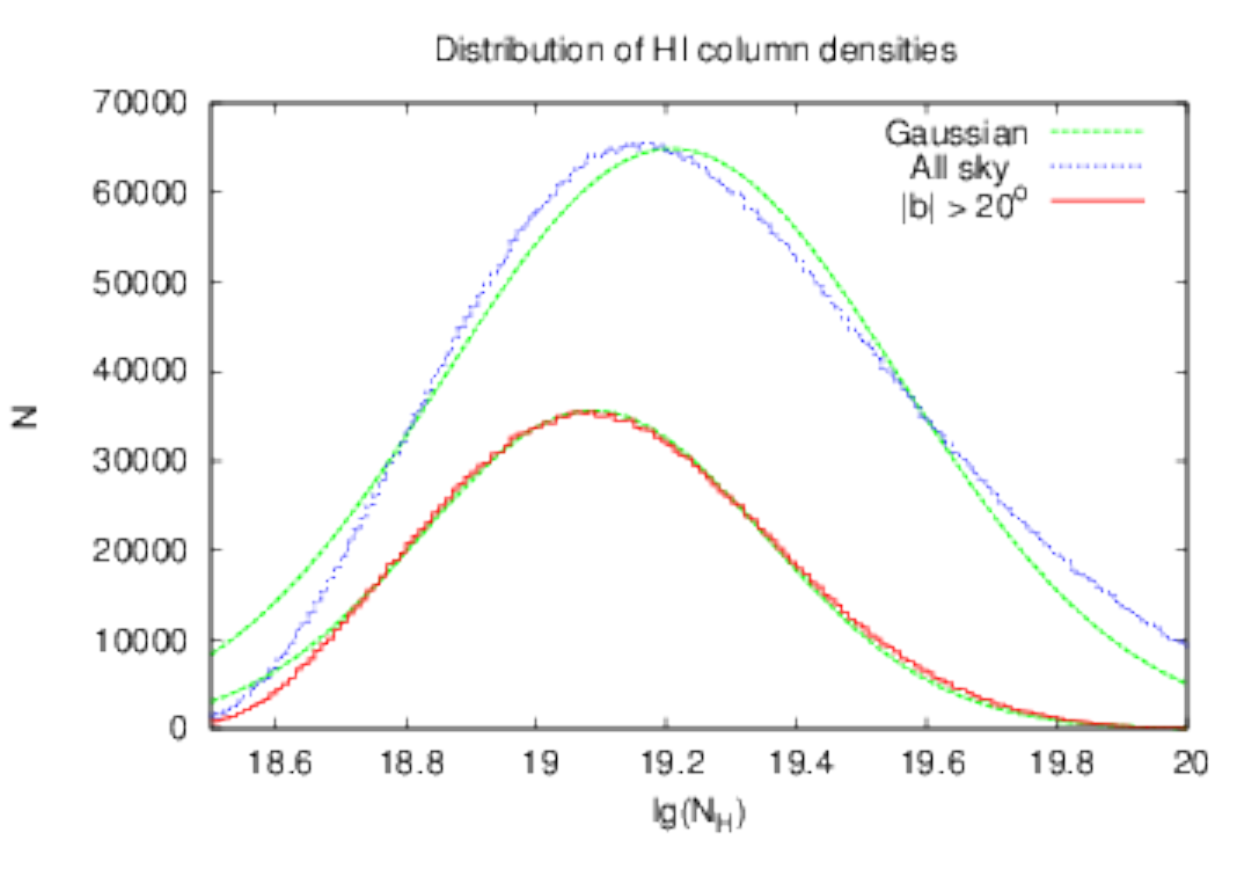}
}
\caption{Column density distribution for the major local USM \hi
  filaments, compared with Gaussian fits. The median 
    for $|b| > 20\degr$ is $\rm{lg}(N_{\rm H}) = 10^{19.1}$. }
   \label{Fig_NH_histo}
\end{figure}

\subsection{Column densities}
\label{column}

Next we derive column densities for the \hi gas in filaments, 
  assuming that the gas is optically thin. Possible biases due to 
  optical depth effects are discussed in Sect. \ref{depth}. We use
a $10 \sigma$ threshold of 1 K.

Figure \ref{Fig_NH_histo} displays the probability density
  distribution (PDF) for the filament column densities. This plot shows
  a log-normal distribution with a well defined maximum at $N_{\rm H} =
  10^{19.087}$ for the gas with $|b| > 20\degr$ with a dispersion
  $\sigma(lg(N_{\rm H})) = 0.265$.  The median\footnote[1]{For a turbulent medium the characterisic PDFs are
    log-normal \citep{Vazquez1994} as a result of the central limit
    theorem applied to self-similar random multiplicative
    perturbations. The geometric mean of a log-normal distribution is
    equal to its median. Reading off peak values from plots of PDFs
    published on a linear scale leads to biased results. } column
  density is $N_{\rm H}
  = 10^{19.093}$. Including the Galactic plane leads to an increase of
the column densities, also the peak shifts somewhat. This shift implies
that confusion, discussed in Sect. \ref{sect_plane}, tends to
increase the apparent column densities of the filaments. This
  distribution deviates from a simple log-normal PDF.

  \citet{Clark2014}, in their analysis of a region that covers about
  1300 degr$^2$, considered only a single characteristic fiber within
  this field. They mention a typical column density $N_{\rm H} \sim
  10^{18.7}$, only 25\% different from our result.

\subsection{Volume densities from distance estimates}
\label{volume}

A standard assumption is that column densities around $N_{\rm H} <
10^{20}$ cm$^{-2}$ are optically thin, but optical depth effects may be
significant (see Sect. \ref{depth}), causing volume density
biases. Another problem is beam dilution. Assuming a distance of
100\,pc, the median distance to the wall of the local cavity
\citep[][from color excess measurements]{Lallement2014}, the FWHM
  of the telescopes corresponds to 0.3 pc and is therefore a limit to
  the spatial resolution.  

  \hi ridges appear to move from channel to channel somewhat less than a
  beam width (Sect. \ref{sheet}). This implies that within the telescope
  beam the line of sight velocities of the \hi cannot differ much more
  than the channel width. For an \hi sheet with isotropic turbulence the
  velocity dispersion across the beam has to be similar to the
  dispersion in perpendicular direction, along the line of sight. In
  turn, since in turbulent media scale length and velocities are related
  to each other, the depth along the line of sight should be comparable
  to the one in perpendicular direction. In presence of velocity gradients
  across the beam the derived depth is an upper limit only.

  For an estimate of the volume density we assume hence a filament with a
  thickness of 0.3 pc. For a typical column density $N_{\rm H} = 10^{19.1}$
  cm$^{-2}$ we obtain a volume density of $n > 14$
  cm$^{-3}$. \citet{Clark2014} derive a similar value from Arecibo
  survey data but later in Sect. \ref{width} we will argue for a
  thickness of $\sim 0.1$ pc and a density $n \sim 47$ cm$^{-3}$. 

  Different environmental conditions were assumed by
  \citet{Planck2016} for the gas associated with dust. Based on
  a local scale height of 100\,pc they assume an average filament
  distance of 430\,pc. This corresponds to a thickness of about 1\,pc,
  adopting spherical symmetry and an angular extent of the filament
  equal to the size of the antenna beam. We obtain in this case a volume
  density of 4\,cm$^{-3}$. Assuming standard gas-to-dust conversion,
  \citet{Planck2016} derive an average \hi volume density of
  300 cm$^{-3}$. At the first glance there appears to be a huge
  discrepancy between both values. But we need to consider that the {\it
    Planck\/} data trace all gaseous species while the USM data selects
  the CNM. Applying the previously determined factor $F_M = 0.09$ we
  obtain a volume density of 27 \,cm$^{-3}$, only a factor of two
  different from our first estimate which was based on a filament
  distance of 100\,pc.

\begin{figure}[tbp]
\epsscale{0.7}
   \centerline{
   \plotone{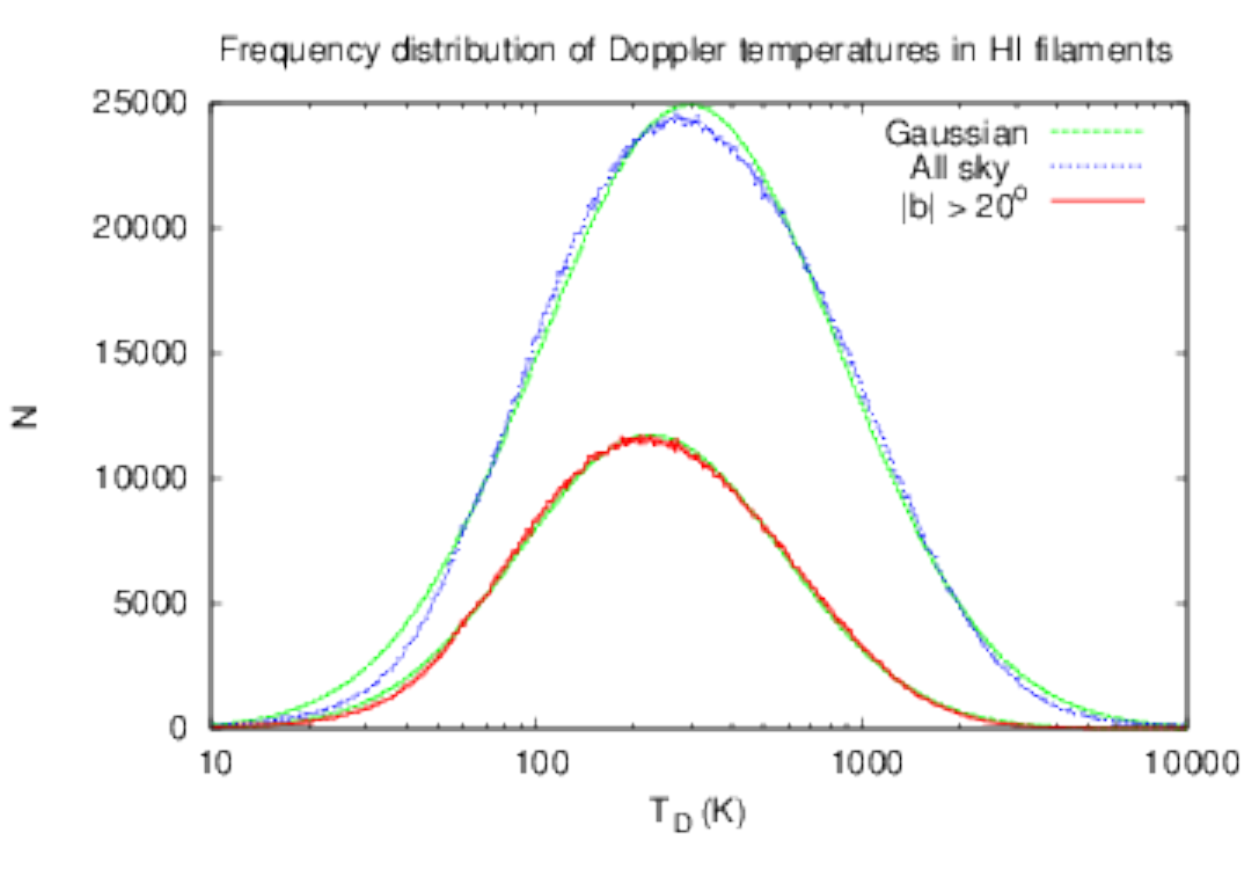}
}
\caption{Distribution function for the Doppler temperatures of the
  major local \hi filaments, compared with Gaussian fits. The median
  Doppler temperature for $|b| > 20 \degr$ is $T_{\rm D} = 223 $ K.  }
   \label{Fig_Tkin_histo}
\end{figure}

\subsection{Doppler temperatures }
\label{Tkin}

It is feasible to estimate roughly the temperatures of the \hi gas from
the $T_{\rm Bon}$ distribution. From the close agreement between dust
emission and filaments in $T_{\rm Bon}$ (top and middle panel of Fig
\ref{Fig_Filaments2}) one may suspect a physical association between the
filaments which implies that these \hi features must be cold. 

From the line-widths of USM filaments we determine Doppler temperatures.
In Fig. \ref{Fig_Tkin_histo} we display the distribution in a
histogram. Restricting our analysis to $|b| > 20 \degr$ (red curve), we
obtain a well defined log-normal PDF.  The median is $T_{\rm D} = 222.6
$ K; from the Gaussian fit we obtain $T_{\rm D} = 223.1 $ K,
corresponding to a FWHM line width of 3.2 \kms.  Including the Galactic
plane leads to somewhat higher Doppler temperatures, with a peak at
$T_{\rm D} = 294 $ K. Apparently, even in case of confusion there are
only slight biases towards higher Doppler temperatures.  A previous
Gaussian decomposition of the LAB survey by \citet{Haud2007} yield for
the narrowest components in a log-normal distribution with $T_{\rm D} =
332 $ K at a FWHM of $3.9 \pm 0.6$ \kms. For a distribution affected by
radial velocity gradients across the beam, the smoothing will increase
the line widths and the Doppler temperatures in case of a larger
telescope beam.

The log-normal Doppler temperature distribution is, similar to the
column density PDF (Fig. \ref{Fig_NH_histo}),  approximated
surprisingly well by a single Gaussian, we observe a remarkable clean
single-phase relation. Also in case of Doppler temperatures we
obtain an excellent agreement between our results and those of
\citet{Clark2014} who derive $T_{\rm D} = 220 $ K. According to the
estimates above, the thermodynamic pressure of the filaments is $
P/k_{\rm B} = nT \la 3100$ {\rm K cm}$^{-3}$, in reasonable agreement with the
standard pressure found in the ISM at the solar circle $ P/k_{\rm B} = 3070$
{\rm K cm}$^{-3}$ \citep{Wolfire2003}. \citet{Clark2014} derive a
pressure of $ P/k_{\rm B} \la  3200$ {\rm K cm}$^{-3}$.

\begin{figure*}[htp]
\epsscale{1.0}
   \centerline{
   \includegraphics[scale=0.44]{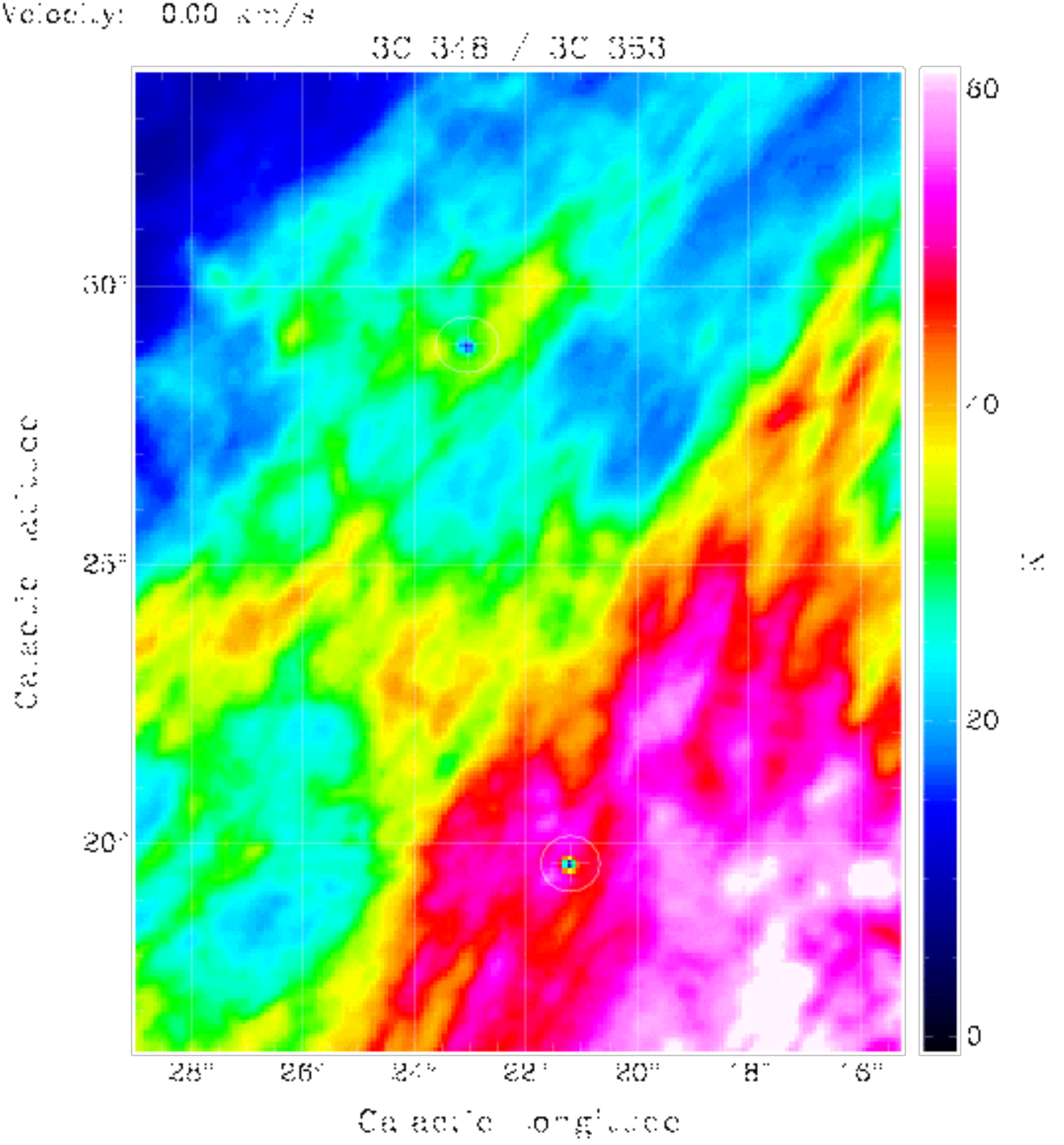}
   \includegraphics[scale=0.44]{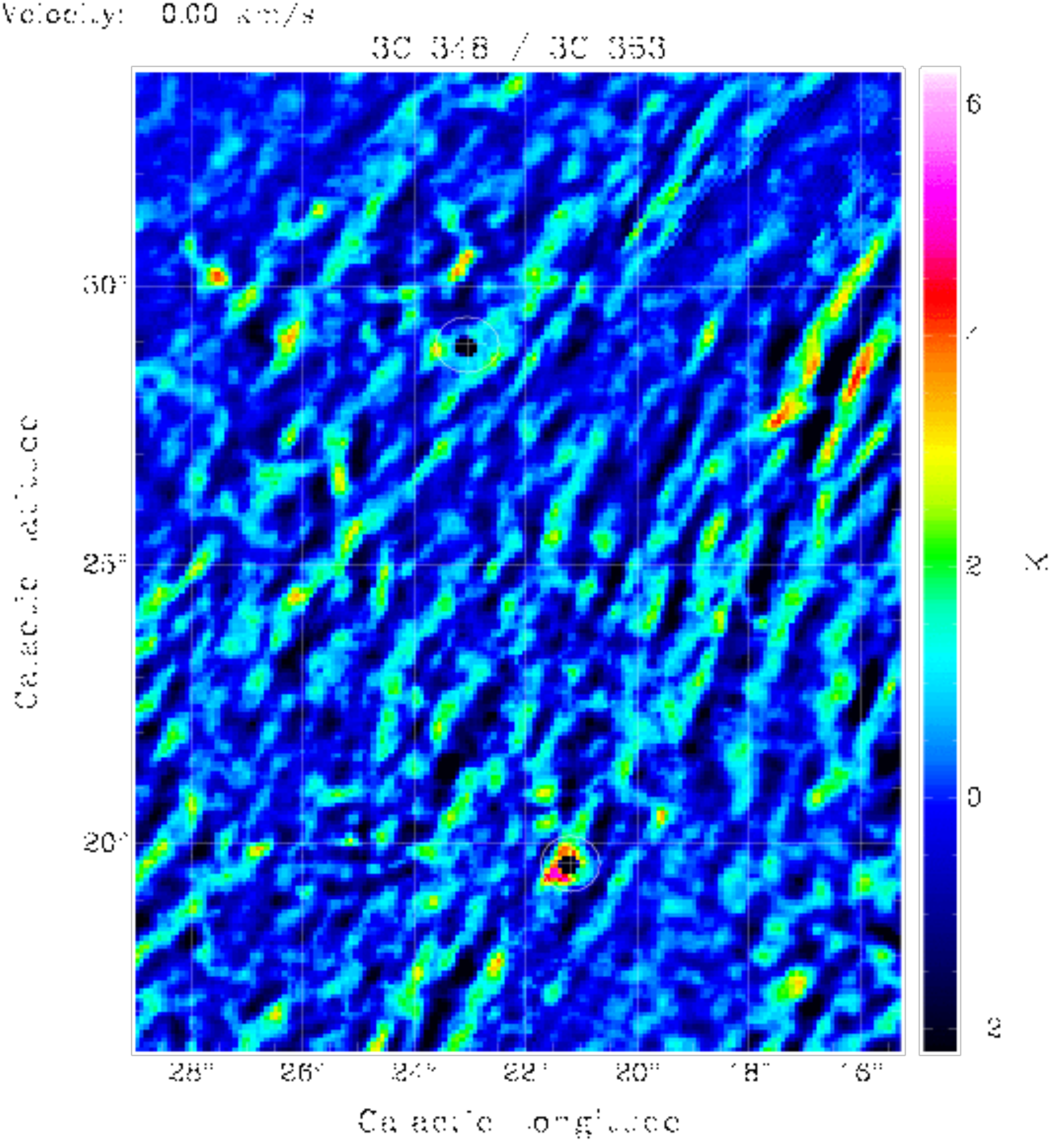}
}
   \caption{$T_{\rm B}$ distribution (left) and USM filaments (right) in the
     vicinity of 3C348 ($ l = 23\fdg05, b = 28\fdg95 $) and 3C353 ($ l =
     21\fdg20, b = 19\fdg65 $). Both sources are seen in absorption, the
     positions are indicated. The length of the crosses 
       correspond to ten times the FWHM beam width of the Arecibo
       telescope. Circles have a diameter of 1 \degr.}
   \label{3C348_353}
\end{figure*}

\begin{figure*}[htp]
\epsscale{1.0}
   \centerline{
   \includegraphics[scale=0.44]{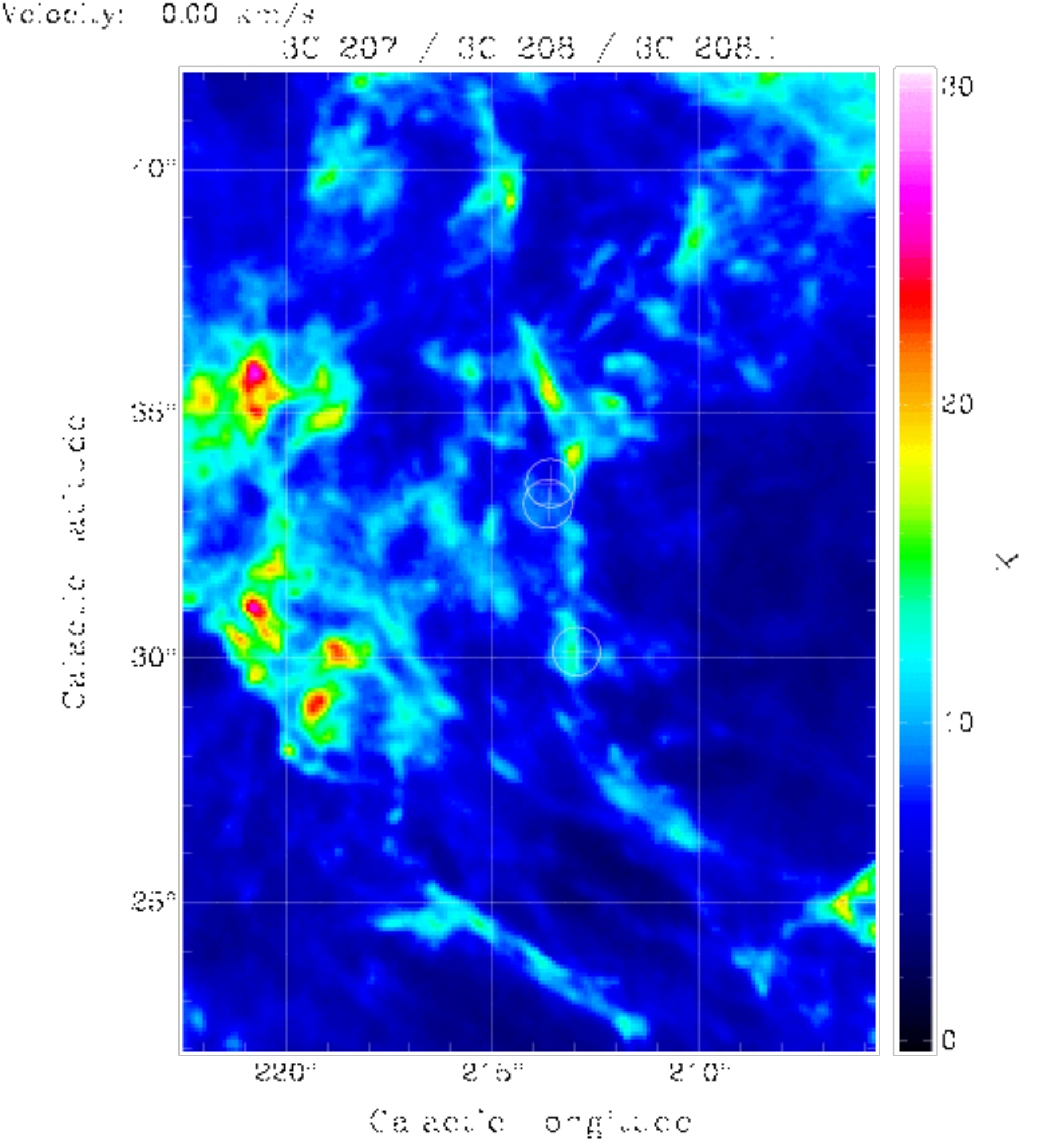}
   \includegraphics[scale=0.44]{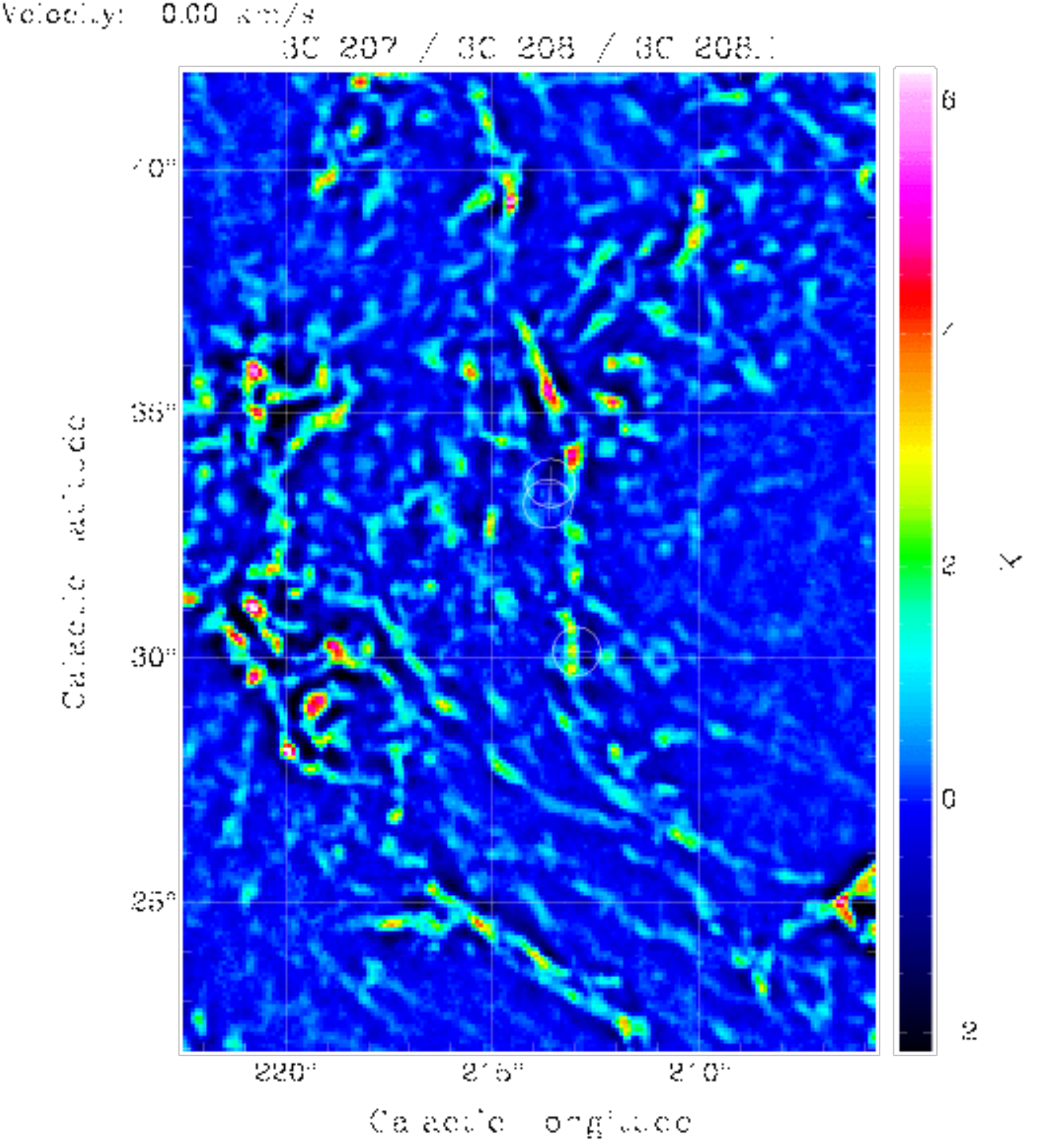}
}
   \caption{$T_{\rm B}$ distribution (left) and USM filaments (right) in the
       vicinity of 3C207 ($ l = 212\fdg97, b = 30\fdg16 $) and 3C208 ($
       l = 213\fdg66, b = 33\fdg16 $), and 3C208.1 ($ l = 213\fdg60, b =
       33\fdg58 $). Absorption from the sources is not obvious, the
       positions are indicated. The length of the crosses 
       correspond to ten times the FWHM beam width of the Arecibo
       telescope. Circles have a diameter of 1 \degr. }
   \label{3C207_208}
\end{figure*}

\subsection{Optical depth effects and self-absorption}
\label{depth}

USM column and volume densities, discussed in Sects. \ref{column} and
\ref{volume} may be affected significantly by optical depth effects.
Having merely \hi emission data, we neither can determine optical depth
nor spin temperatures for the filaments. However at a few positions,
absorption data are available from \citet{Heiles2003a}. Here we will
compare their data (column densities $N_{\rm H}$, optical depths $
\tau$, spin temperatures $T_{\rm spin}$, and upper limits to kinetic
temperatures $T_{\rm k,max}$) with our results (Doppler temperatures
$T_{\rm D}$ and column densities $N_{\rm Husm}$). Note that
\citet{Heiles2003b} use $T_{\rm k,max}$ in place of $T_{\rm D}$. Our
estimates are derived at positions close to the continuum sources along
the filaments. We study accordingly two fields with strong background
sources.

Covered by the area of interest displayed in Fig. \ref{3C348_353} we
have 3C348 and 3C353 and \hi filaments are crossing these sources.

For 3C348 we estimate close to the source at $v = 0~ $\kms~ along the
filament a Doppler temperature of $T_{\rm D} = 30 $ K and a column
density $N_{\rm Husm} = (2.5 \pm 0.45)\cdot 10^{19}$ cm$^{-2}$. From Arecibo data
\citet{Heiles2003a} obtain an optical depth of $ \tau = 0.604 \pm 0.004$
and temperatures $T_{\rm spin} = 32.54 \pm 5.82 $ K and $T_{\rm k,max} =
98 $ K. We miss almost a factor of 8 in the measured column
densities. 

For 3C353 we estimate at $v = 0~ $\kms~ a Doppler temperature of
$T_{\rm D} = 150 \pm 10 $ K. From Arecibo data \citet{Heiles2003a}
determine $T_{\rm k,max} = 170$ K but a spin temperature as low as $T_{\rm
  spin} = 36.89 \pm 10.12 $ K. The optical depth $ \tau = 1.209 \pm
0.007$ is high. Correspondingly, the column density is $N_{\rm H} = 2.4\cdot
10^{20}$ cm$^{-2}$, but we derive only $N_{\rm Husm} = (2.5 \pm 0.5)\cdot
10^{19}$ cm$^{-2}$. % T_{\rm B} = 50 \pm 3 K 

Common for both is that the CNM gas shows up with spin temperatures
around 30\,K. Both are located in regions with a similar background
temperature, $T_{\rm B} \sim 31 $ K for 3C348, and $T_{\rm B} \sim 50 $
K for 3C353. Some self-absorption might be present
\citep{Gibson2000}. In absence of a continuum background source the
difference in brightness temperature due to the CNM cloud is given by
\begin{equation}
\label{eq:selfabs}
\Delta T_{\rm B} = (T_s -f\, T_{bg}) (1 - e^{-\tau}). 
\end{equation}
Here $T_s$ is the spin temperature and $T_{bg}$ the brightness of the
WNM background. $f < 1$ is the unknown fraction of the gas that is
beyond the CNM filament. For $ T_s \sim f\, T_{bg}$ self-absorption
becomes important, thus the HI emission of the filament may be
attenuated strongly. Thus we expect that for $T_{bg} > 30$ K
self-absorption may be recognizable. 20\% of the CNM filaments at
latitudes $|b| > 20 \degr$ are considered to be affected. However,
checking these data we find no indications for obvious self-absorption,
rather the average CNM column densities {\it increase} by 23\%. This
implies, that in regions with luminous WNM emission also the CNM column
densities tend to be higher. Hence, at latitudes $|b| > 20 \degr$ no
indications for obvious self-absorption effects are evident from our
analyses, opposite to the examples at low latitudes as shown in
Fig. \ref{plane}. Checking whether $T_{\rm D}$ depends on $T_{bg}$ we
find no significant trends with $T_{bg}$. We also applied an optical
depth correction for the column densities according to
\citet[][Eq. 8]{Martin2015}. Testing for spin temperatures ranging
between 20 and 80\,K, again no significant effects are found for the
distribution plotted in Fig.\ref{Fig_NH_histo}.

In Fig. \ref{3C207_208} we have 3 continuum sources, 3C207, 3C208, and
3C208.1. There are no clear filaments but a number of \hi fragments,
perhaps originating from a disintegrated filament. At a velocity of $v =
4.2~ $\kms~ the \hi in front of 3C207 \citet{Heiles2003a} determine an
optical depth $ \tau = 0.25 \pm 0.002$ and a spin temperature of $T_{\rm
  spin} = 24.65 \pm 4.74 $ K, the Arecibo limit to the kinetic
temperature is $T_{\rm k,max} = 602 $ K.  We determine a $T_{\rm D} = 250
\pm 50 $ K and a column density $N_{\rm Husm} = (4 \pm 0.8)\cdot 10^{19}$
cm$^{-2}$, compared to an Arecibo column density of $N_{\rm H} = 6.3\cdot
10^{19}$ cm$^{-2}$. 
% 17 K background

The \hi in front of 3C208, and 3C208.1 has at a velocity of $v = 4~
$\kms~ optical depths below $ \tau = 0.1$. But for 3C208 at $v = 6~
$\kms~ our Doppler temperature $T_{\rm D} = 95 \pm 5$ K compares to an
Arecibo spin temperature of $T_{\rm spin} = 90 $K, the column densities
are within an uncertainty of 40\% comparable. For 3C208.1 at $v = 6~
$\kms~ we have $T_{\rm D} = 120 $, compared to $T_{\rm spin} = 350
$K and our column density is 40\% higher than that from Arecibo. 
% 12 K background 

Self-absorption for the 3C207/208 field are considered to be
insignificant because the low WNM temperatures close to the continuum
sources are only $ 7 \la T_{\rm B} \la 17 $K.

We conclude that  Doppler temperatures and column densities for the
filaments, derived by us, are quite uncertain but roughly consistent
with parameters from  \citet{Heiles2003a} as long as the optical depth
is low. We cannot measure the optical depth and for values 
$ \tau \ga 0.5 $ we may seriously underestimate the column densities.  

The question arises what fraction of the filamentary \hi gas at
latitudes $|b| > 20 \degr$ is optically thick. In our examples we
considered two cases with quite different morphologies, regular parallel
filaments and less ordered fragments. In both cases the dust emission
from the HFI Sky Map at 353 GHz is rather low. Is there, concerning
optical depth effects, a systematical difference between both cases, 
  regular or broken filaments?

\subsection{Filaments within a warm environment}
\label{environment}

 Figures \ref{Fig_Filaments1}, \ref{Fig_Filaments1b}, and
  \ref{Fig_Filaments1c} suggests that the filaments with low
intensities are embedded in a warmer environment that dominates the
observed $T_{\rm B}$ distribution. We use an all-sky Gaussian
decomposition \citep{Haud2000,Haud2007,Kalberla2015} of the observed
brightness temperature distribution to search for this surrounding
gas. At each position for $|b| > 20\degr$ we select components with
center velocities $v_{\rm Uon} - \sigma_v < v < v_{\rm Uon} + \sigma_v
$, where $\sigma_v = \delta v_{\rm Uon} /{\sqrt{8 {\rm ln}(2)}}$ is the
dispersion corresponding to FWHM line width $\delta v_{\rm Uon}$. Note
that this condition is very stringent concerning the center
velocities. Narrow ($\delta v < 2$ \kms) and insignificant low intensity
components ($T_{\rm B} < 0.5$ K) are disregarded. We use the best
fitting component with the lowest velocity dispersion and exclude this
way broad features that might be unrelated to the filaments. Using
$T_{\rm D} = 21.86~ \delta v^2$ (Sect. \ref{deriving}) we convert line
widths to Doppler temperatures.

\begin{figure}[thb]
\epsscale{0.5}
   \centerline{
   \plotone{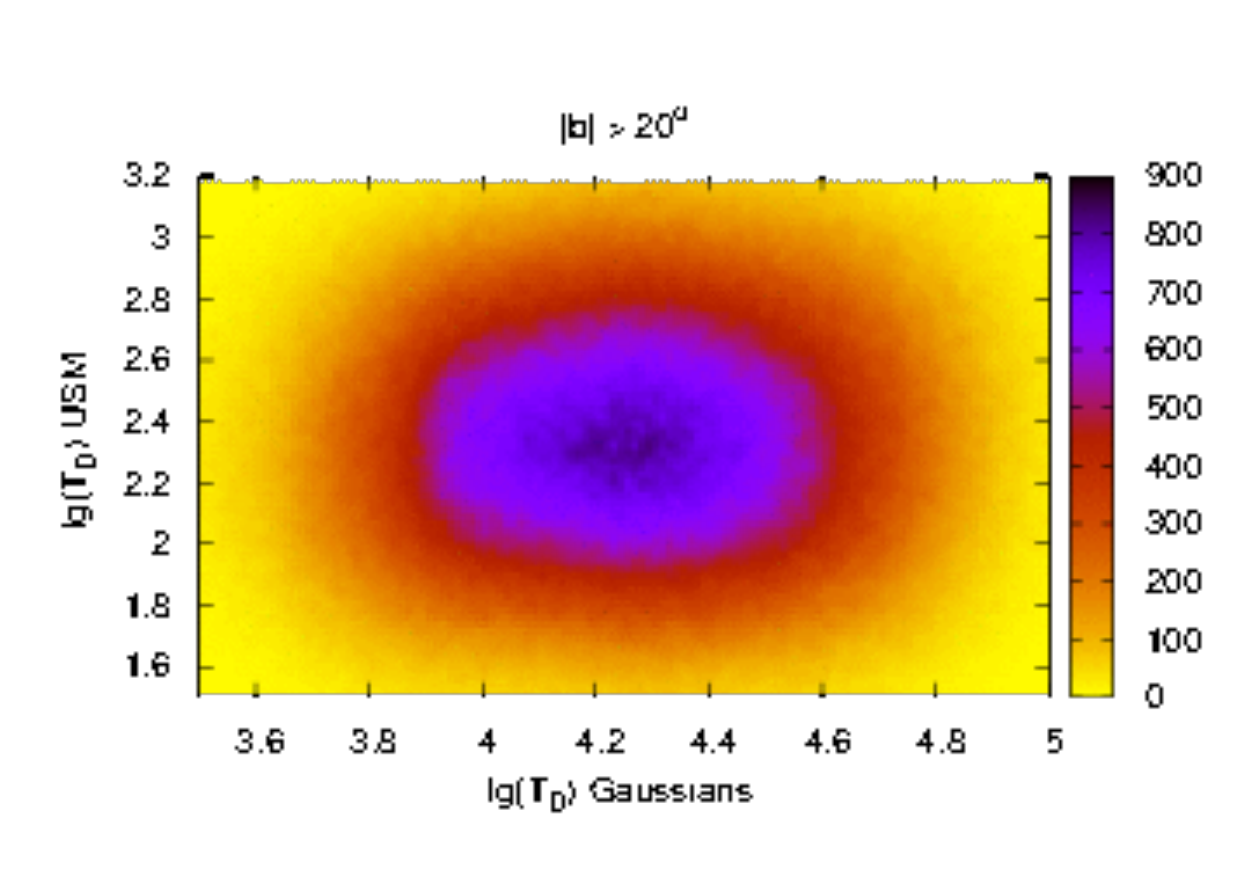}
}
\caption{2-D density distribution for Doppler temperatures within
  \hi filaments and the surrounding WNM gas from a Gaussian analysis. }
   \label{Gauss_Tkin}
\end{figure}

Figure \ref{Gauss_Tkin} shows the 2-D density distribution for 
  Doppler temperatures within \hi filaments and the surrounding WNM
gas. The horizontal spread corresponds to the range of the WNM Doppler
  temperatures lg($T_{\rm D}$), while vertically
  lg($T_{\rm D}$) for the USM filaments is displayed.  This plot  is
consistent with a two-phase distribution, cold filaments are embedded
within a warm environment with a median $T_{\rm D} \sim 18374$ K.

\begin{figure}[htb]
\epsscale{0.5}
   \centerline{
   \plotone{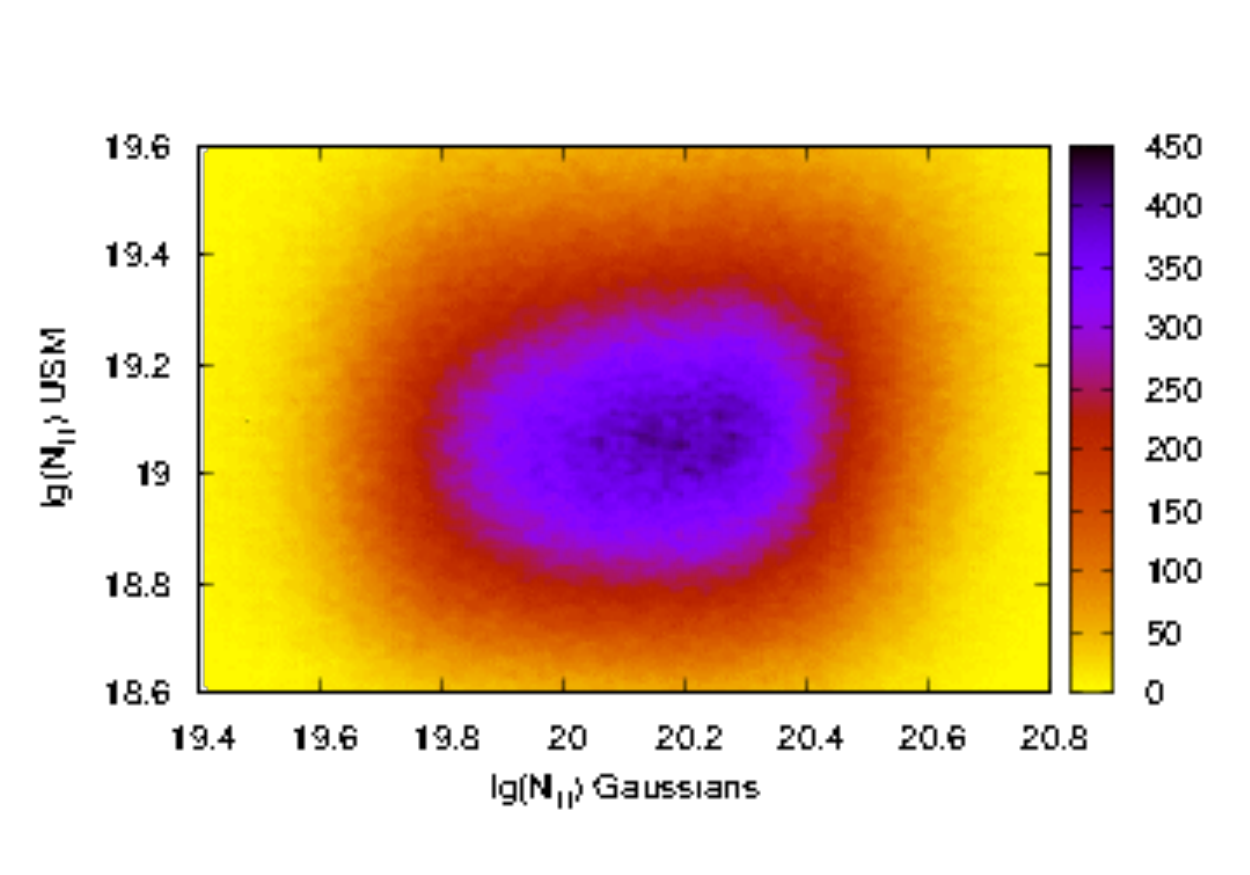}
}
   \caption{2-D density distribution for column densities of \hi
     filaments and the surrounding WNM gas from a Gaussian analysis. }
   \label{Gauss_NH}
\end{figure}

We also determined the two-phase distribution of column
densities. Figure \ref{Gauss_NH} shows that the filaments with typical
column densities $N_{\rm H} \sim 10^{19.1}$ cm$^{-2}$ are embedded in a WNM
with typical column densities of $N_{\rm H} \sim 10^{20.2}$ cm$^{-2}$. From
the column densities we derive consistent with Sect. \ref{surface} a
mass fraction of $F_M = 0.1$, yielding $F_{\rm tot}= 0.2$ including
selection effects for multiple filaments along the line of sight as
discussed in Sect. \ref{surface}.

\subsubsection{Latitude dependence}
\label{latitude}

\begin{figure*}[tbp]
   \centerline{
   \includegraphics[scale=0.4]{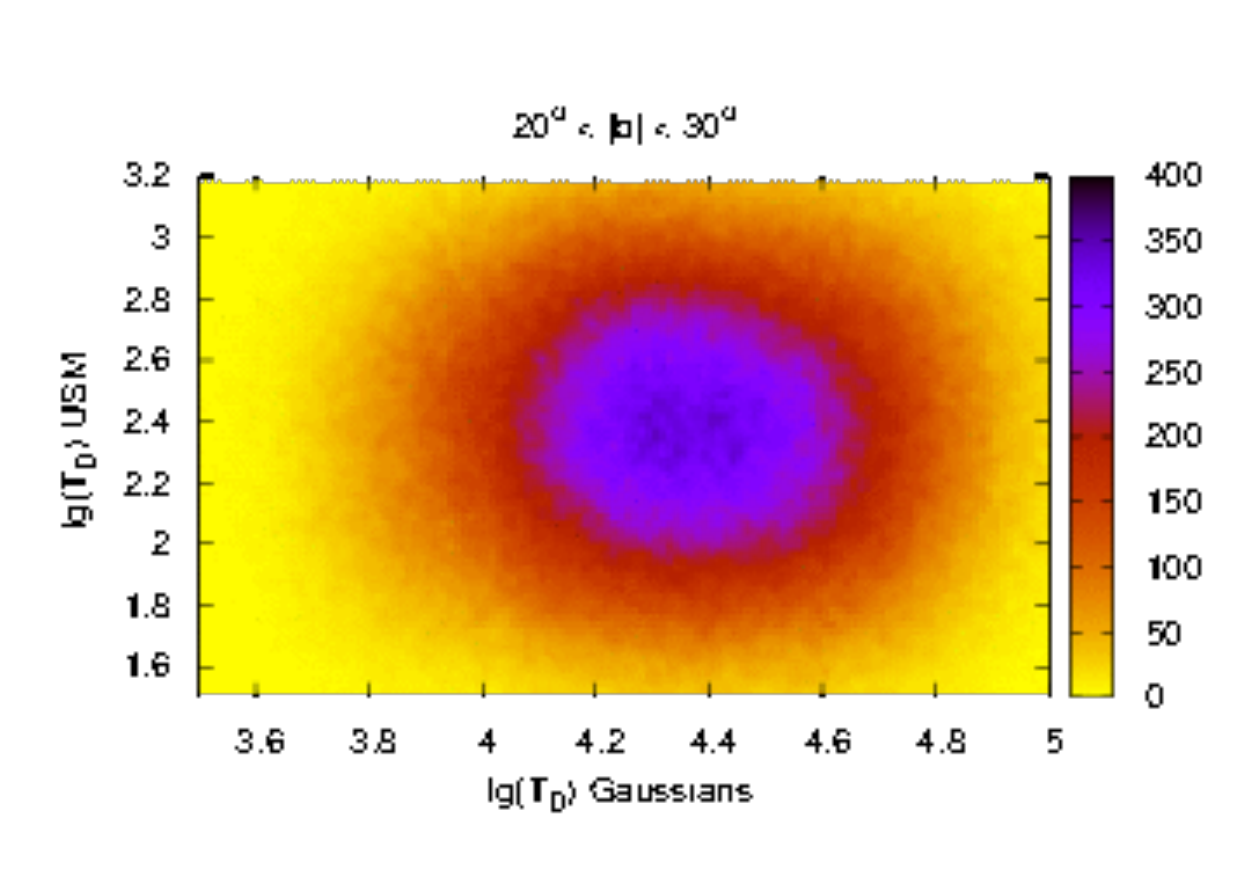}
   \includegraphics[scale=0.4]{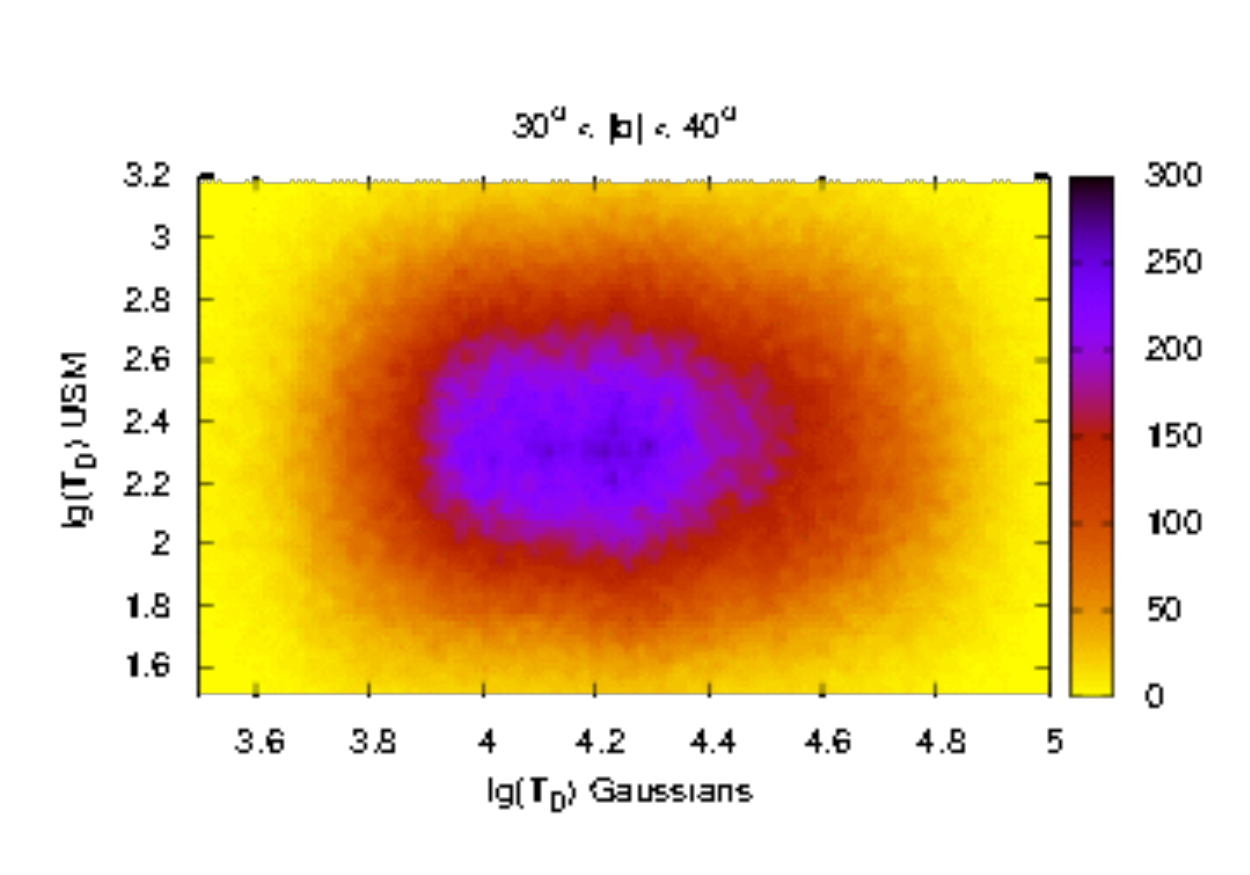}
   \includegraphics[scale=0.4]{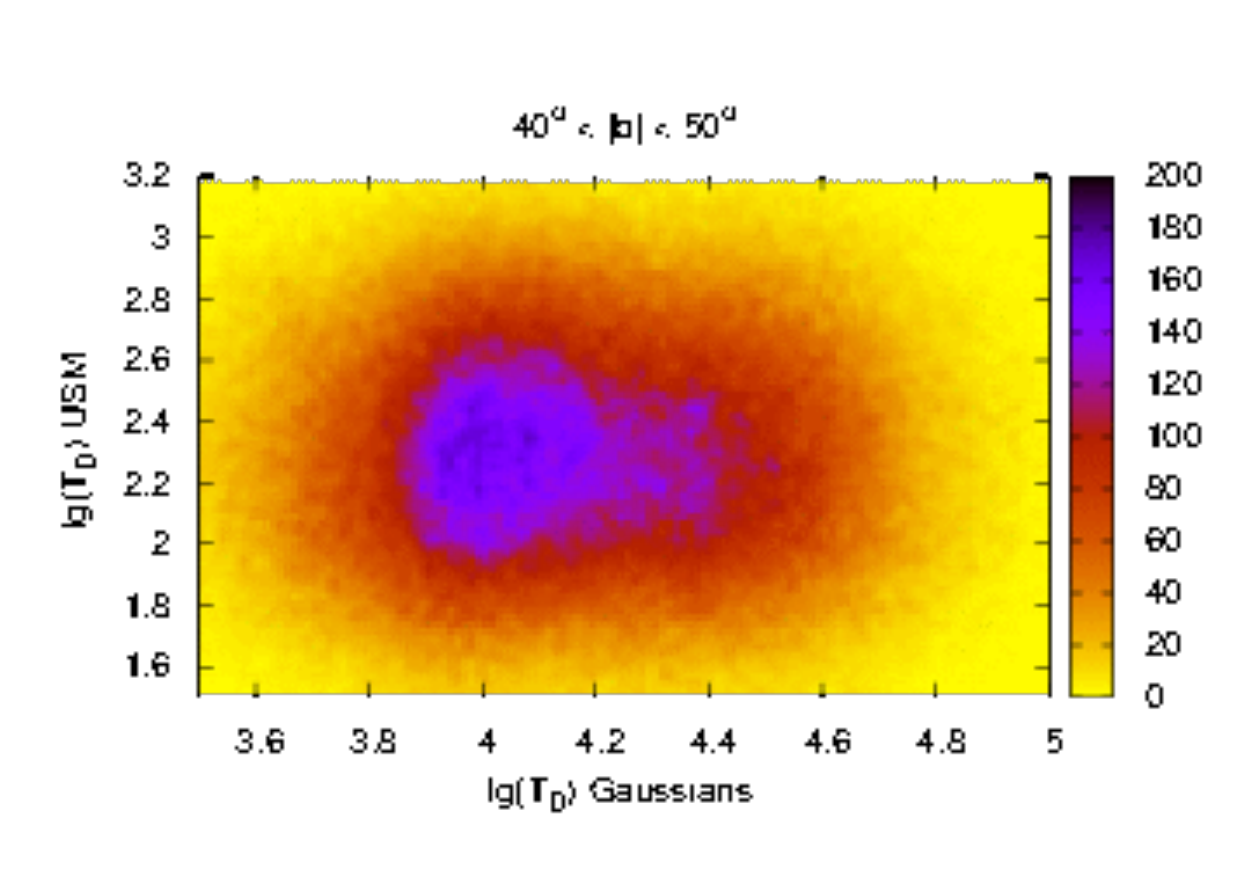}
}
   \centerline{
   \includegraphics[scale=0.4]{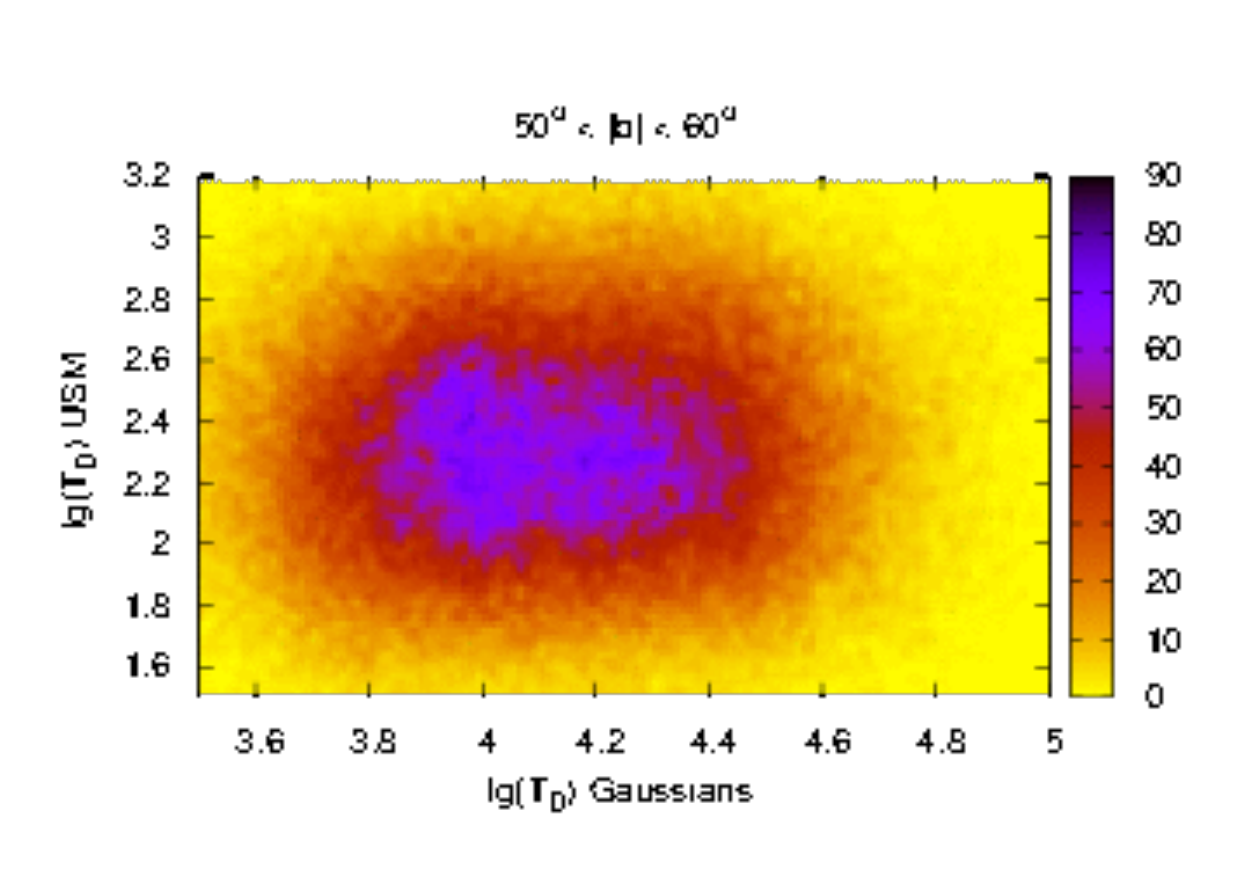}
   \includegraphics[scale=0.4]{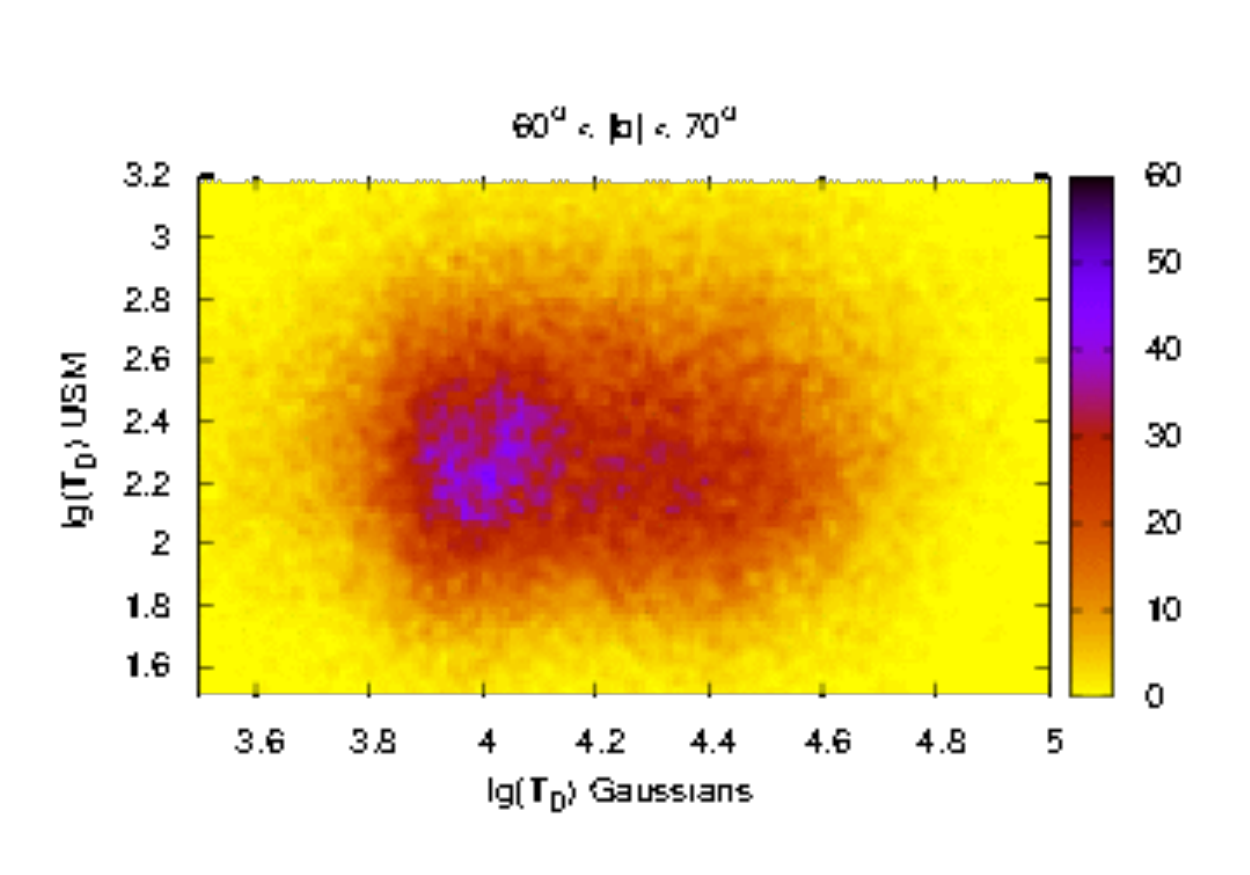}
   \includegraphics[scale=0.4]{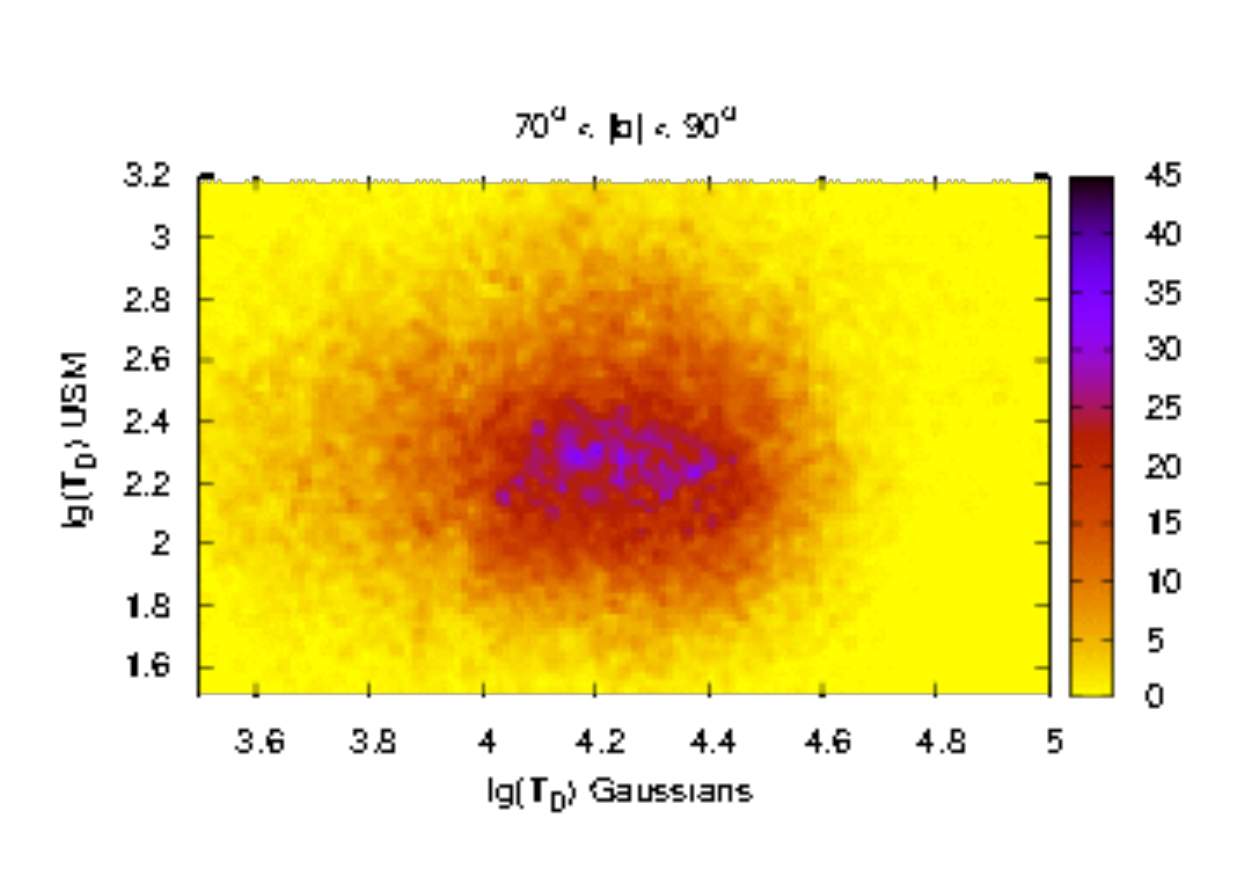}
}
\caption{Latitude dependence of 2-D density distributions for 
    Doppler temperatures within \hi filaments and the surrounding gas
  from a Gaussian analysis. }
   \label{Gauss_Tkin_all}
\end{figure*}

Next we study whether derived Doppler temperatures may depend on
  Galactic latitude. In Fig. \ref{Gauss_Tkin_all} we display the
  2-D density distributions of Doppler temperatures for several latitude
  ranges. We find no significant changes in the distribution of the
  Doppler temperatures for the filaments, the peak at lg($T_{\rm D}$)
  $\sim 2.35$, corresponding to $T_D = 223$ K, remains almost
  constant. However the $T_{\rm D}$ distribution for the warm envelopes
  shows some variations between $4 \la {\rm lg}(T_{\rm D}) \la 4.4$,
  corresponding to $10 000 \la T_D \la 25 000$ K.

\begin{figure*}[tbp]
   \centerline{
   \includegraphics[scale=0.4]{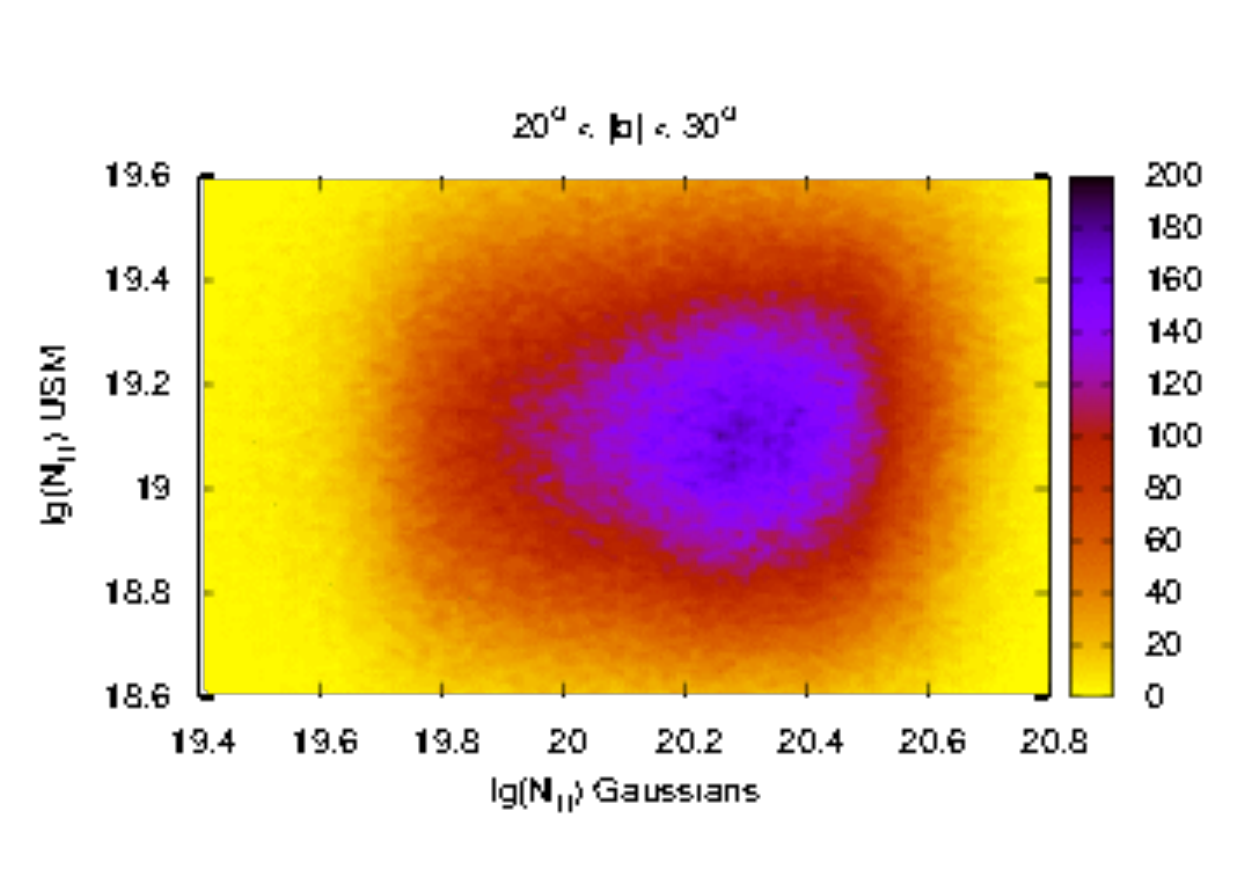}
   \includegraphics[scale=0.4]{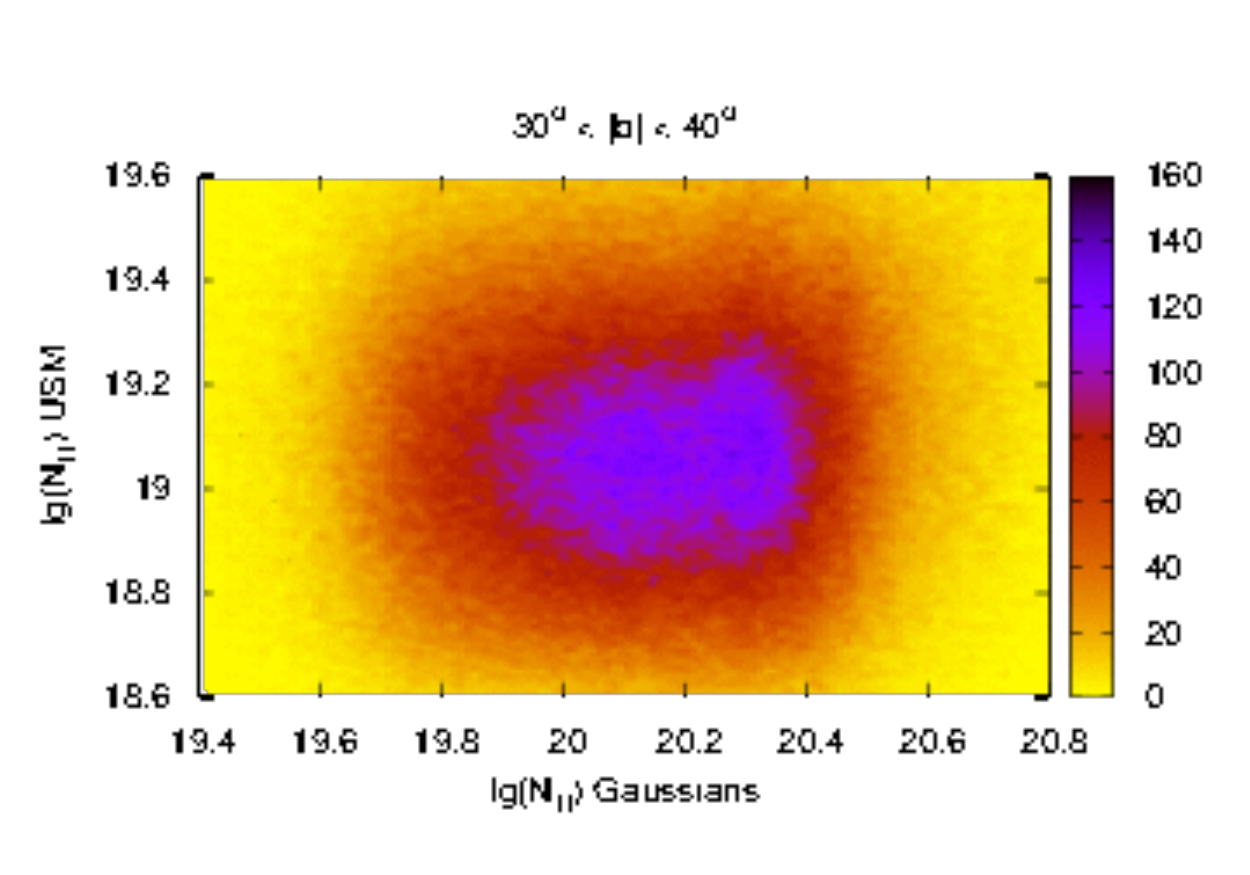}
   \includegraphics[scale=0.4]{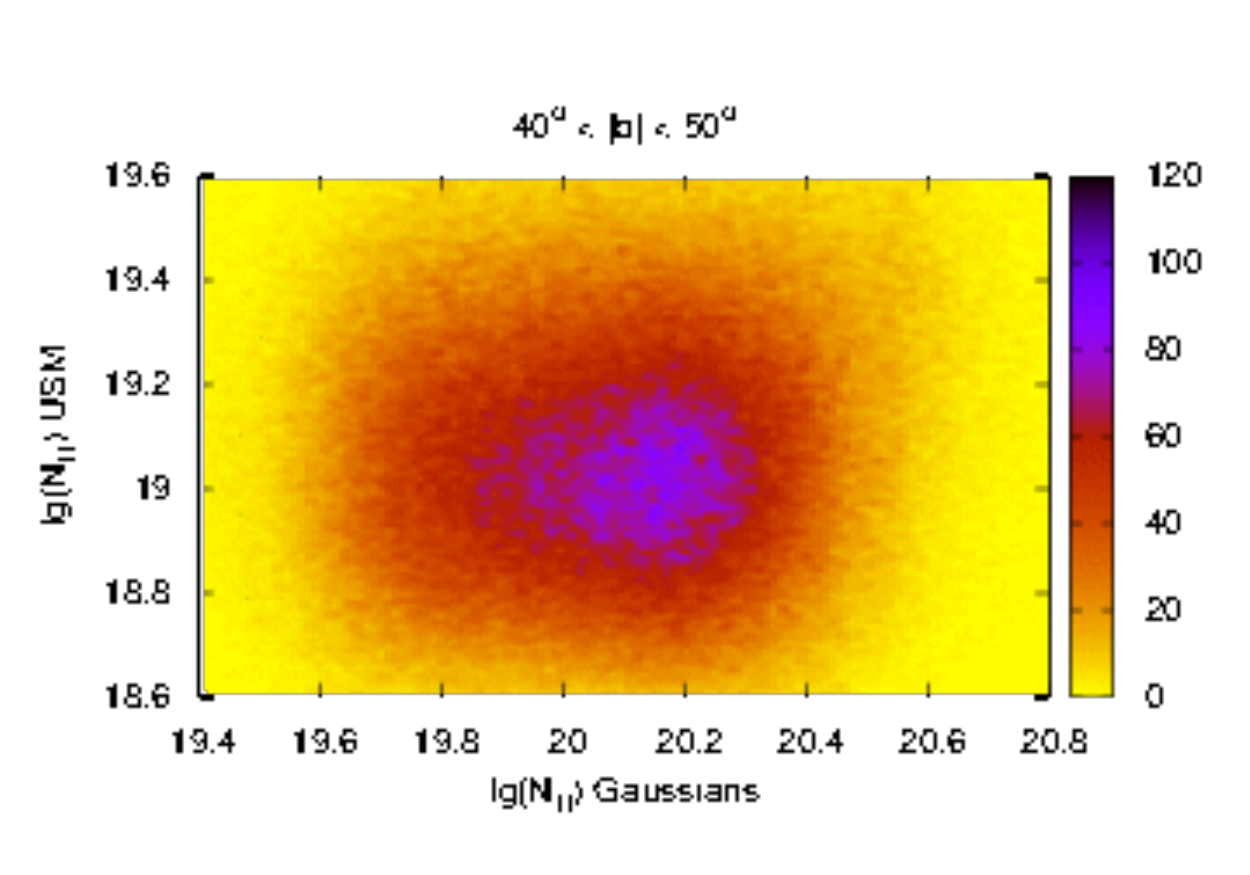}
}
   \centerline{
   \includegraphics[scale=0.4]{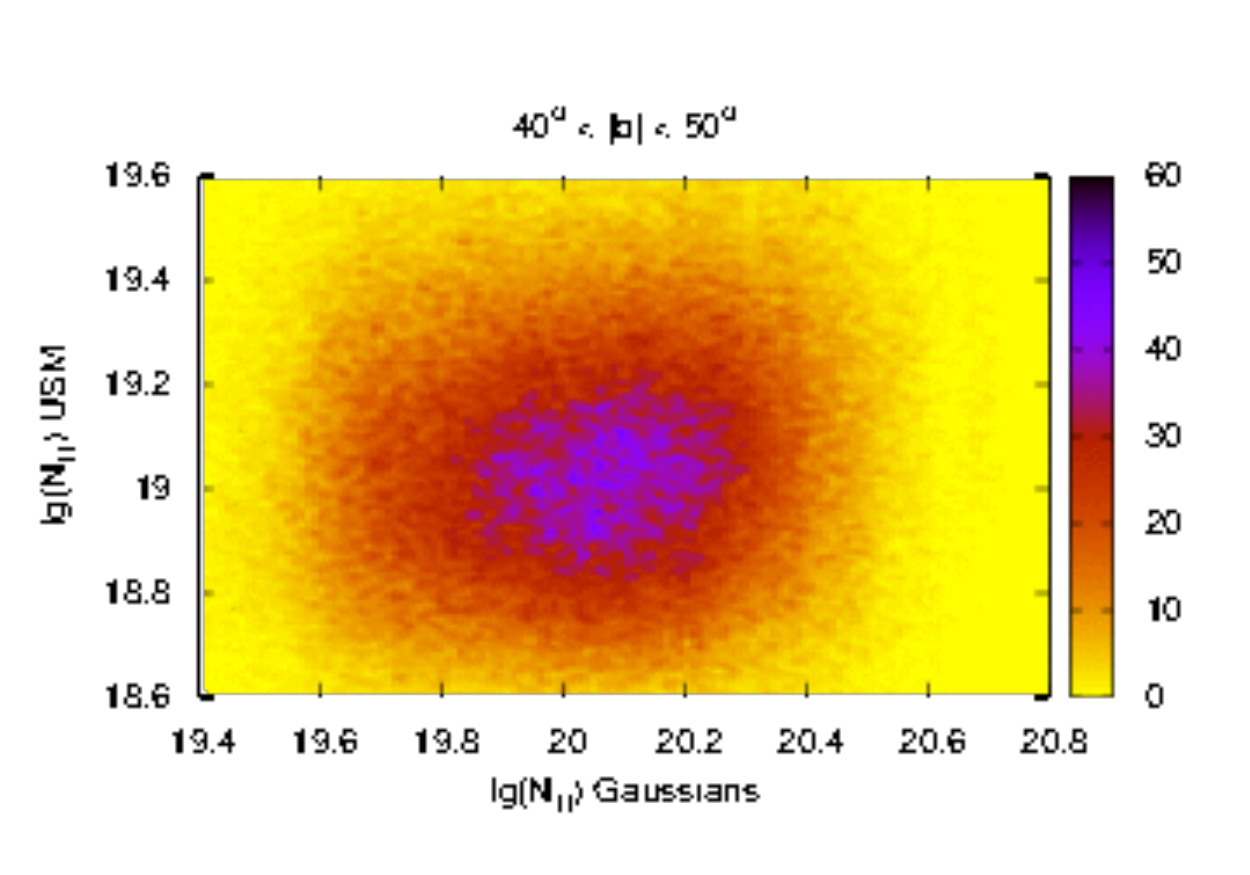}
   \includegraphics[scale=0.4]{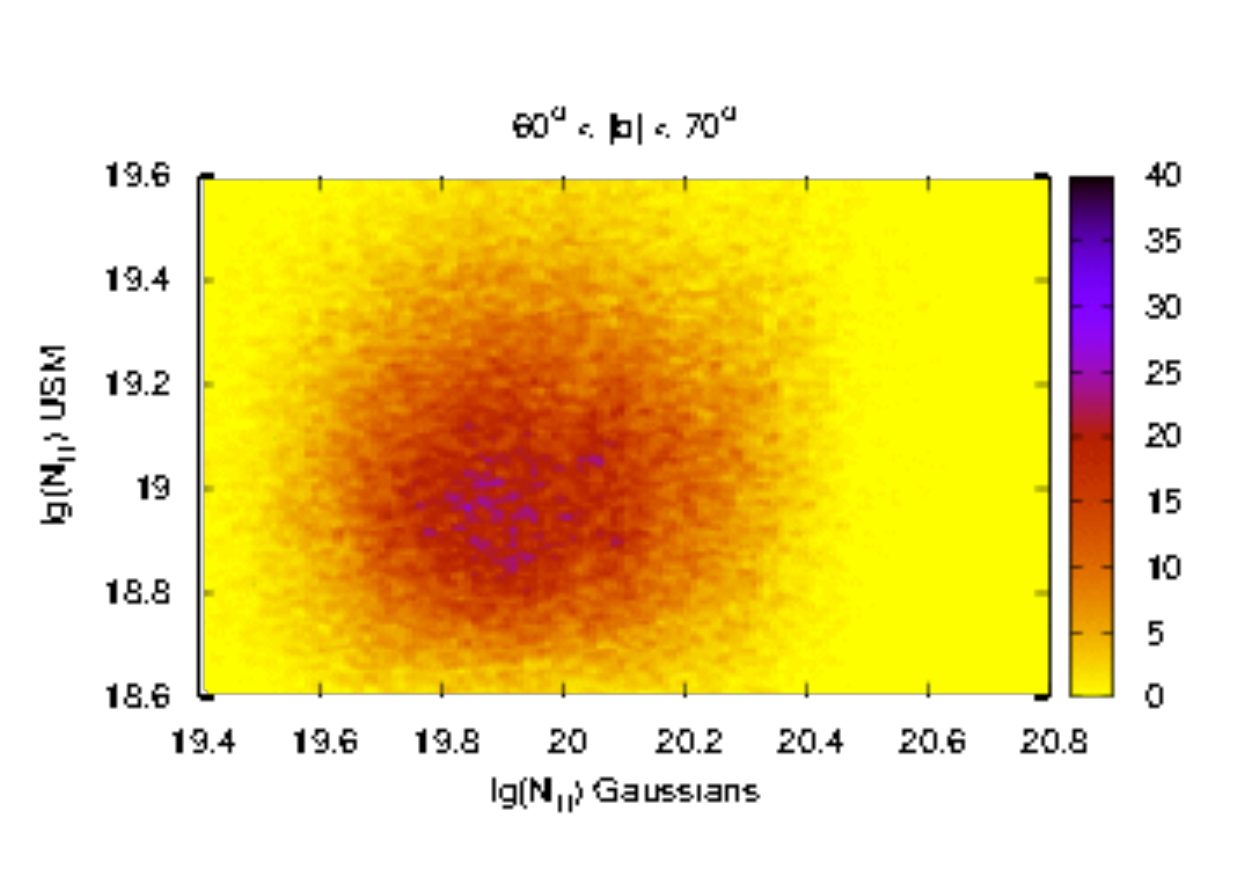}
   \includegraphics[scale=0.4]{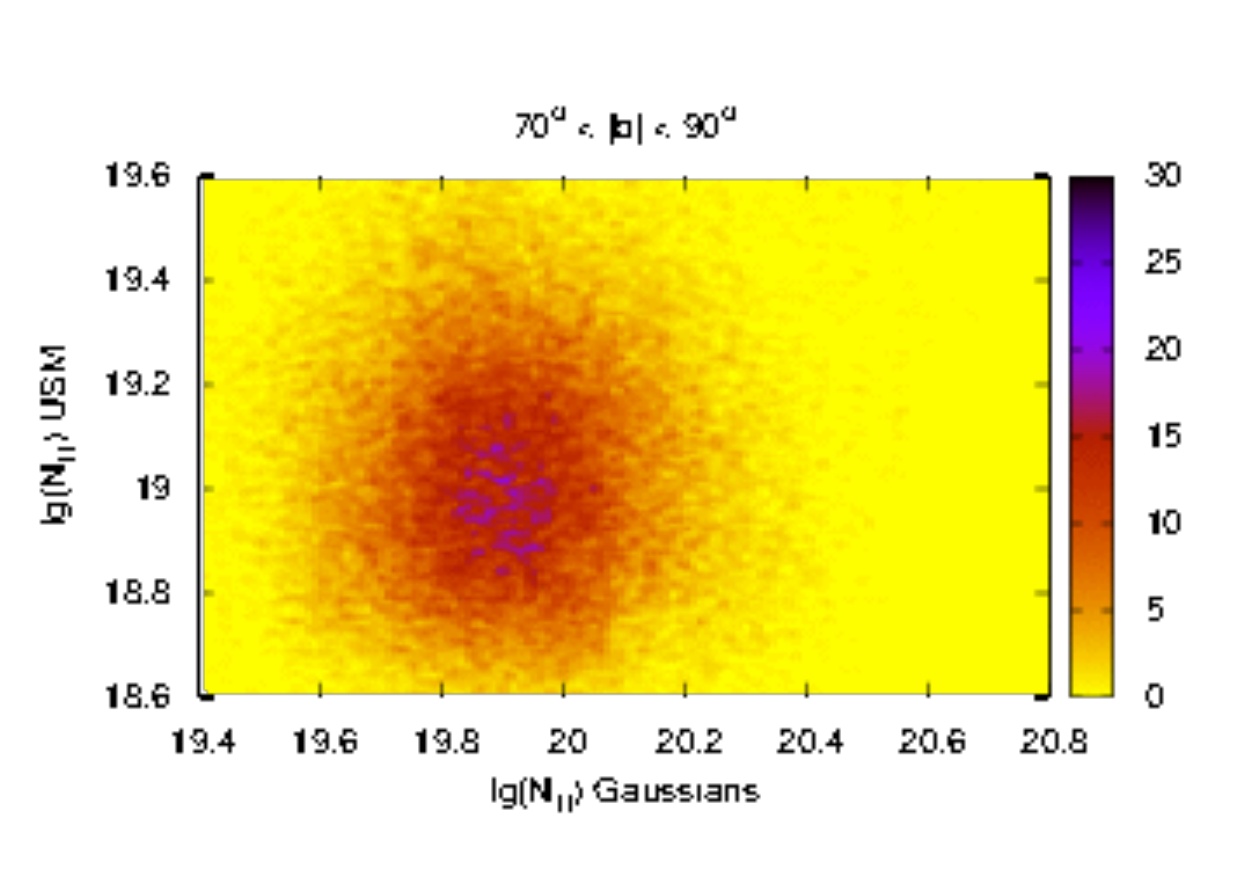}
}
   \caption{Latitude dependence of 2-D density distributions for column
     densities of \hi filaments and the surrounding gas from a Gaussian
     analysis. }
   \label{Gauss_NH_all}
\end{figure*}

  Figure \ref{Gauss_NH_all} displays latitude dependent 2-D density 
  distributions of the column densities. Filament column densities change
  only marginally, with a peak from $N_{\rm H} \sim 10^{19.1}$ cm$^{-2}$ at low
  latitudes to $N_{\rm H} \sim 10^{19}$ cm$^{-2}$ at high latitudes. For the
  WNM envelopes the dependency is much more significant, $N_{\rm H}(b=25\degr) /
  N_{\rm H}(b=90\degr) \sim 2.5 $.

  The \hi vertical distribution in the solar neighbourhood can be
  described as a layer with approximately constant scale height
  \citep{Kerr1969}. The path along the line of sight through this layer
  follows $1/{\rm sin}|b|$. We obtain from this relation for the WNM a
  factor of 2.4 in column density while we observe an increase by a
  factor of 2.5, consistent with the hypothesis that the WNM
  distribution can be approximated by an extended layer.
  
  The filaments do not show any significant latitude dependence, neither in
  Doppler temperatures, nor in column densities. We conclude
  that these parameters do not depend on latitude and are unaffected by
  other line-of-sight effects.

\subsection{Two-phase stability}
\label{Twophases}

Models of the ISM assume that on average the WNM and the CNM are in
  pressure equilibrium.  Our current knowledge about heating and cooling
  in the Solar vicinity is based on a fiducial column density of $N_{\rm H}
  \sim 10^{19}$ cm$^{-2}$ at the WNM/CNM boundary.  This implies a local
  pressure $ P/k_{\rm B} = nT = 3070$ {\rm K cm}$^{-3}$ \citep[][
  Sect. 6.1]{Wolfire2003}. Accordingly the CNM can exist for thermal
  temperatures $ 62 \la T \la 258$ K. The allowed temperature range for
  the NM is $5040 \la T \la 8310$ K. \hi gas at temperatures $ 258 \la
  T \la 5040$ K is thermally unstable.

  This model is in conflict with our determination of a WNM column
  density of $N_{\rm H} \sim 10^{20}$ cm$^{-2}$ in Sects. \ref{column} and
  \ref{latitude}. Phase diagrams, showing the dependence of $P/k_{\rm B}$ and
  hydrogen nucleus density for different values of the WNM column
  density, are given in Fig.\,9 of \citet{Wolfire2003}. According to
  Table 3 of the same paper the column densities determined here
  imply an average pressure at the solar circle of $P/k_{\rm B} = 1690$ {\rm K
    cm}$^{-3}$. Stability for CNM clouds is thus expected at a somewhat
  higher temperature range $86 \la T \la 324$ K. The low pressure $P/k_{\rm B}
  = 1690$ {\rm K cm}$^{-3}$ is however in conflict with a recent determination
  of the local thermal pressure of $P/k_{\rm B} \sim 3700$ {\rm K cm}$^{-3}$
  by \citet{Jenkins2011}. According to a private communication
  by Mark Wolfire the theory of the two-phase ISM demands a revision
  and the WNM column density of $N_{\rm H} \sim 10^{20}$ cm$^{-2}$ appears to
  be today consistent with a pressure around $3070 \la P/k_{\rm B} \la 3700$
  {\rm K cm}$^{-3}$.

  To finally proof stability it would be necessary to determine kinetic
  temperatures of the CNM filaments under investigation. From our
  observations we can only determine Doppler temperatures that are
  affected by turbulent motions of the gas. However, it is feasible to
  constrain them by some ensemble properties.

  Turbulent motions in the CNM can be characterized by the turbulent
  Mach number 
\begin{equation}
\label{eq:Mach}
M_{\rm t} = \sqrt{ 4.2 (T_{\rm D}/T_{\rm kin} - 1)}
\end{equation}
  \citep[][Sect. 6.2.4]{Heiles2003b}. CNM clouds tend to be strongly
  supersonic and high Mach numbers are common
  \citep[][Fig. 12]{Heiles2003b}.

  The all-sky maps, shown in Sect. \ref{allsky}, imply that CNM
  filaments are sheets, aligned to the magnetic fields. For such an
  alignment \citet{Heiles2005} determine from the Millennium Arecibo
  21-cm absorption-line survey a median magnetic field strength of $(6
  \pm 1.8) \mu$G. They find that turbulence and magnetism are in
  approximate equipartition with a characteristic turbulent Mach number
  $M_{\rm t} = 3.7$ at a median thermal CNM temperature of 50 K.

  Our data are consistent with this proposal; for $M_{\rm t} = 3.7$ the
  median Doppler temperature $T_{\rm D} = 223 $ K of the filaments
  corresponds to a thermal temperature of 52 K. The upper limit for a
  stable CNM with a temperature of 258 K corresponds then to a Doppler
  temperature of 1100 K; Fig. \ref{Fig_Tkin_histo} shows only few
  components in excess of that limit.

  \citet{Heiles2005} did not consider a latitude dependence of the
  derived parameters. Using data from the Arecibo Millennium survey
  \citep{Heiles2003a}, restricting their sources to latitudes $|b| >
  20\degr$ and spin temperatures to $T_{\rm spin} \la 300 $K, we obtain
  a median Mach number $M_{\rm t} = 3.4$. In this case we derive an
  upper limit of $T_{\rm D} = 968 $ K for the stable CNM phase and
  already 5.5 \% of the components plotted in Fig. \ref{Fig_Tkin_histo}
  may be unstable. For $M_{\rm t} = 3.2$, discussed in
  Sect. \ref{width}, this fraction increases to 6.6\%.

  We conclude that in presence of a magnetic field the CNM population
  derived by us can be stable against thermal instabilities. Without
  magnetic field support some fraction of the filaments would probably
  be unstable. According to \citet{Heiles2005} the magnetic pressure is
  $P_{\rm mag}/k_{\rm B} \sim 10400 $, significantly higher than the
  thermal pressure. In Sect. \ref{width} we will show that the thickness
  of the CNM filaments is dictated by the magnetic field, leading to a
  median CNM volume density of $n = 46.6\,{\rm cm^{-3}}$.

  For the WNM at $|b| > 20\degr$ we find, if we disregard turbulence,
  that only 3\% of the components have a Doppler temperature below 5040
  K, the lower limit required for stability. But adopting also for the WNM
  a correction for turbulent motions, with a typical Mach number of
  $M_{\rm t} = 3.7$, we denote that almost 60\% of the components would be
  below the temperature limit required for stability. This exercise
  implies that the amount of unstable WNM gas depends critically on the
  turbulent Mach number which is today only poorly known. 

  The median $T_{\rm D} \simeq 18.400$\,K (Sect.\,5.10 and
    Fig.\,15) in combination with the WNM gas temperature of $T =
    8000$\,K \citep{Wolfire2003} yields a WNM Mach number of $M_{\rm t}
    \simeq 2.3$.  In this case 27\% of the WNM gas would be
    unstable. \citet{Heiles2003b} derived a lower limit of 48\% for the
    CNM-associated WNM gas in the thermally unstable region but the
    lower limit of 28\% to 30\%, determined by \citet{Roy2013,Roy2015},
    is in better agreement with our result.

\subsection{Magnetic pressure confinement for the CNM}
\label{width}

To probe the role of magnetic field confinement for the CNM filaments, we
evaluate in the following the relevant physical parameters. We assume a
homogeneous sheet with a volume density $n$, hosting the turbulent CNM gas
with a Doppler temperature $T_{\rm D}$. Consequently the
gaseous pressure is $P_{\rm int} = n\,{\rm k_B}\,T_{\rm D}$.
 For a thickness $d$ along the line of sight we observe a
column density $ N_{\rm H} = n / d$, hence $ P_{\rm int} = N_{\rm H} k
T_{\rm D} / d$. Assuming now that the shape of the \hi sheet is entirely
confined by a magnetic field $B$ with a magnetic pressure $ P_{\rm mag}
= B^2 / (8 \pi) $, we obtain for pressure balance the condition
\begin{equation}
\label{eq:Pmag}
P_{\rm mag} = P_{\rm int} = N_{\rm H} k_{\rm B} T_{\rm D} / d. 
\end{equation}
According to \citet[][Sect. 3.1.3 \& 9]{Heiles2005}, internal turbulent
motions have in this case to be two-dimensional, restricted to
directions perpendicular to the mean magnetic field. For a turbulent
Mach number $M_{\rm t} = 3.7$ and an Alfv\`en velocity $v_{\rm Alf} =
1.5$ \kms, turbulent motions along the magnetic field lines are thought
to cause shock waves, and such oscillations are strongly damped.
Gravitational instabilities are not expected for the CNM
\citep[][Sect. 7.2]{Heiles2005}.

\begin{figure}[tbp]
\epsscale{0.7}
   \centerline{
   \plotone{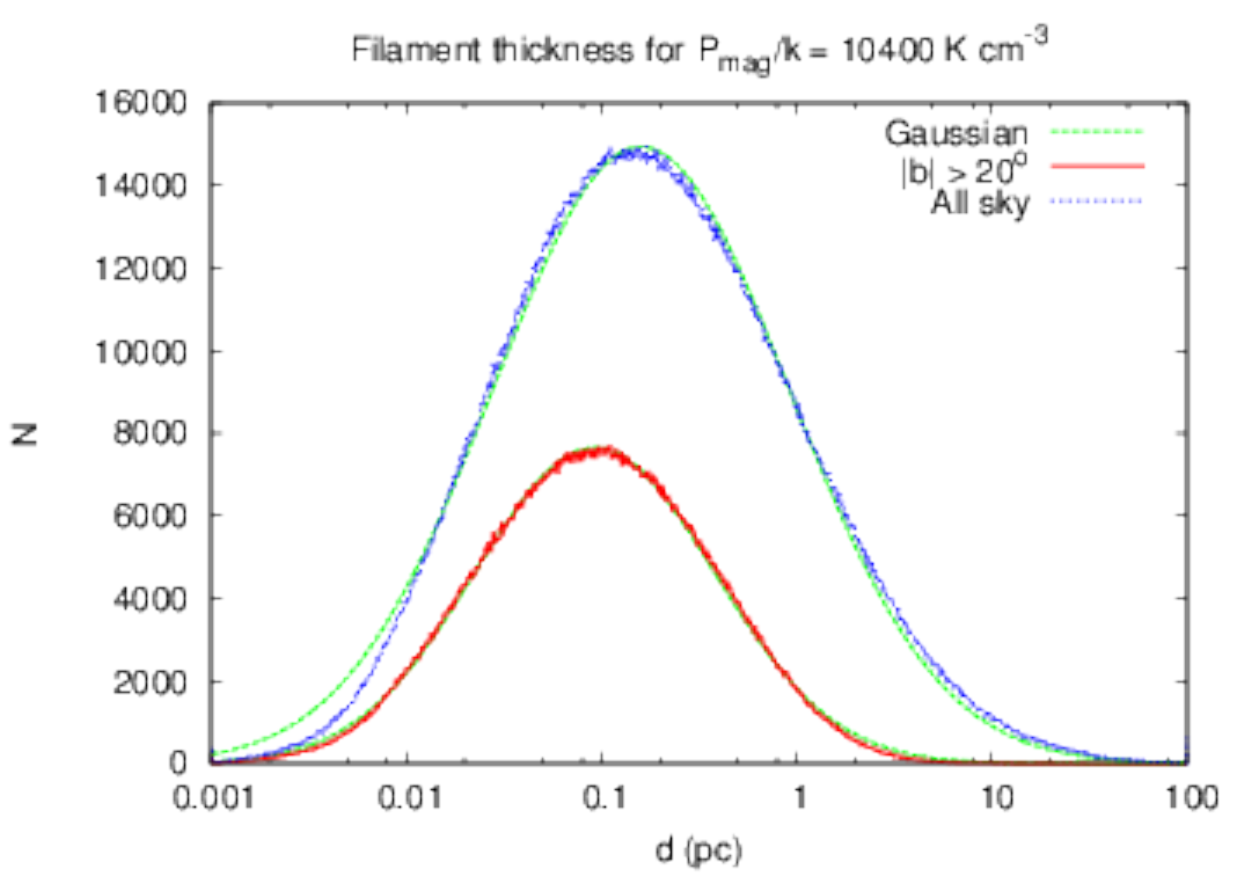}
}
\caption{Distribution function of filament thickness $d$ along the line
  of sight, assuming that the filament shapes are dominated by magnetic
  pressure $P_{\rm mag}$. The median for $|b| > 20\degr$ is $d = 0.09$
  pc.}
   \label{Fig_thickness}
\end{figure}

From our observations we deduce $N_{\rm H}$ and $T_{\rm D}$. With a
magnetic pressure $P_{\rm mag} = 10400 $ {\rm K cm}$^{-3}$ from
\citet{Heiles2005}, we can derive $d$.  Figure \ref{Fig_thickness} shows
the PDF for the derived filament depths $d$.  We obtain a well-defined
log-normal distribution, for $|b| > 20\degr$ with a peak at $d = 0.0927$
pc, the median filament thickness is $d = 0.0929$ pc. The all sky
distribution has a maximum at $d = 0.160$ pc with a median $d =
0.166$ pc.

We obtain here only 1/3 of the upper limit for $d$ derived in Sect.
\ref{volume} from estimates based on the median distance to the wall of
the local cavity \citep{Lallement2014}. Consistently, for a median
$T_{\rm D} = 223$\,K we get also a characteristic volume density $n =
46.6\,{\rm cm^{-3}}$ that is three times as large as determined
previously.  The single dish telescopes used are limited in angular
resolution to structures of 0.3\,pc size at 100\,pc distance. Systematic
velocity gradients below this angular resolution limit can be traced
because of the low gas temperatures. Velocity gradients are observed
frequently on larger scales perpendicular to the main axis of the
filament (see e.g. Figs. \ref{FIG_filament}, \ref{FIG_Cfilament}, and
\ref{FIG_Dfilament}) and sometimes along the filament
(Fig. \ref{FIG_Bfilament}).  Note, the determination of the filament
  thickness according to Eq. \ref{eq:Pmag} is not a function of distance
  or geometry and thus independent on the angular resolution of the HI
  data analyzed.  The derived thickness $d = 0.09$ pc (from Eq.
  \ref{eq:Pmag}) corresponds only to 1/3 of our spatial resolution, so
  it is a lower limit for the filament thickness with an uncertainty of
  about 30\% due to uncertainties of the average magnetic field strength
  \citep{Heiles2005}.

  The derived median filament thickness of 0.09 pc depends stongly on
  the hypothesis of magnetic pressure confinement and it may be
  questionable whether such a case applies. The Arecibo telescope has a
  far better spatial resolution but \citet[][Sect. 8.1]{Clark2014} claim
  that the filaments are {\it ``largely unresolved and therefore correspond
  to $<0.12$ pc''} at a distance of 100 pc. Within the uncertainties this
  estimate is compatible with our result and strengthens the case for a
  magnetic pressure confinement. The Square Kilometre Array (SKA) is
  needed to resolve the detailed spatial structure of the filaments.  

In case of a thermal equilibrium the median volume density of $n =
46.6\,{\rm cm^{-3}}$ yields at a pressure of $P={n \rm k_B} T =
3070\,{\rm K cm^{-3}}$ a gas temperature of $T = 65$\,K.  The
  corresponding median Mach number is $M_{\rm t} = 3.2$. Using data from
  the Arecibo Millennium survey \citep{Heiles2003a}, restricting their
  sources to latitudes $|b| > 20\degr$ and spin temperatures to $T_{\rm
    spin} \la 300 $K, we obtain medians $M_{\rm t} = 3.4$ and $T_{\rm
    spin} = 61 $ K. Both values are also within the uncertainties
  consistent with our results. The derived gas parameters, adopting
  magnetic pressure confinement, describe a stable CNM gas phase and fit
  best to the model assumptions \citep[][Sect. 6.1]{Wolfire2003} with an
  average temperature $T = 85 $ K and density $n = 32.9\,{\rm
    cm^{-3}}$. 

Confinement of gaseous phases is not limited to the atomic phase only,
it is applicable also to molecular gas. In this respect latest {\it
  Herschel} measurements \citep{Andre2014} are of particular
interest. Consistent with the values derived for the atomic phase
\citet{Andre2014} find filaments in dense star forming region with
linear extents of $d = (0.09\,\pm\,0.04)$\,pc.
Moreover, \citet{Federrath2016} concludes
from high-resolution MHD simulations that follow the evolution of
molecular clouds and the formation of filaments and stars, that there is
a remarkably universal filament width of 0.10 pc which is independent of
the star formation history of the clouds. 

Because of saturation effects, \hi filaments in regions with molecular
gas are not well defined but we find a general trend that filaments in
such regions are more extended. This is expected since molecular gas
condensations are mostly located interior to \hi filaments.  Close to
the Galactic plane, for $|b| < 20\degr$, we find $d \sim 0.28$ pc but it
is unclear, how far this result is biased by confusion. 

\subsection{Global velocity distribution of the filaments}
\label{velocities}

\begin{figure}[tbp]
\epsscale{0.7}
   \centerline{
   \plotone{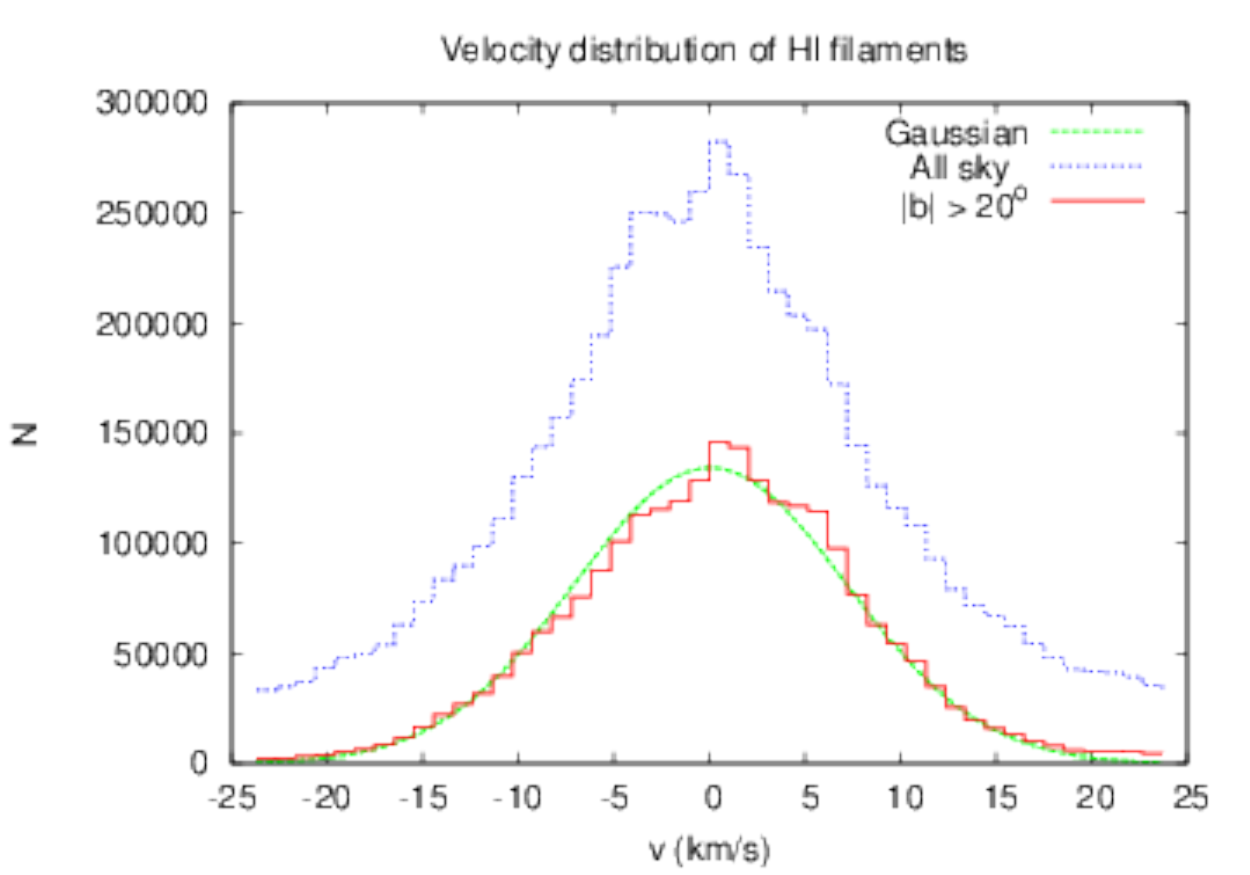}
}
   \caption{Velocity distribution for the major local \hi filaments,
     compared with a Gaussian fit. }
   \label{Fig_vel_histo}
\end{figure}

The histogram in Fig. \ref{Fig_vel_histo} displays the global
distribution of center velocities for all USM filaments. The red curve
gives the distribution for latitudes $|b| > 20 \degr$ and can be
approximated by a Gaussian with a FWHM width of $\Delta v = 16.8$ \kms,
centered at $v = 0.05$ \kms. 

The Gaussian shape of the velocity distribution is remarkable since in
theory there is no consensus that the probability density distribution
of the velocity has to be Gaussian \citep{Elmegreen2004}. According to
the central limit theorem the arithmetic mean of a sufficiently large
number of independent random variables, each with a well-defined mean
and well-defined variance, will be approximately normally
distributed. Around the Sun we have a random distribution of filaments,
triggered by random events. Apparently, the presence of distinct
filaments, each with well defined velocity centroids, does not preclude
that we have {\it in total} a random distribution. The channel maps
displayed in Figs. \ref{Fig_Filaments1}, \ref{Fig_Filaments1b}, and
  \ref{Fig_Filaments1c}  are chosen to be at the velocities of the peak
and half width points, hence they are characteristic for local
filaments.

From the Gaussian shape of the velocity distribution in
Fig. \ref{Fig_vel_histo} we infer that our sample of CNM clouds, which
is complete for $|b| > 20 \degr$, represents a well defined and unbiased
ensemble of objects in turbulent motion.  The width of the velocity
distribution is then a measure of the kinetic energy. We derive a formal
turbulent temperature of 6500 K. This value is low in comparison to the
median Doppler temperature of the WNM gas, $T_{\rm D} \sim 18374$ K
(Sect. \ref{environment}) but fits to the allowed thermal temperature
range of $5040 \la T \la 8310$ K for the WNM \citep{Wolfire2003}. Our
result is consistent with numerical simulations of \citet{Saury2014} on
the structure of the thermally bistable and turbulent atomic gas in the
local interstellar medium.  They found that {\it ``the structure of the
  CNM and WNM are tightly interwoven: the two media share the same
  velocity fields and the CNM cloud-cloud velocity dispersion is close
  to the WNM sound speed''}.

  Including the Galactic plane in our analysis we find significantly
  more filaments at high radial velocities (blue curve in
  Fig. \ref{Fig_vel_histo}). This is consistent with
  Fig.\,\ref{Fig_Filaments3} (botom) that shows at low latitudes some signs of
  Galactic rotation. Once more we find indications that our definition
  of major local filaments appears to be ill-defined at low Galactic
  latitudes, the ensemble properties are significantly affected by
  non-local gas.

\subsection{Kinetic energy distribution}

\begin{figure*}[tbp]
\epsscale{1.}
   \centerline{
   \plotone{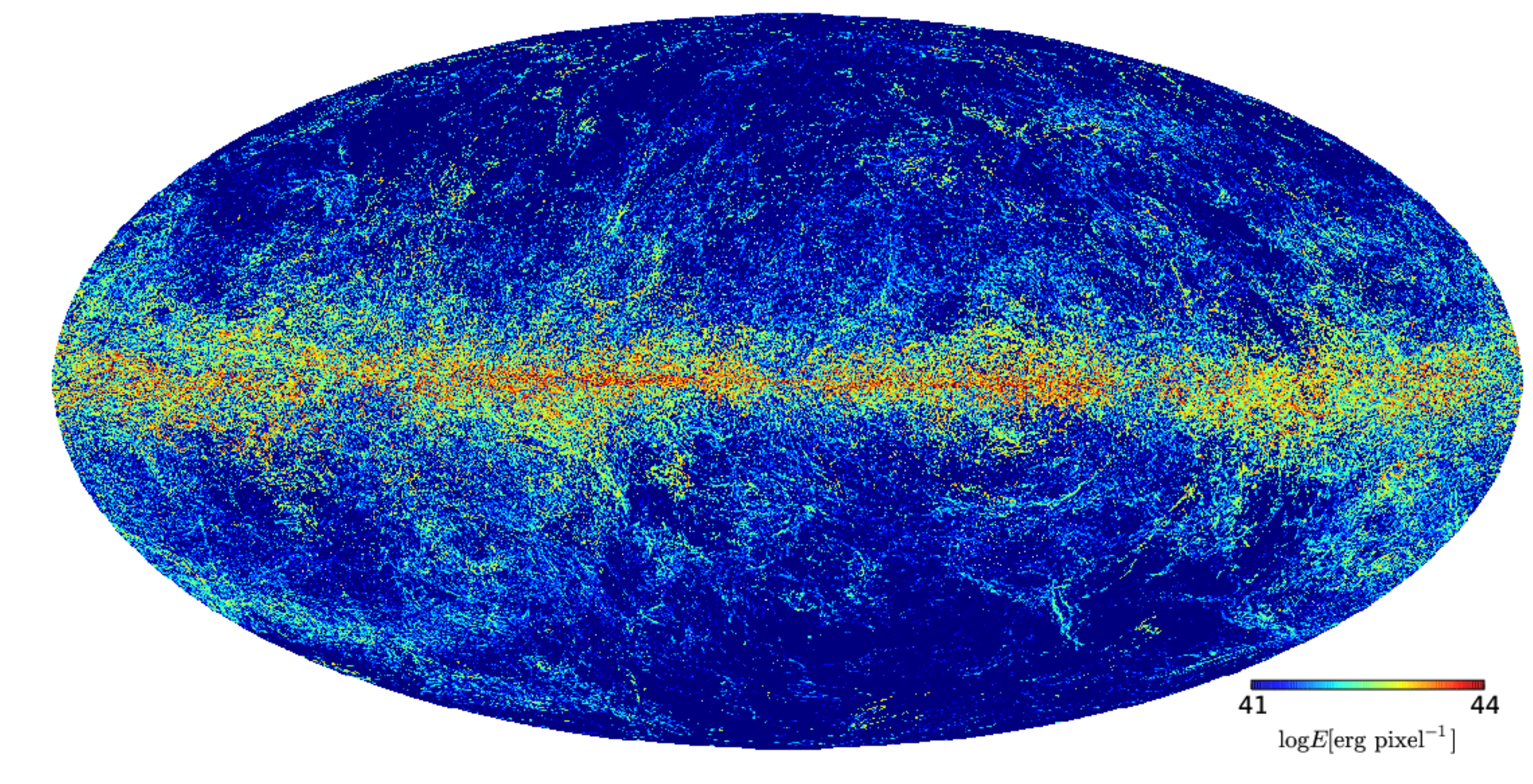}
}
   \caption{All sky distribution of kinetic energy (ergs/pixel) for
     filaments on a logarithmic scale. We assume that the \hi gas is at
     a distance of 100 pc. }
   \label{Fig_NH_VV}
\end{figure*}

Filaments, if caused by blast waves originating from supernovae, are
thought to resemble the dynamical properties of their origin. We expect
that these sources are located around us close to the Galactic
plane. Filaments should tell us then about the mechanical energy that
was injected into the ISM. For observed radial velocities $v$ of
filaments and USM column densities $N_{\rm Huon}$ we calculate an all
sky map of the kinetic energy distribution. Here we assume for all
  \hi filaments a distance of $D = 100$ pc. We apply a brightness
  temperature threshold 0.3 K.

  For an individual pixel in the nside = 1024 map we obtain a kinetic
  energy
\begin{equation}
\label{eq:mvv}
\frac{E}{[{\rm erg}]} = 7.96\, 10^{20}\,\frac{N_{\rm Huon}}{[{\rm
    cm^{-2}}]} 
\frac{v^2}{[({\rm km\, s^{-2}})^2]}
\frac{D^2} {[({\rm 100 \,pc})^2]}
\end{equation}

Figure \ref{Fig_NH_VV} shows this estimate for the energy distribution
in filaments, the look up table is logarithmic in ergs/pixel. The
  total energy, integrating over the sphere, is $10^{49.3}$ ergs. 
For low latitudes, $ |b| < 20\degr$, we find high energies but this is
subject to confusion, also most probably significantly affected by
non-local sources. The distribution at high latitudes deviates
significantly from Figs. \ref{Fig_Filaments2} and
\ref{Fig_Filaments3}. The reason is that the kinetic energy in
  Fig. \ref{Fig_NH_VV} depends on $v^2$ while Figs. \ref{Fig_Filaments2}
  and \ref{Fig_Filaments3} are not weighted by radial velocity.

Our estimate for the energy distribution suffers certainly from unknown
distances. But we do not see obvious structures that show signatures of 
the walls from the local cavity \citep{Lallement2014}. The large
scale filaments at intermediate negative latitudes in the anti-center are
probably at larger distances \citep[][Figs. 11 \& 12]{Lallement2014}.

\section{Summary and conclusions } 
\label{discussion}
We explored the data of the first Galactic EBHIS data release and
merged them with the GASS III survey. Using USM methods we extracted
the \hi distribution on scales of $11'$ to $16'$ and derived
all sky maps of unprecedented quality. Filamentary \hi gas structures
are visible nearly everywhere. Analyzing the most prominent structures
(major filaments) we demonstrate that these filaments show a good 
agreement with the polarized Galactic dust emission observed by {\it
  Planck\/}.

All the sky is full of filaments but restricting our analysis to the
most prominent local structures for USM temperatures $T_{\rm Uon} > 1$K,
corresponding to a $10 \sigma$ threshold, we obtain a distribution that
is in surprisingly close agreement with the dust emission observed by
{\it Planck\/} at 353 GHz (Figs. \ref{Fig_Filaments2},
\ref{Fig_Filaments3}, and \ref{Fig_Filaments3b}). Individual filaments
can be traced in detail, but in the Galactic plane we find confusion. We
draw the conclusions of our analysis for latitudes $|b| > 20\degr$ but
demonstrate also that in the Galactic plane most of the derived
parameters are not seriously affected by confusion.

We do not attempt to correlate \hi filaments with {\it Planck\/}
polarization data at 353 GHz. From a direct comparison between \hi and
{\it Planck\/} 353 GHz maps it is obvious that we trace similar features
(our Figs.  \ref{Fig_Filaments2} to 
\ref{Fig_Filaments3}). Also the Hessian analysis gives identical results
(Sect. \ref{Hesse}, see also Fig. 3 of
\citet{Planck2016}). From {\it Planck\/} data it is evident
that for low \hi column densities $N_{\rm H} < 10^{20}$ cm$^{-2}$
filaments lie along magnetic field lines
\citep{PlanckXIXa,PlanckXX2015b,Planck2016}. Such a correlation
was also shown by \citet{Clark2014,Clark2015} who used Arecibo data.  So
we can well assume that \hi filaments studied by us are associated with
magnetic fields.

The full-sky single dish HI surveys allow for the first time to
establish a complete census of the local CNM. All high Galactic latitude
filaments, but also the the bulk of the gas located close to the
Galactic plane, share a common set of physical parameters describing the
ISM conditions.  The gas is characterized as a CNM with a median Doppler
temperature of $T_{\rm D} = 223$ K; it has column densities around
$N_{\rm H} = 10^{19.1}$ cm$^{-2}$.  The probability density
distributions are very well approximated by a log normal one and
independent of position.  The maxima of the PDFs agree with the
medians. Filaments appear isolated but may be accompanied by other
filamentary structures, similar to striations, in the vicinity.

The widhts of the CNM filaments are unresolved by our
telescopes. Adopting a typical filament distance of 100 pc to the wall
of the local cavity \citep{Lallement2014} we derive for the CNM sheets a
thickness of $d \la 0.3$ pc and correspondingly a volume density of $n
\ga 14$ cm$^{-3}$ (Sect. \ref{volume}).  \citet{Clark2014} claim that
filamentary structures, observed by them, are unresolved with the
Arecibo telescope. This implies a sheets thickness of $d \la 0.12$ pc.
Considering in addition pressure confinement from a magnetic field of
$B_{\rm tot} \sim (6.0 \pm 1.8)\mu$G \citep{Heiles2005} we derive a
distance independent lower limit of $ d =0.09$ pc for the median sheet
thickness, consistent with the Arecibo results. Such a sheet depth
implies an upper limit for the median volume density of $n \la 47$
cm$^{-3}$.  Our analysis discloses that the CNM is organized in cold
filaments that are aligned with, and dominated by, magnetic fields.

These filaments are embedded in a more extended warm gas phase with
column densities that are an order of magnitude higher
(Sect. \ref{environment}). From our data alone it is difficult to
comment on the stability of the two phases.  In Sect. \ref{Twophases} we
consider the CNM and conclude that for a turbulent Mach number of $3.2
\la M_{\rm t} \la 3.7$ at least 93.4 \% of the CNM filaments should
populate the stable part of the phase diagram \citep{Wolfire2003}.  In
case of magnetic pressure confinement (Sect. \ref{width}) all CNM
filaments can be safely considered to be stable. The stability issue is
less clear for the WNM. We argue in Sect. \ref{Twophases} that 30\% of
the WNM that is in contact with the CNM filaments may be unstable.
 
Our analysis contains a few parameters affected by systematic
biases. From emission data alone it is not feasible to determine optical
depth effects. A rough comparison with Arecibo absorption observations
(Sect. \ref{depth}, data from \citet{Heiles2003b}) indicates, within the
accuracy of both analyses, consistent results except for optical depths
$ \tau \ga 0.5$. For optically thick gas we seriously underestimate the column
densities, hence also derived volume densities are biased. We determine
for the optical thin filaments a mass fraction $F_{\rm tot}= 0.2 \pm
.01$. This is low in comparison to the canonical value of 0.40 
\citep{Heiles2003b} but we estimate that if only about 10\% of the
observed filaments are optically thick the apparent discrepancy 
may cease. 

The velocity structure of the local gas filaments is remarkable. We find
for the whole sample a Gaussian velocity distribution, centered at $v
\sim 0$ \kms, with a FWHM spread of $\Delta v = 16.8$ \kms. On large
scales filaments have frequently companion features at similar
velocities. We find some USM structures that can not be described as
obvious filaments. Partly such fragments look like disintegrated
features from filaments, partly they may be caused by projection
effects, as a result of apparently crossing filaments. In fact, any
geometry is possible; i.e. the spider \citep[][Fig. 9 \& 13]{Planck2011} is
caused by several crossing filaments, easily traceable in the USM
channel maps. 

Using USM with a fixed Gaussian smoothing kernel (effective FWHM of
$30'$) may cause biases in the analysis because spatial frequencies
corresponding to the FWHM beam sizes of the Parkes and Effelsberg
telescopes are preferred. All derived parameters were compared with
those by \citet{Clark2014,Clark2015} who used predominantly Arecibo data
for their analysis. In all cases we found excellent agreement. The
methods that have been used by this team are very different from
ours. This implies that neither our analysis nor that of
\citet{Clark2014,Clark2015} are strongly affected by systematical
effects. Also there is no recognizable dependence on telescope size or
spatial resolution. Arecibo is far ahead with respect to the beam width
and velocity resolution but unfortunately restricted with respect to the
observable portion of the sky.

The most striking result from our investigation is the large number of
extended cold coherent filamentary structures that we observe. A number
of these features reach angular scales of $\ga 20\degr$, implying a
total length of $\ga 36$ pc at an assumed distance of 100 pc. CNM sheets
of this kind were previously observed and discussed by
\citet{Heiles2003b}. A median sheet width of 0.09 pc implies
length-to-thickness aspect ratios up to 400 or more. For comparison, the
edge-on aspect ratio of a CD is 100. The two-phase model is based on
observational evidence that the CNM is embedded in the WNM, giving rise
to the so called ``raisin pudding'' model \citep{Field1969}. Our data
imply that the CNM clouds are very different from standard clouds
\citep{Spitzer1968,McKe1977} or ``raisins'' \citep{Clark1964}. What we
observe are rather ``steamrollered raisins'', flatter than a CD, and
warped by a turbulent ISM. The magnetic field appears to play a major
role in this process.

Initially we intended to explore the quality of an all-sky survey. The
combined EBHIS/GASS III survey shows an excellent performance, acting
well as a pathfinder for future SKA observation.

   \begin{acknowledgements}

     We acknowledge the referee’s careful reading and detailed criticism
     and Mark Wolfire for comments on the two-phase medium. The authors
     thank the Deutsche Forschungsgemeinschaft (DFG) for support
     under grant numbers KE757/7-1, KE757/7-2, KE757/7-3 and
     KE757/9-1. L.F. was partially funded by the International Max
     Planck Research School for Astronomy and Astro- physics at the
     Universities of Bonn and Cologne (IMPRS Bonn/Cologne). L.F. is also
     a member of IMPRS Bonn/Cologne. D.L. is a member of the
     Bonn–Cologne Graduate School of Physics and Astronomy (BCGS). U.H.
     was supported by institutional research funding IUT26-2 of the
     Estonian Ministry of Education and Research and by Estonian Center
     of Excellence TK120.  This research has made use of NASA’s
     Astrophysics Data System.  EBHIS is based on observations with the
     100-m telescope of the MPIfR (Max-Planck-Institut f\"ur
     Radioastronomie) at Effelsberg. The Parkes Radio Telescope is part
     of the Australia Telescope, which is funded by the Commonwealth of
     Australia for operation as a National Facility managed by CSIRO.
   \end{acknowledgements}

%\Online

\newpage

%\begin{appendix} %First online appendix
\appendix
\label{appendix}
%-----------------------------------------------------------------------

\section{Supplementing $T_{\rm B}$ and USM maps}

To demonstrate changes in brightness temperature and derived USM maps we
supplement here Fig. \ref{Fig_Filaments1} with data at velocities of
$v=$ -8, and 8 \kmss.

%-----------------------------------------------------------------------
\begin{figure*}[hbp]
\epsscale{.8}
   \centerline{
   \plotone{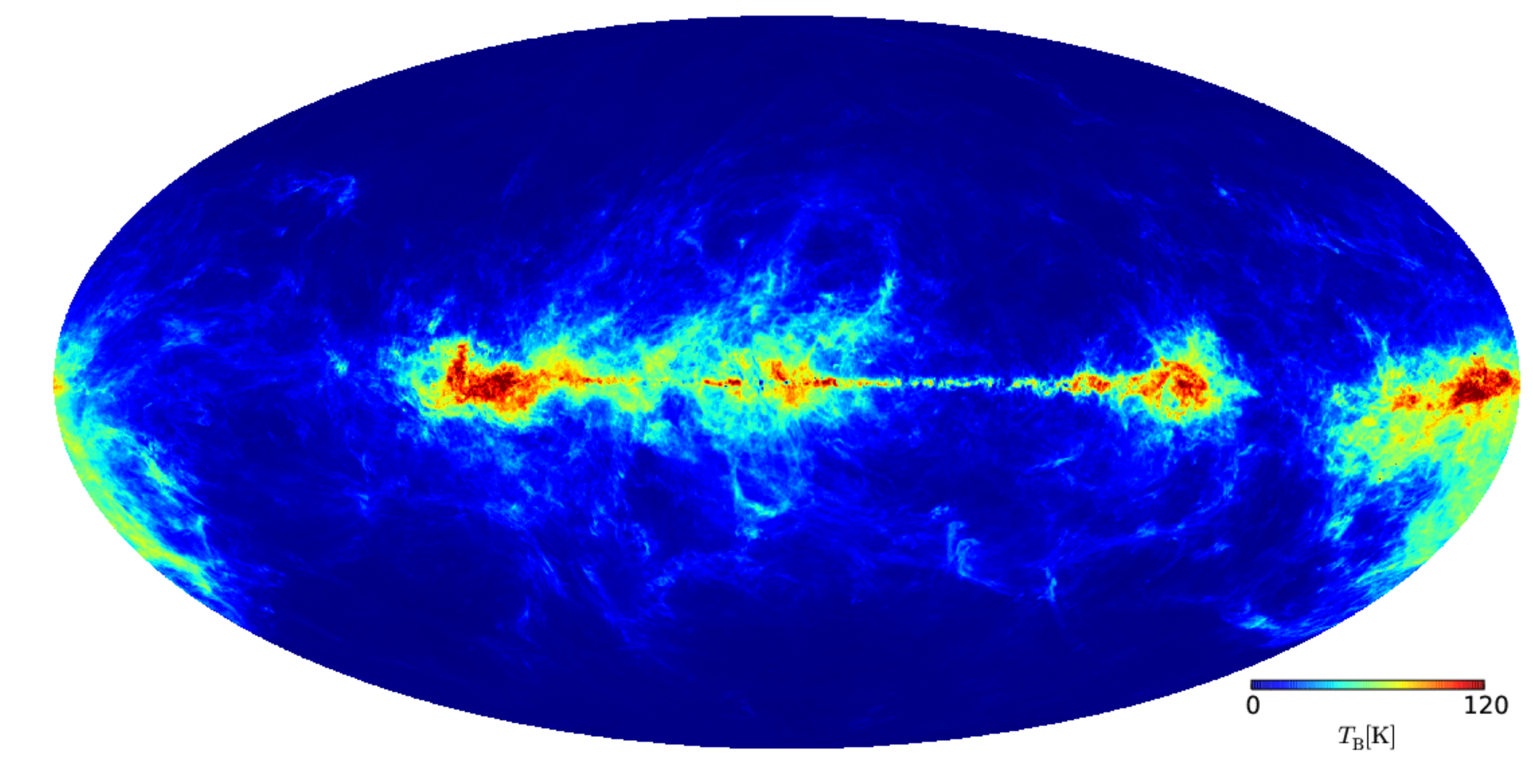}
}
   \centerline{
   \plotone{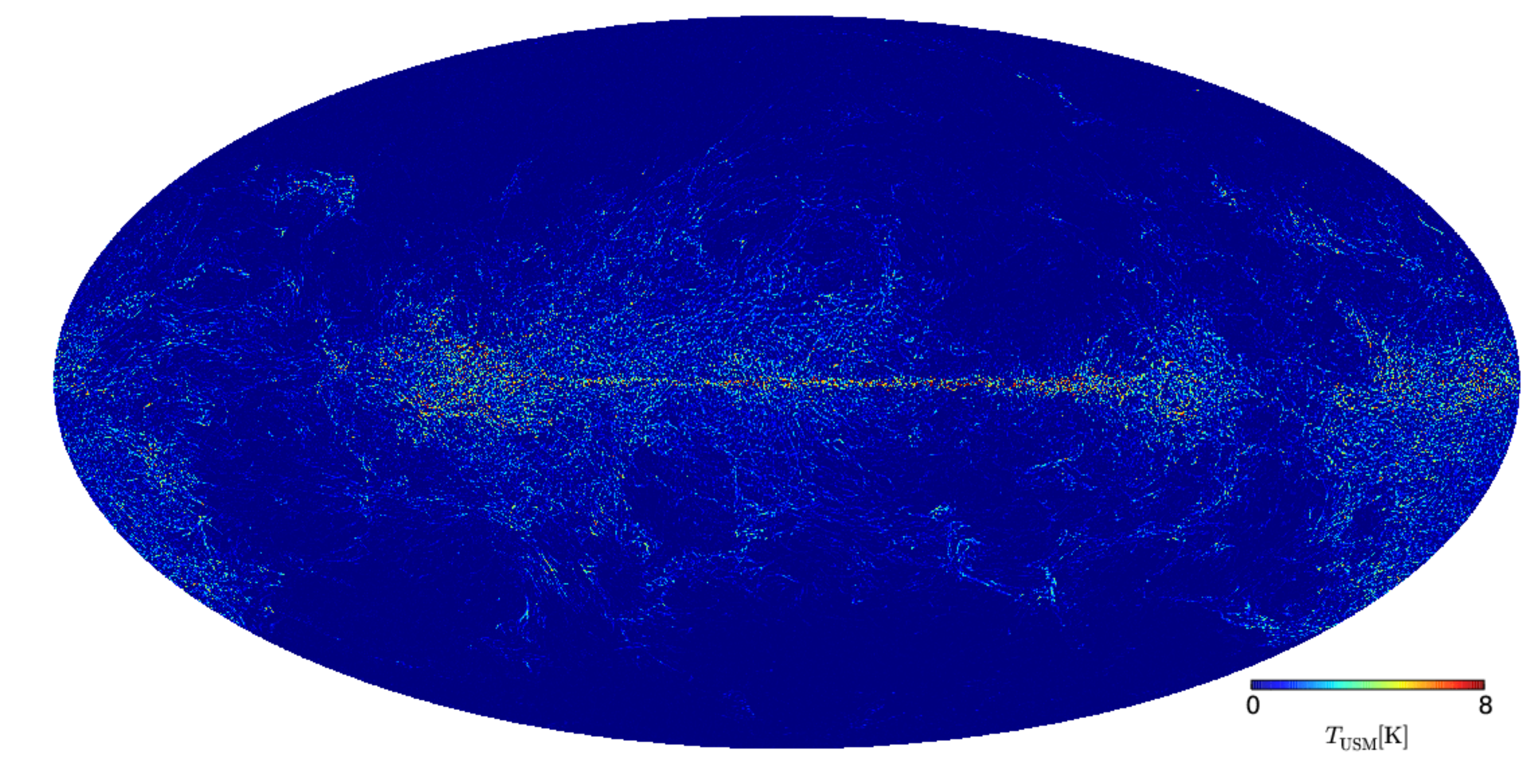}
}
\caption{All sky Mollweide display of the observed \hi
  brightness temperature distributions at a velocity of $v=-8$ 
  \kmss{\it (top)} and filamentary structures, derived by unsharp
  masking, at the same velocitiy (bottom).}
        \label{Fig_Filaments1b}
\end{figure*}
%-----------------------------------------------------------------------

%-----------------------------------------------------------------------
\begin{figure*}[tbp]
\epsscale{.8}
   \centerline{
   \plotone{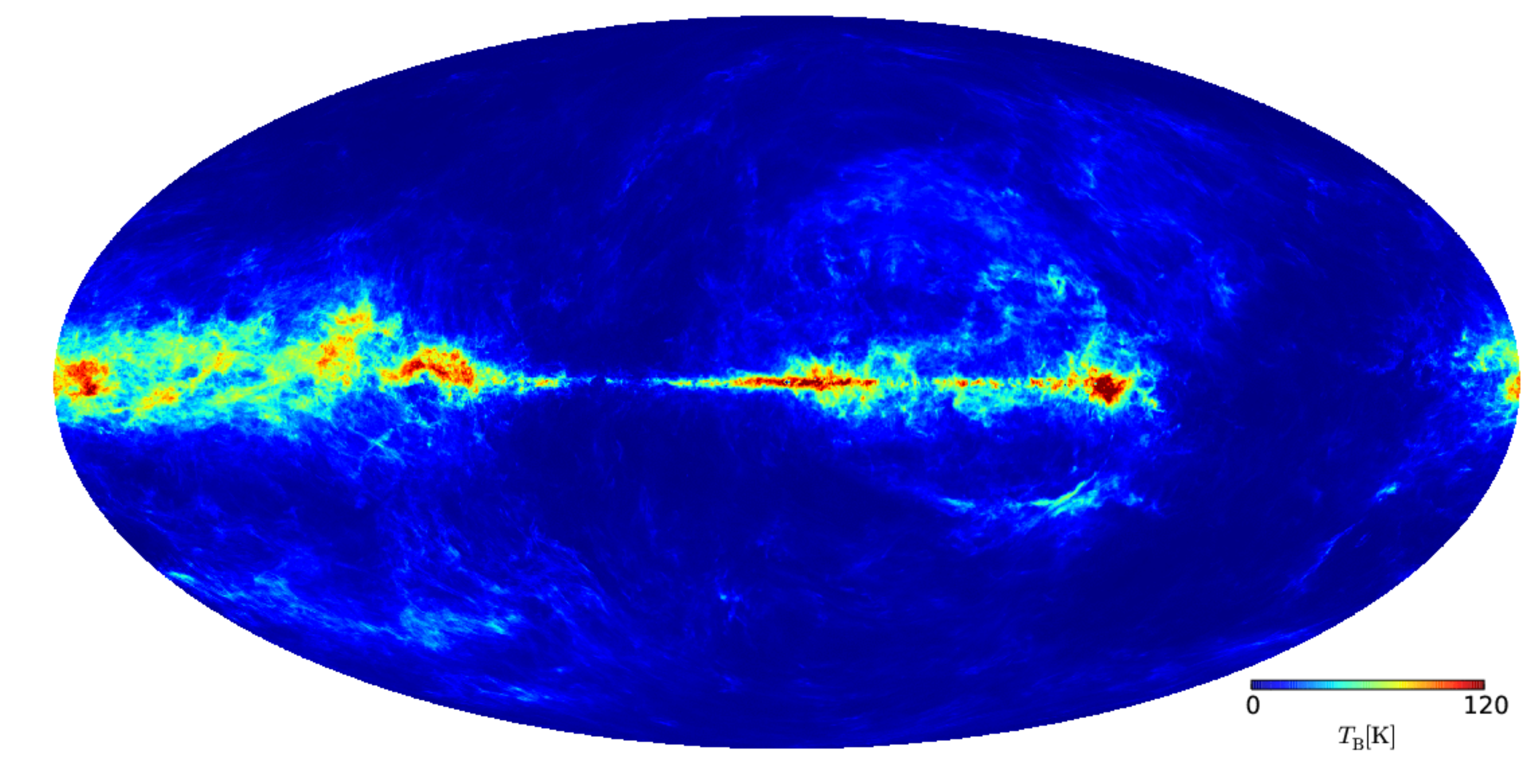}
}
   \centerline{
   \plotone{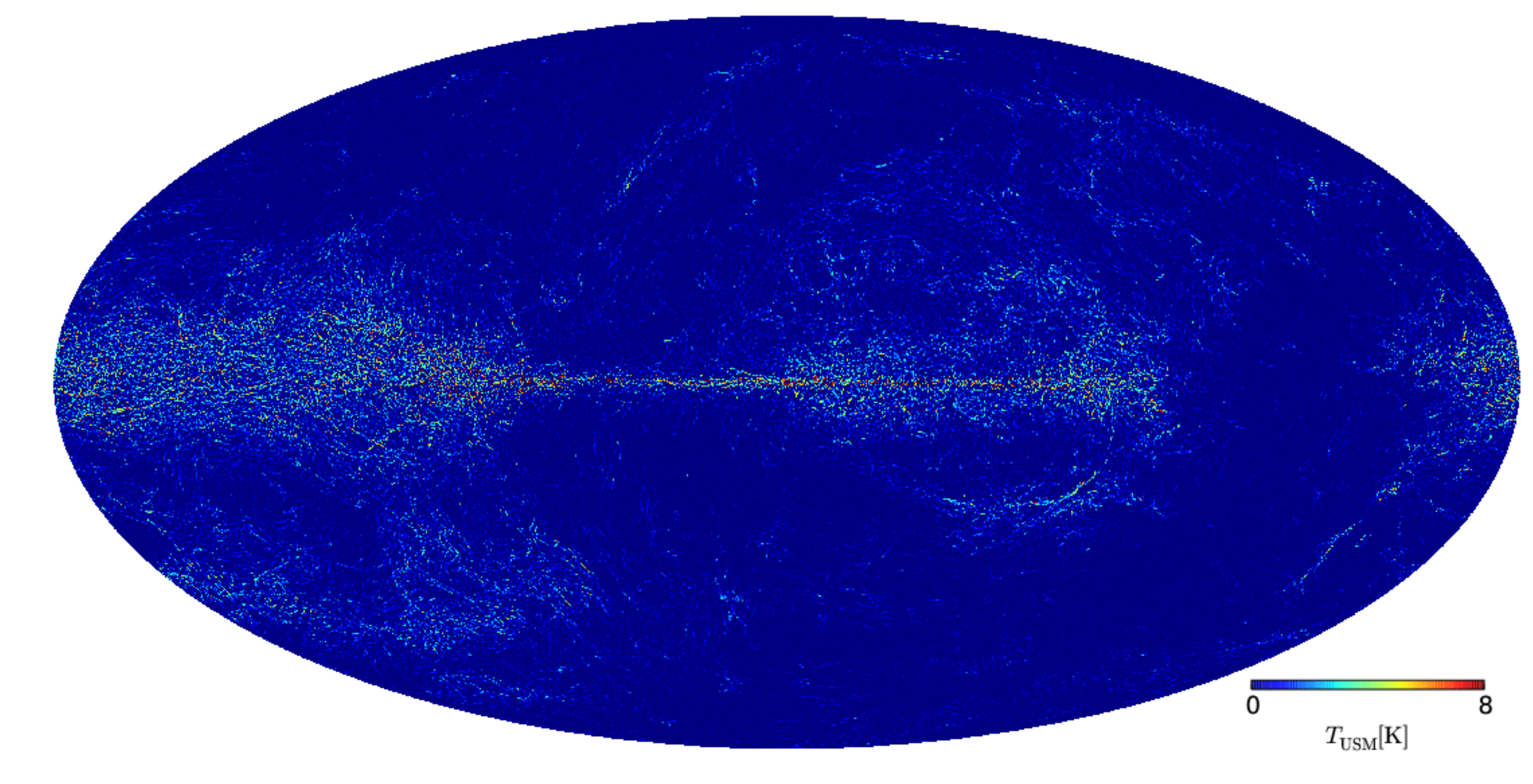}
}
\caption{All sky Mollweide display of the observed \hi
  brightness temperature distributions at a velocity of $v=+8$ 
  \kmss{\it (top)} and filamentary structures, derived by unsharp
  masking, at the same velocity (bottom).}
        \label{Fig_Filaments1c}
\end{figure*}
%-----------------------------------------------------------------------

\newpage

%-----------------------------------------------------------------------

\section{Prominent filamentary structures}

\begin{figure*}[hbp]
   \centerline{
   \includegraphics[scale=0.3, angle=270]{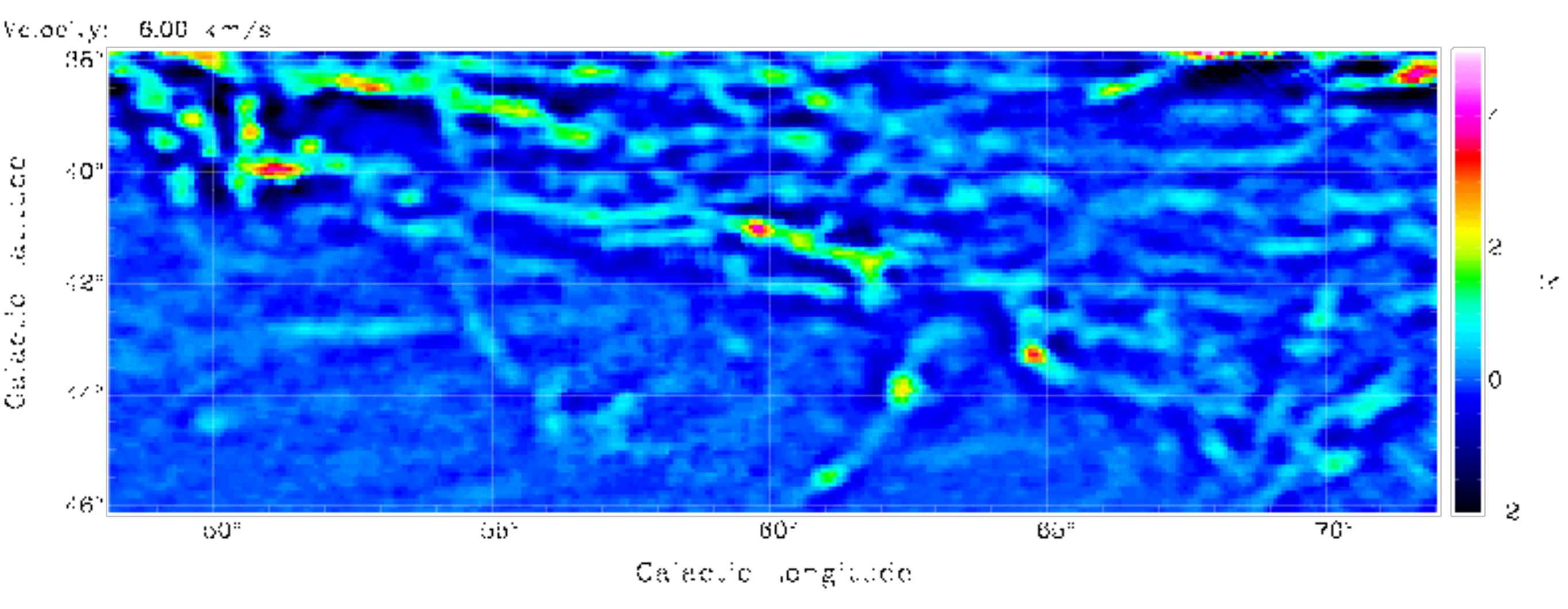}
   \includegraphics[scale=0.3, angle=270]{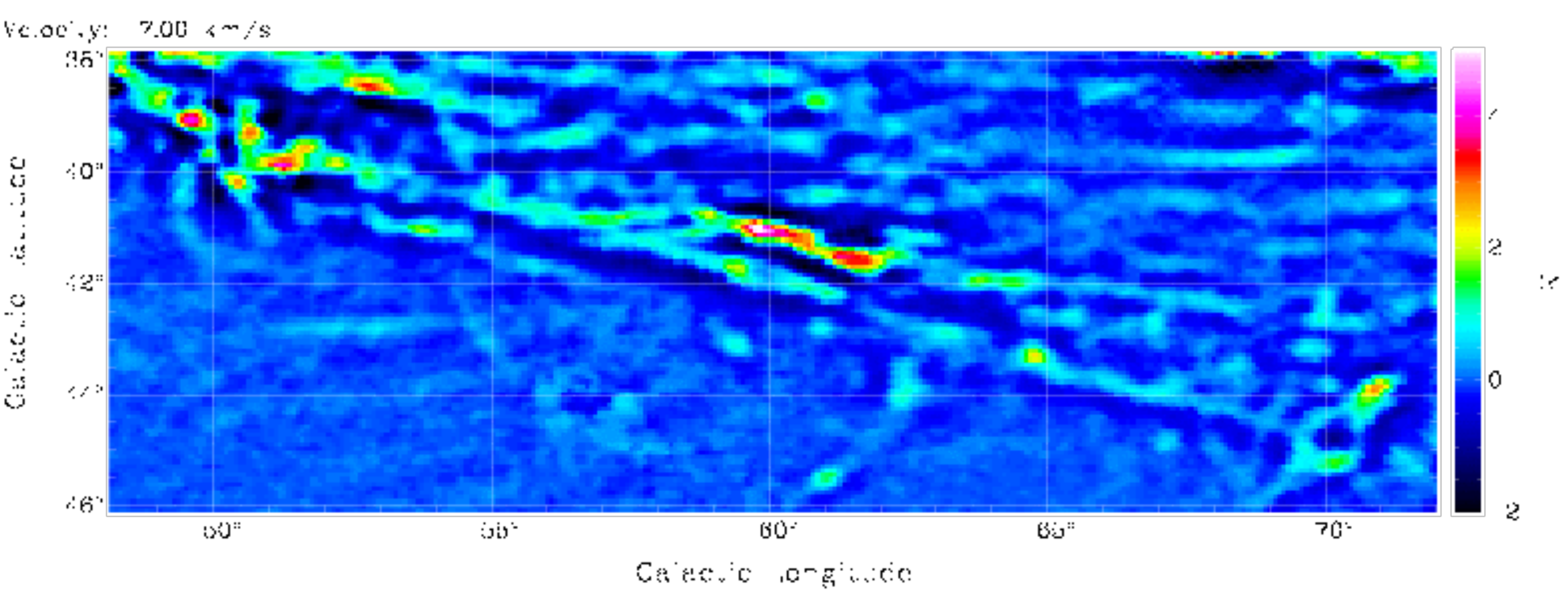}
}
   \centerline{
   \includegraphics[scale=0.3, angle=270]{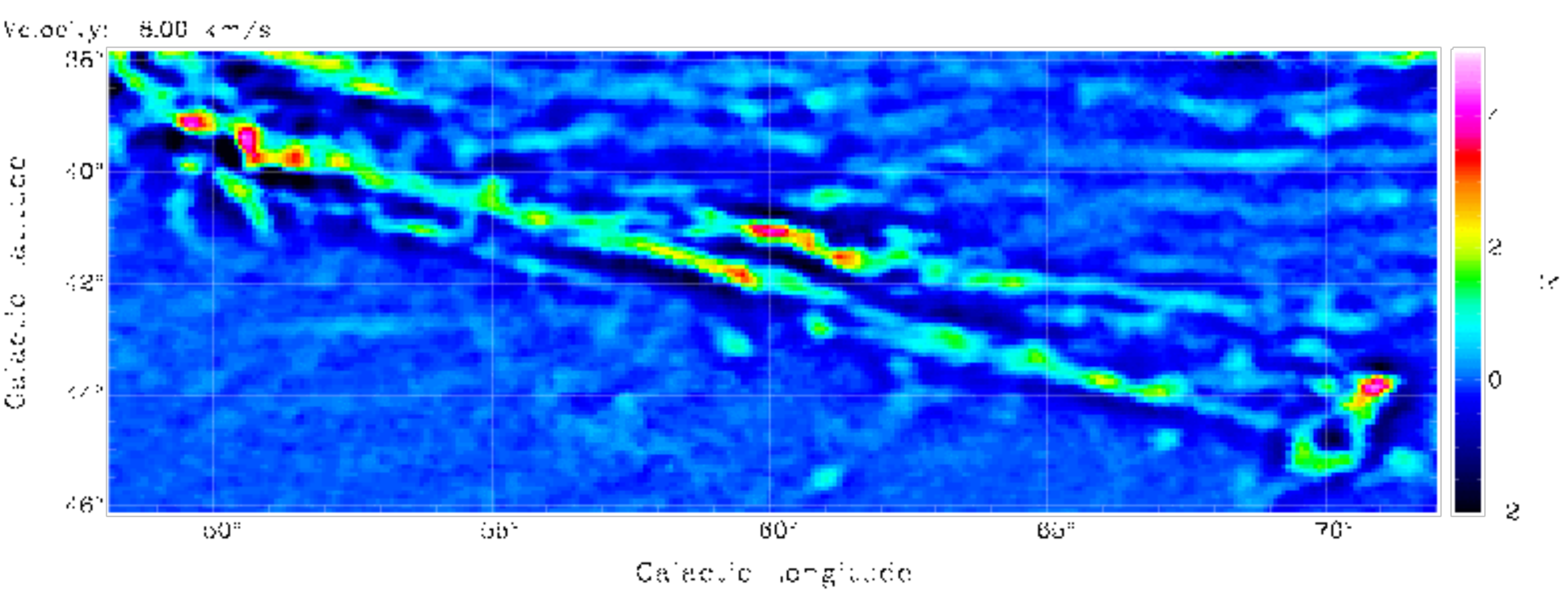}
   \includegraphics[scale=0.3, angle=270]{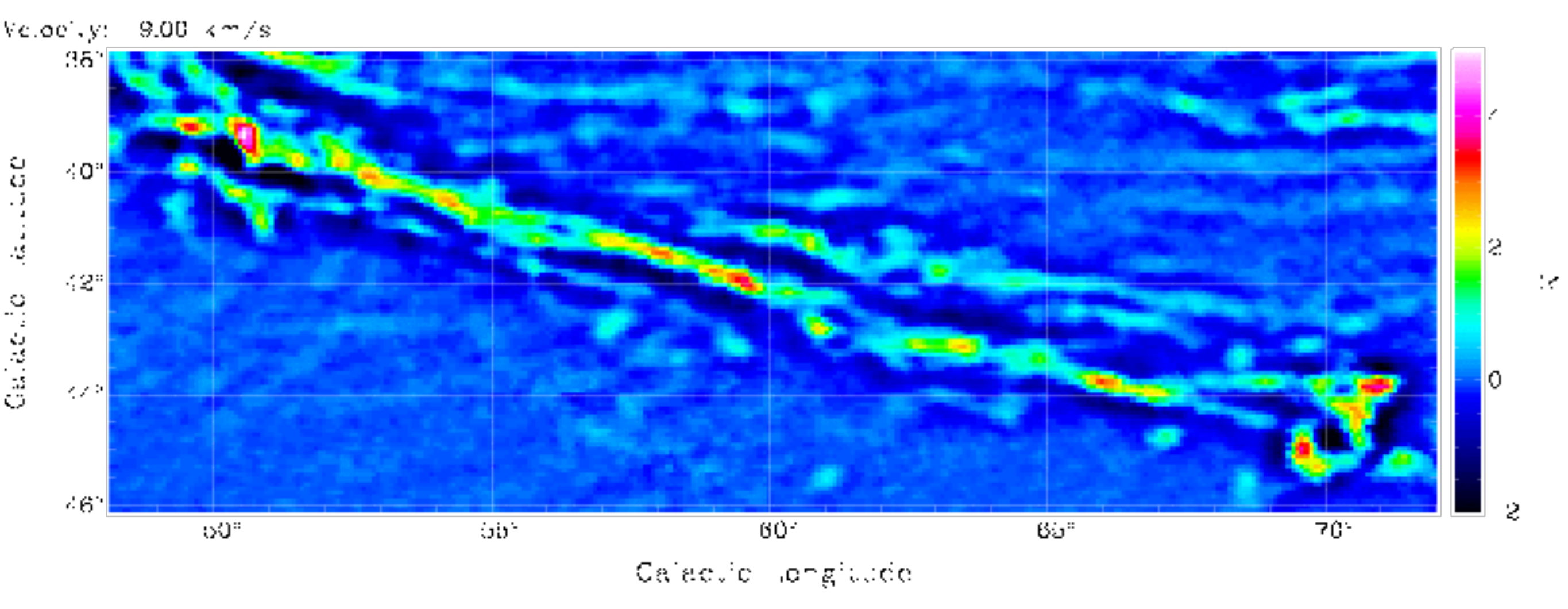}
}
   \centerline{
   \includegraphics[scale=0.3, angle=270]{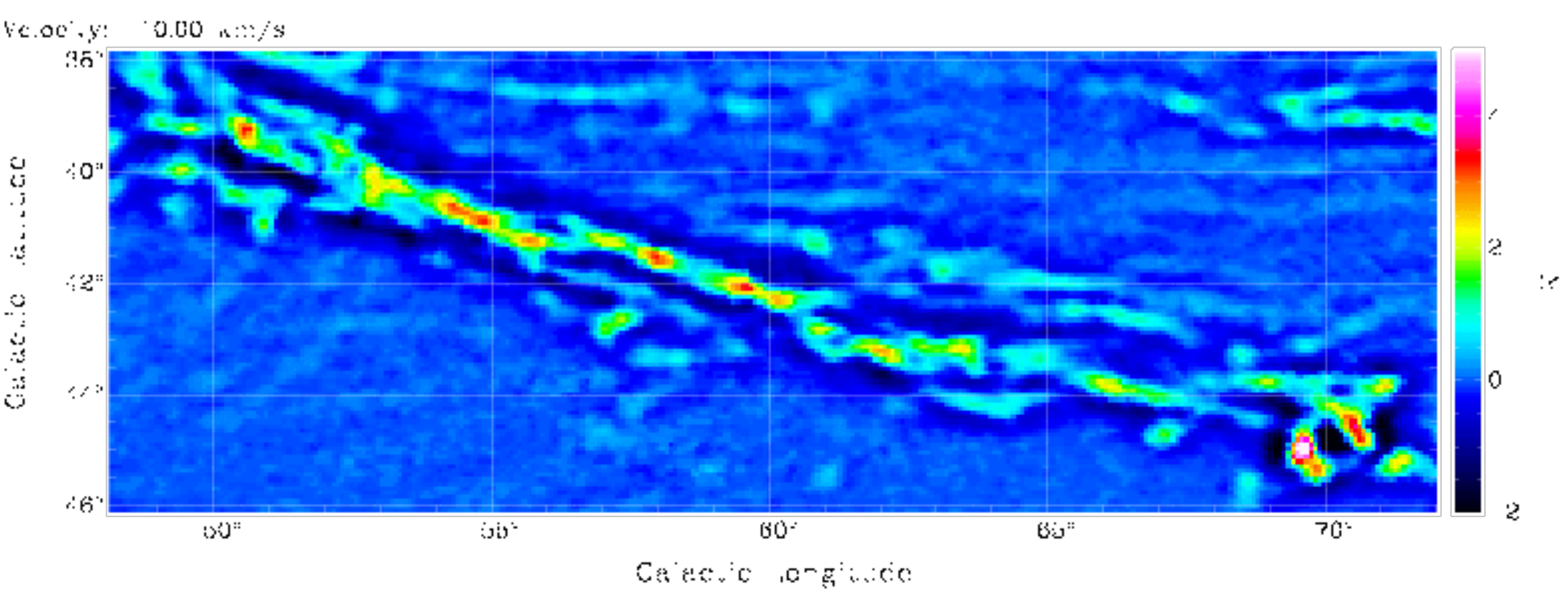}
   \includegraphics[scale=0.3, angle=270]{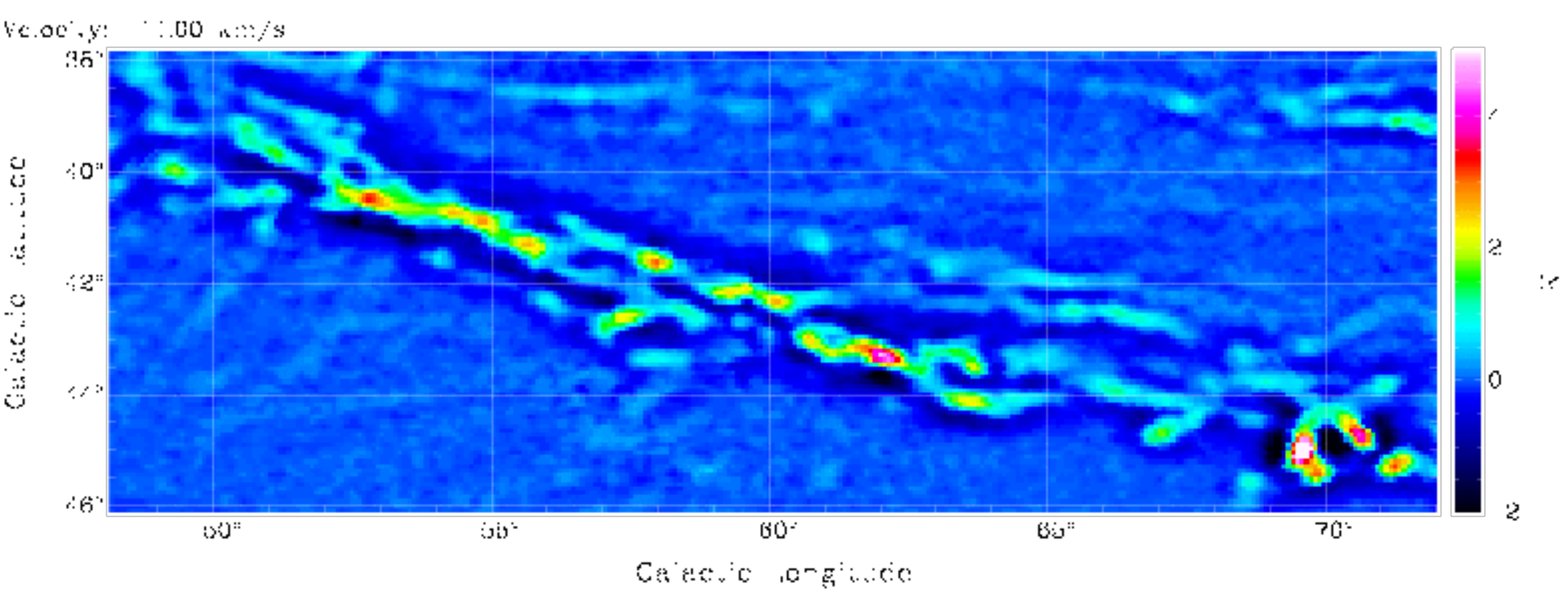}
}
   \centerline{
   \includegraphics[scale=0.3, angle=270]{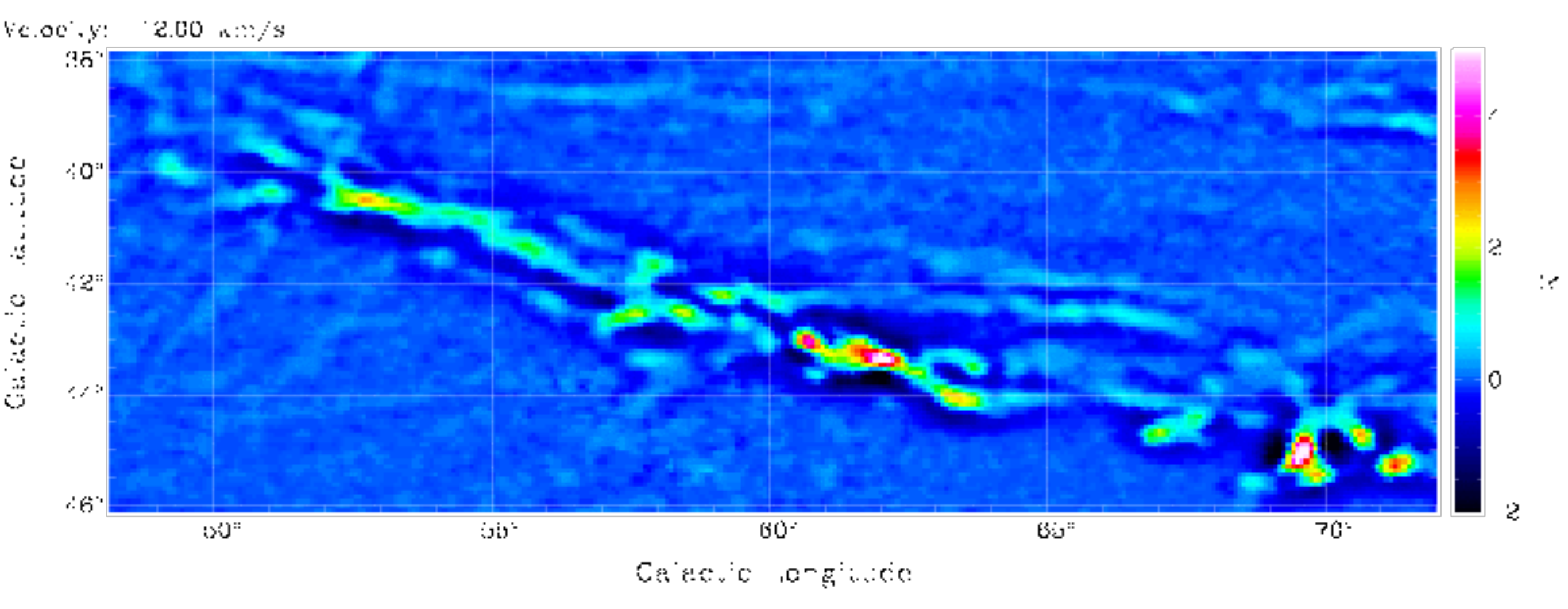}
   \includegraphics[scale=0.3, angle=270]{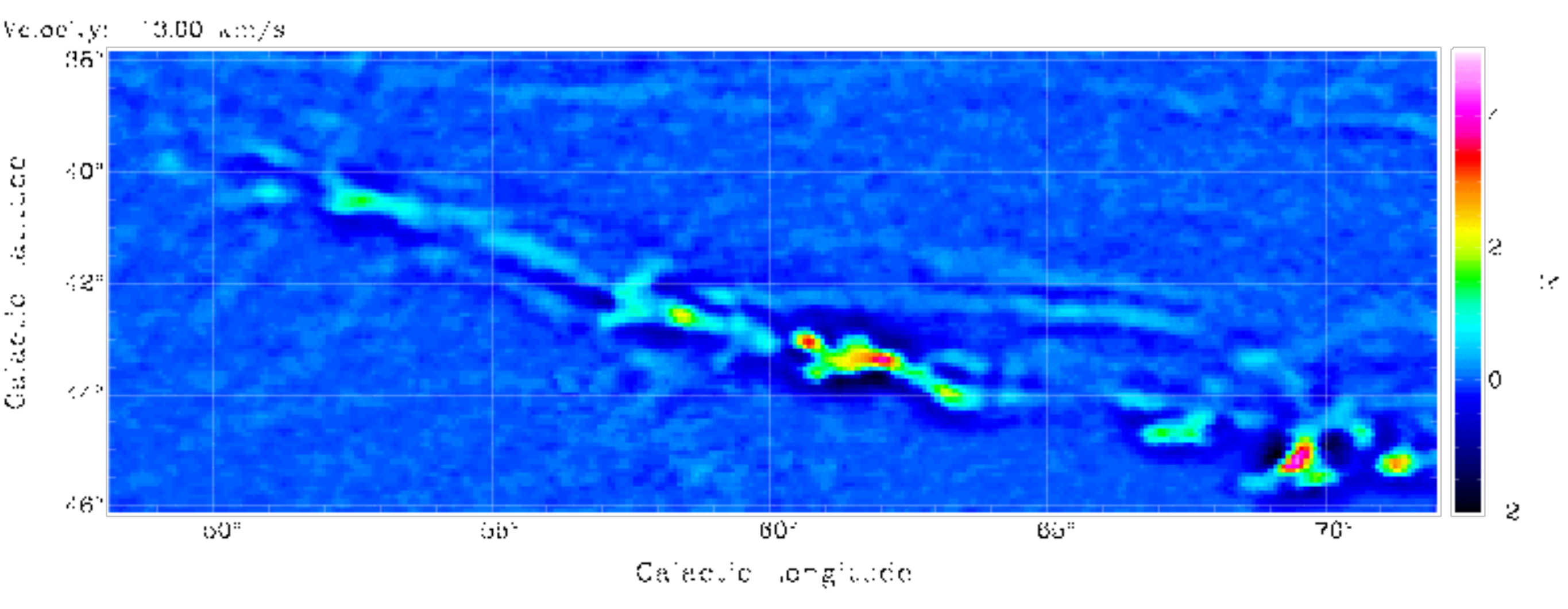}
}
   \caption{USM channel maps for  filaments at  $l \sim 300\degr, b \sim
     -42$.  }
   \label{FIG_Cfilament}
\end{figure*}

In Sect. \ref{sheet} we discussed apparent position shifts of the
filaments in Fig. \ref{FIG_filament} with changes in the observed radial
velocity. Such shifts are frequent. In Fig. \ref{FIG_Cfilament} we show
another case at $l \sim 300\degr, b \sim -42\degr$. This filament has at
$ v = 8$ \kms~ almost no bending. It is not possible to argue for a
convex shape and a blast wave origin. 

\begin{figure*}[hbp]
   \centerline{
   \includegraphics[scale=0.3, angle=270]{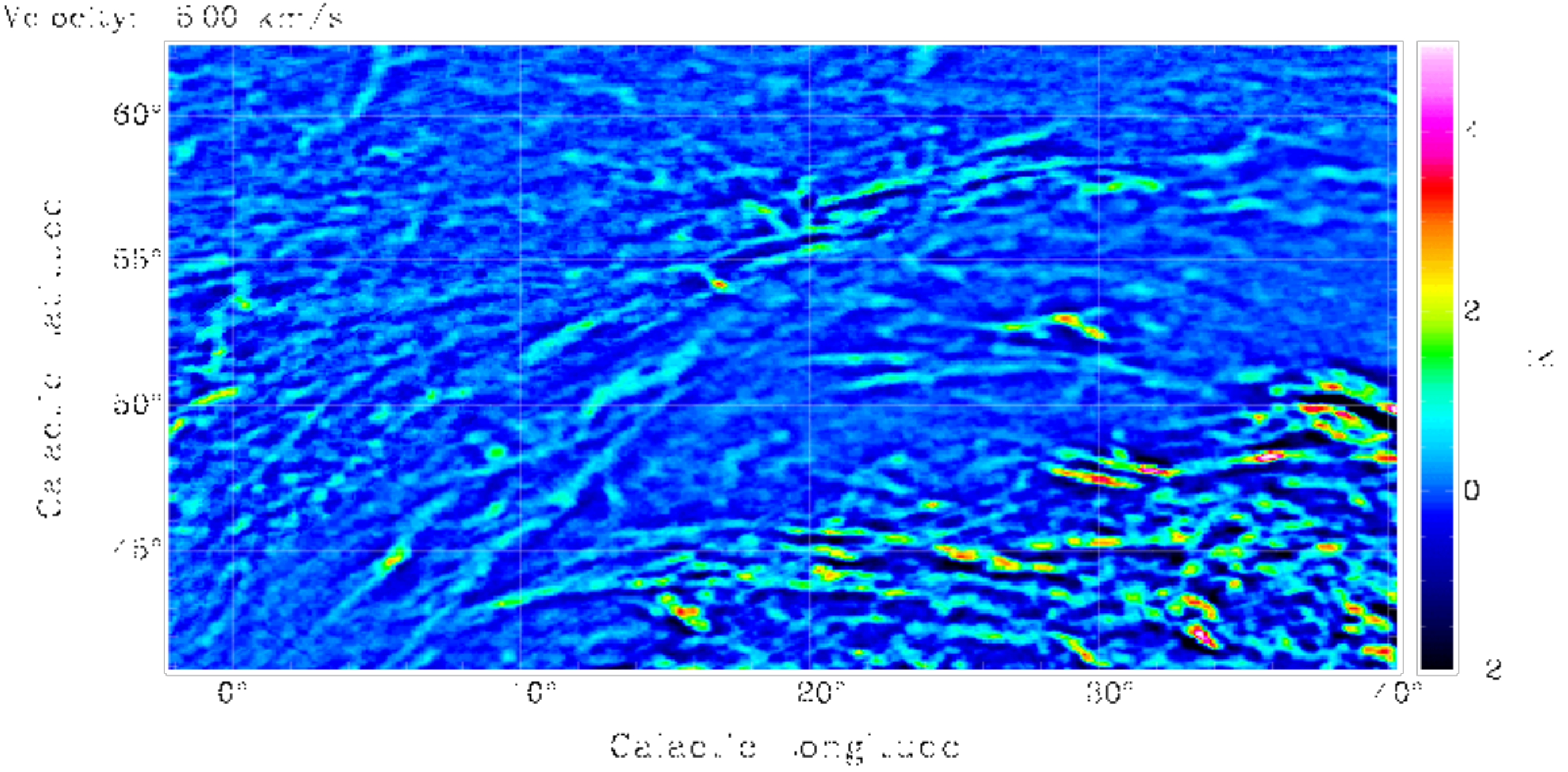}
   \includegraphics[scale=0.3, angle=270]{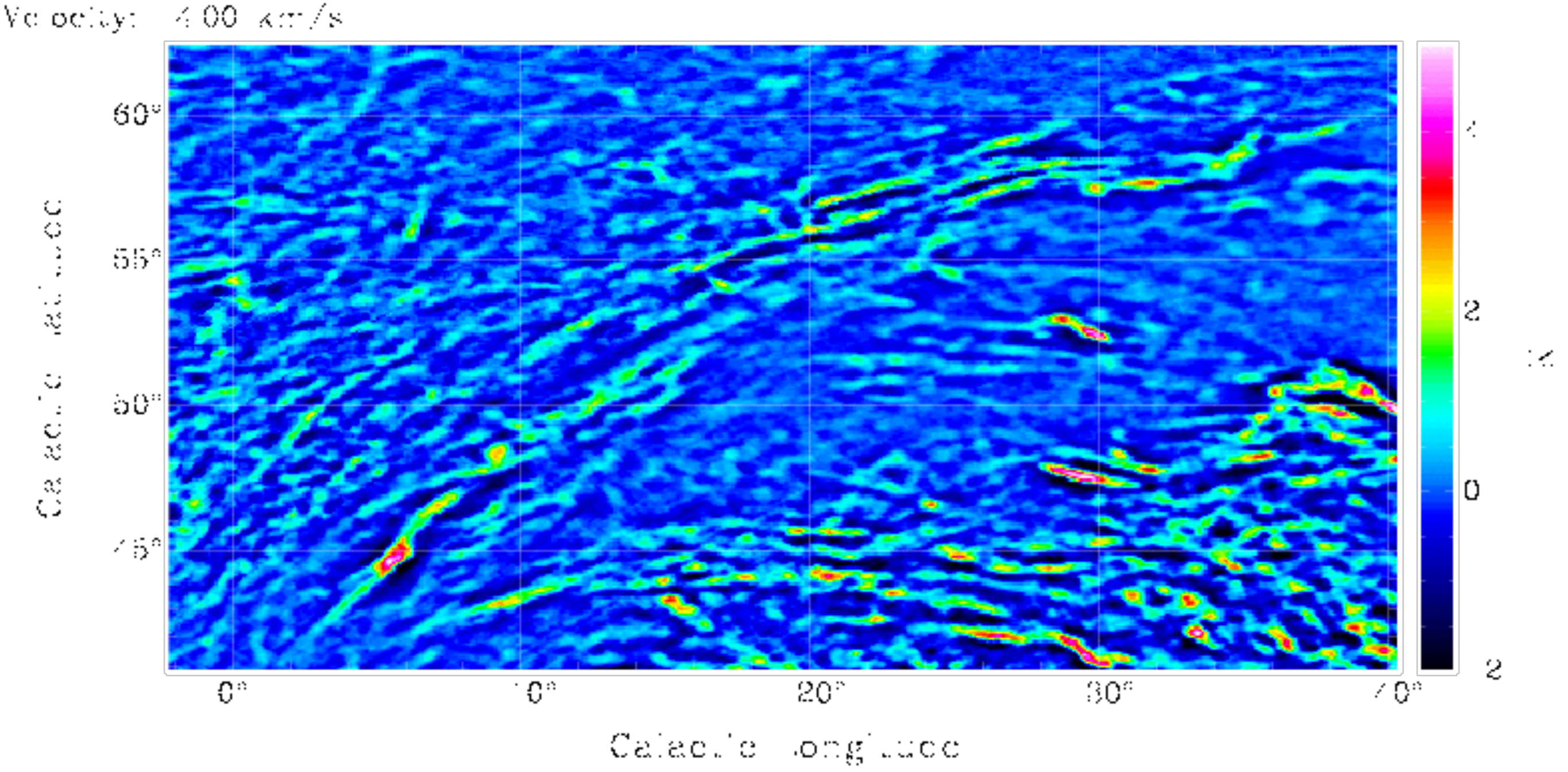}
}
   \centerline{
   \includegraphics[scale=0.3, angle=270]{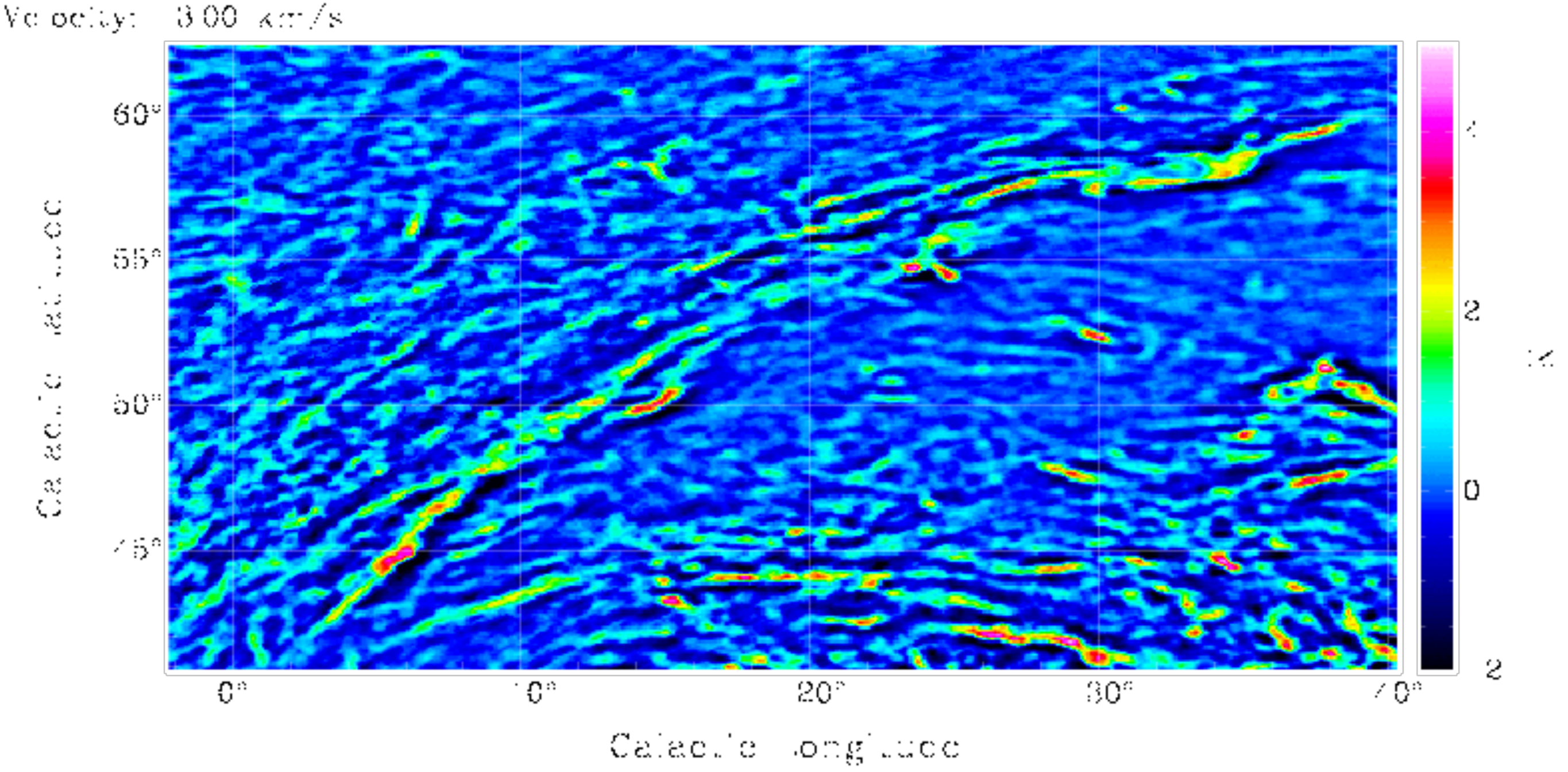}
   \includegraphics[scale=0.3, angle=270]{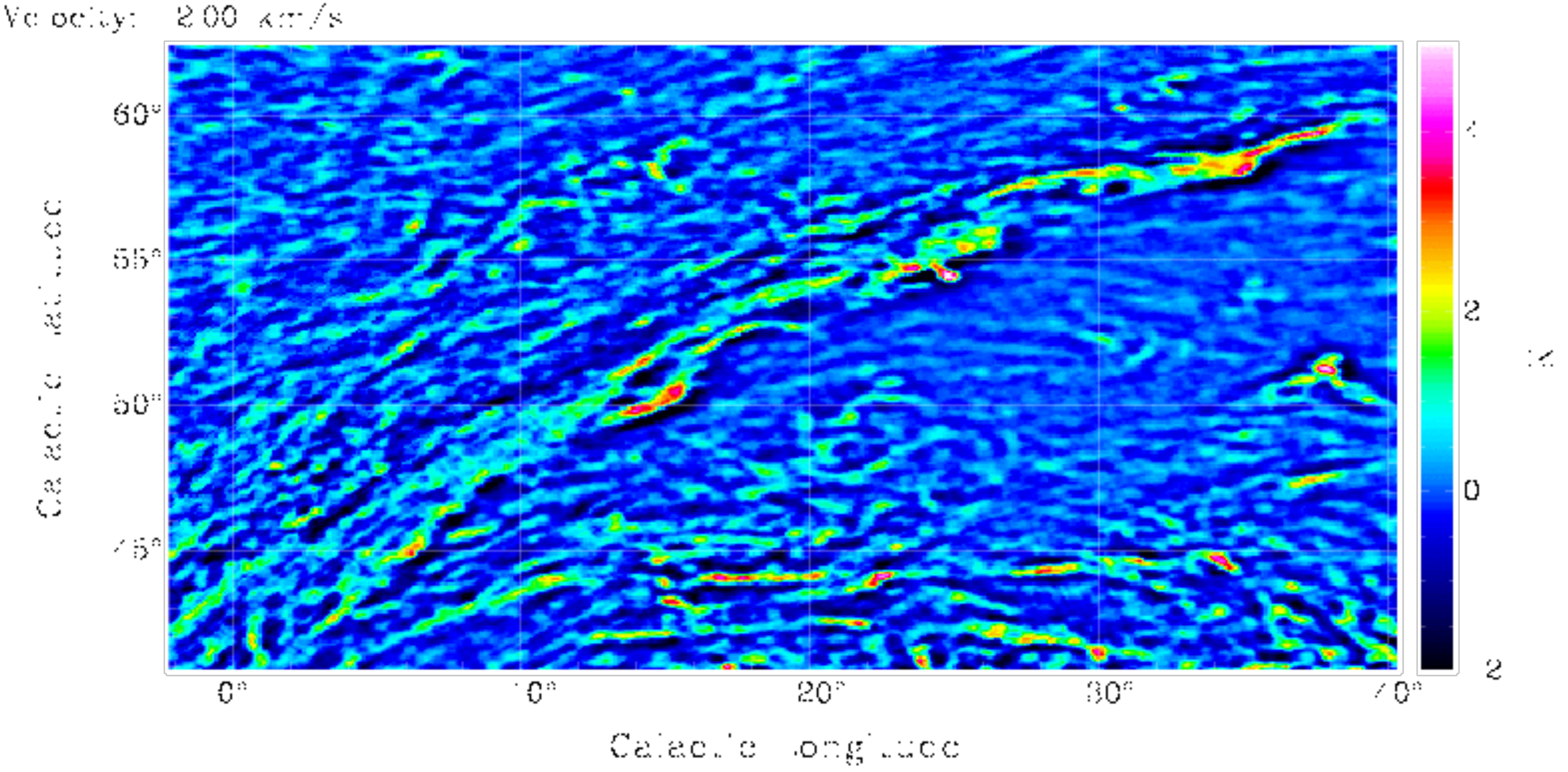}
}
   \centerline{
   \includegraphics[scale=0.3, angle=270]{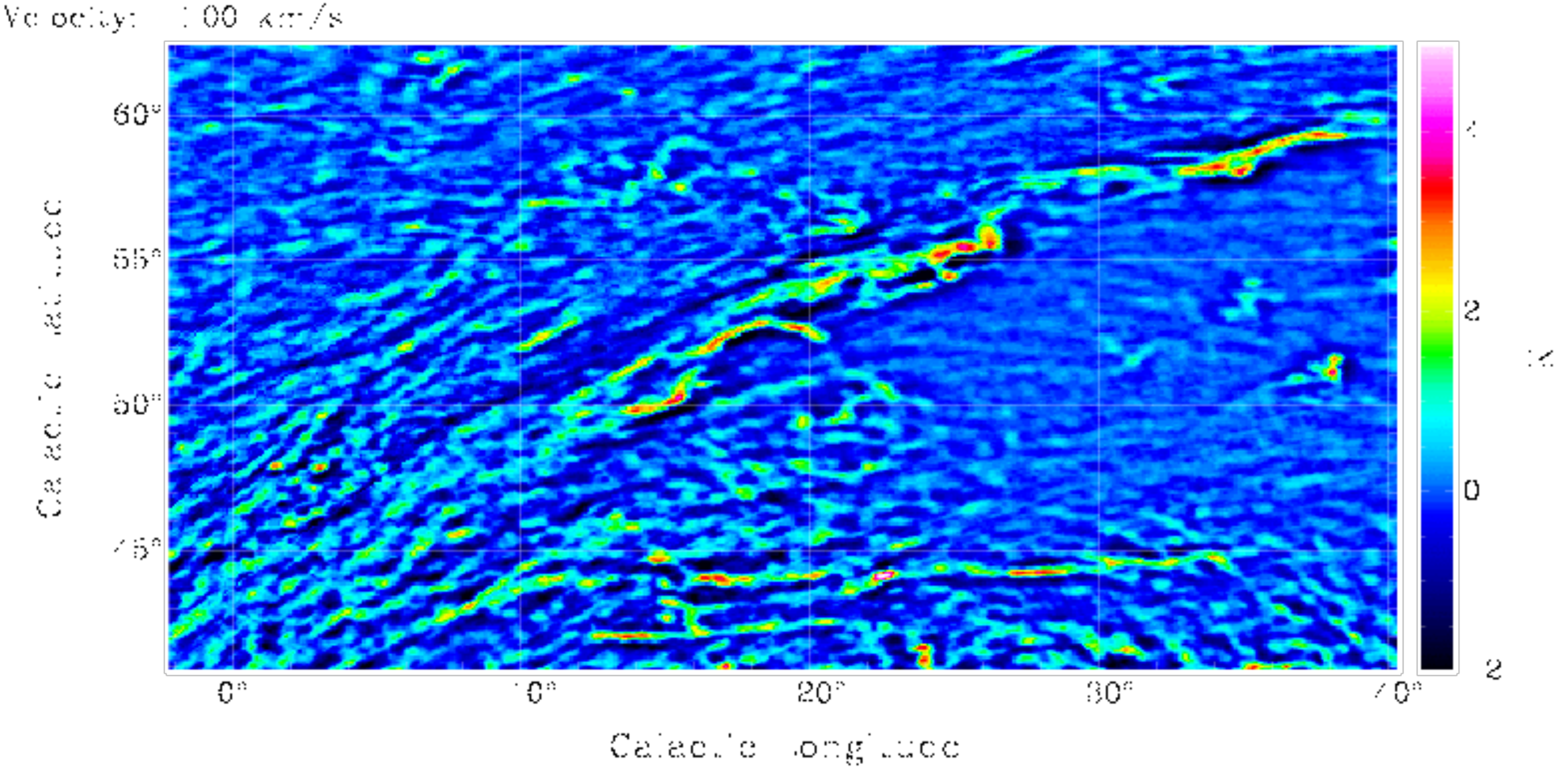}
   \includegraphics[scale=0.3, angle=270]{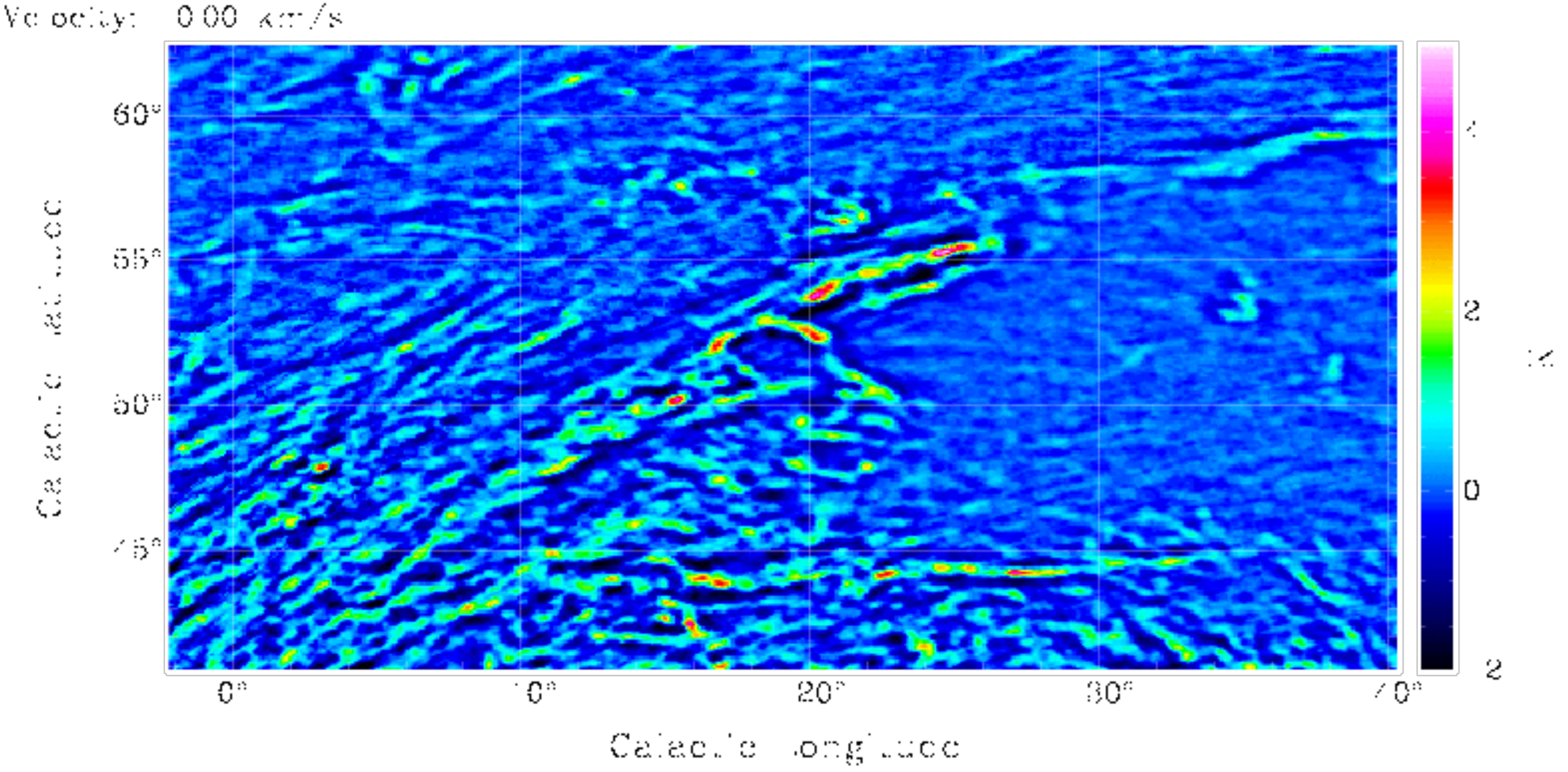}
}
   \centerline{
   \includegraphics[scale=0.3, angle=270]{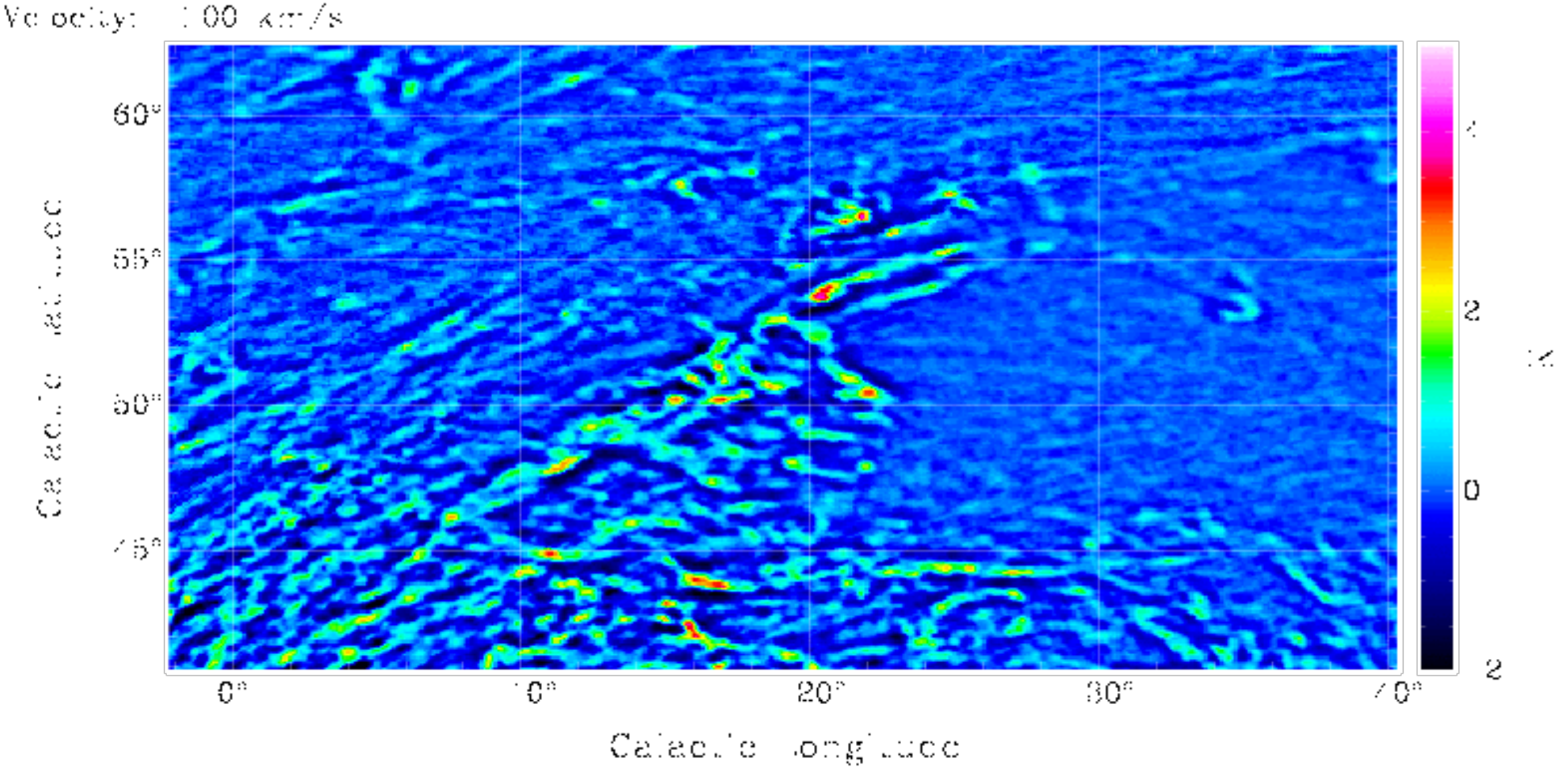}
   \includegraphics[scale=0.3, angle=270]{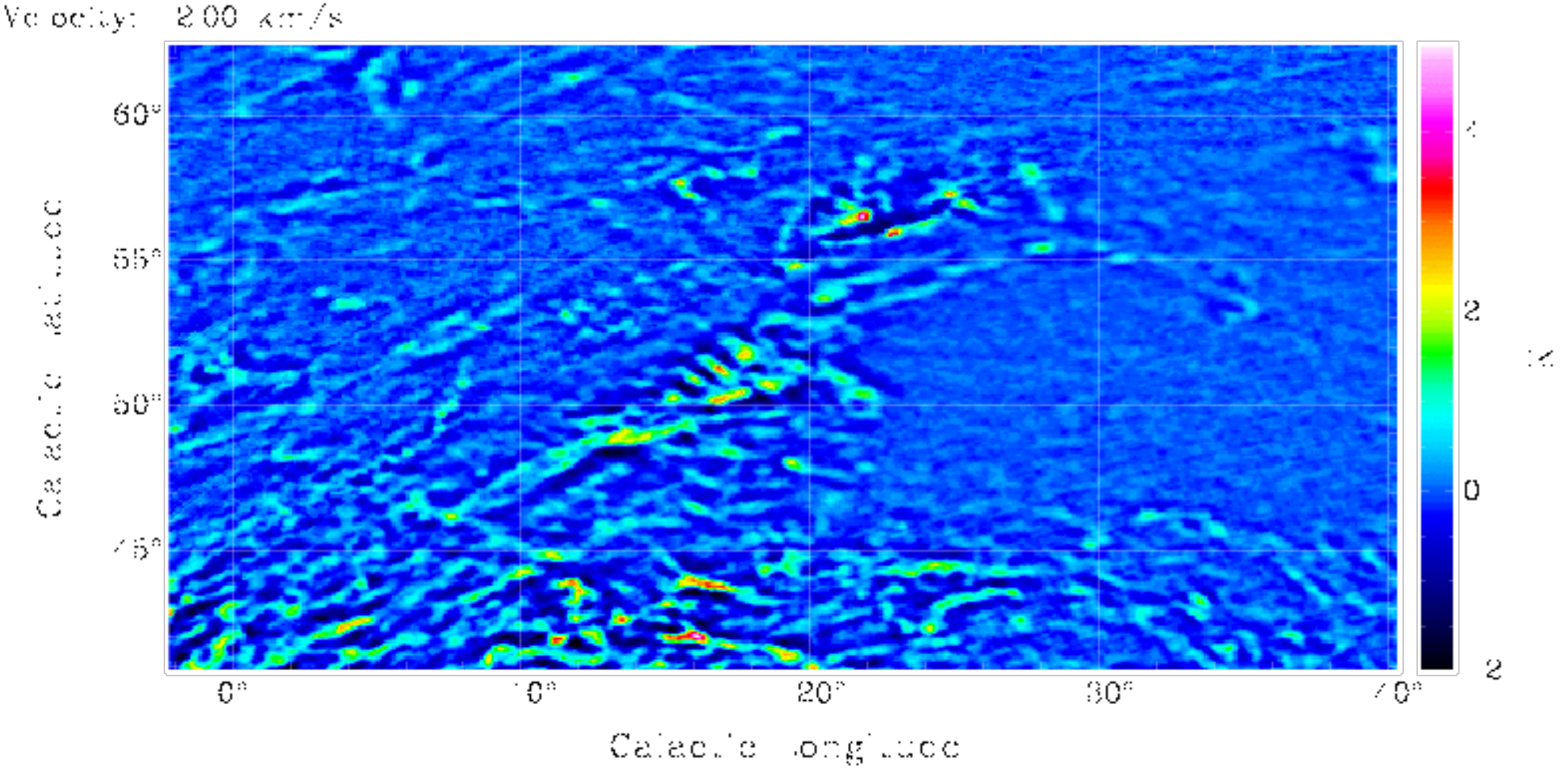}
}
   \caption{USM channel maps for a loop at $l \sim 340\degr, b \sim
     65\degr$.  }
   \label{FIG_Dfilament}
\end{figure*}

In Fig. \ref{FIG_Dfilament} we give details of a prominent loop at $l
\sim 340\degr, b \sim 65\degr$. Significant shifts in position are
easily visible in the USM channel maps. At more positive velocities the
filaments tend to break into substructures.

\begin{figure*}[tbp]
   \centerline{
   \includegraphics[scale=0.3, angle=270]{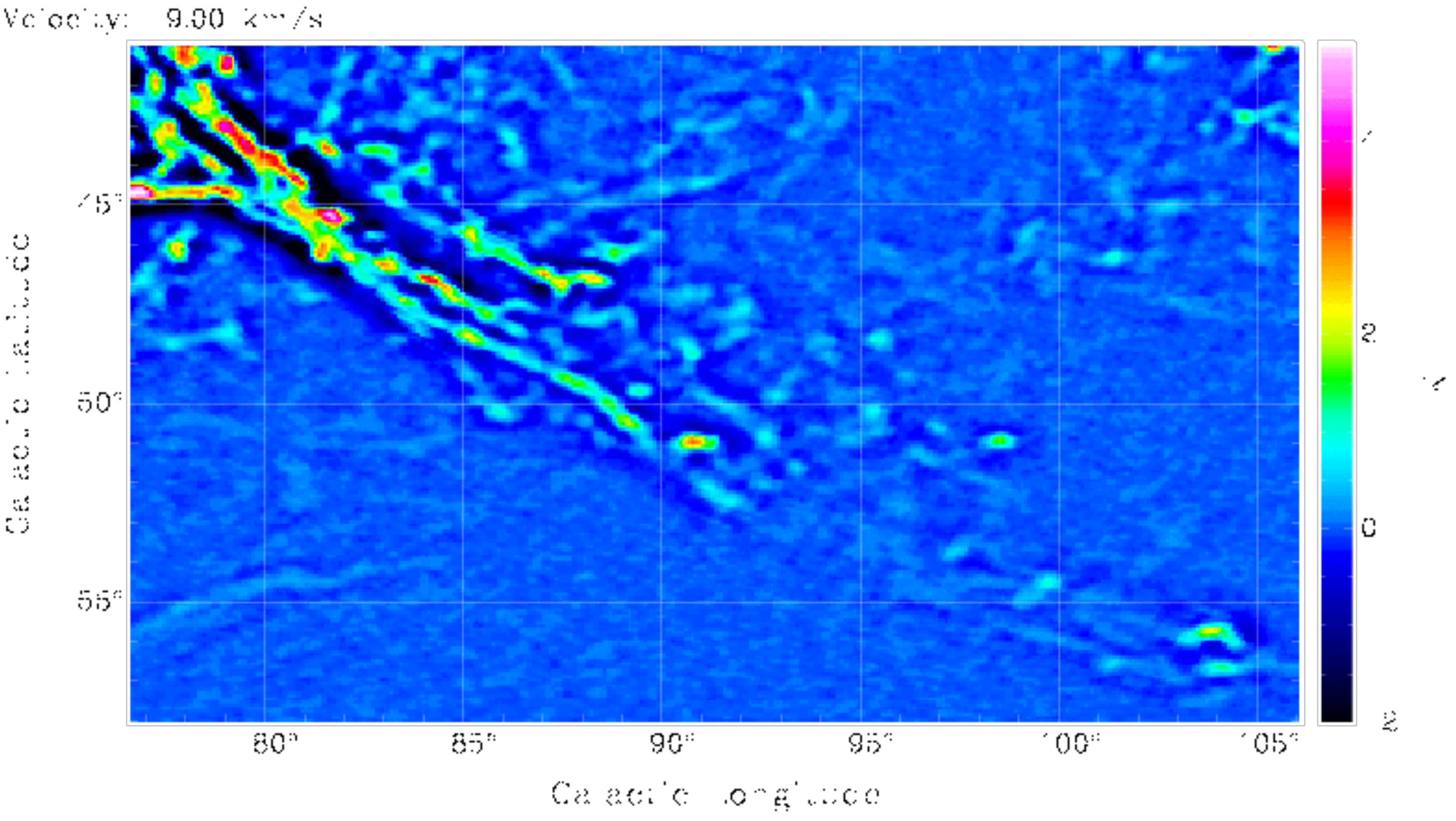}
   \includegraphics[scale=0.3, angle=270]{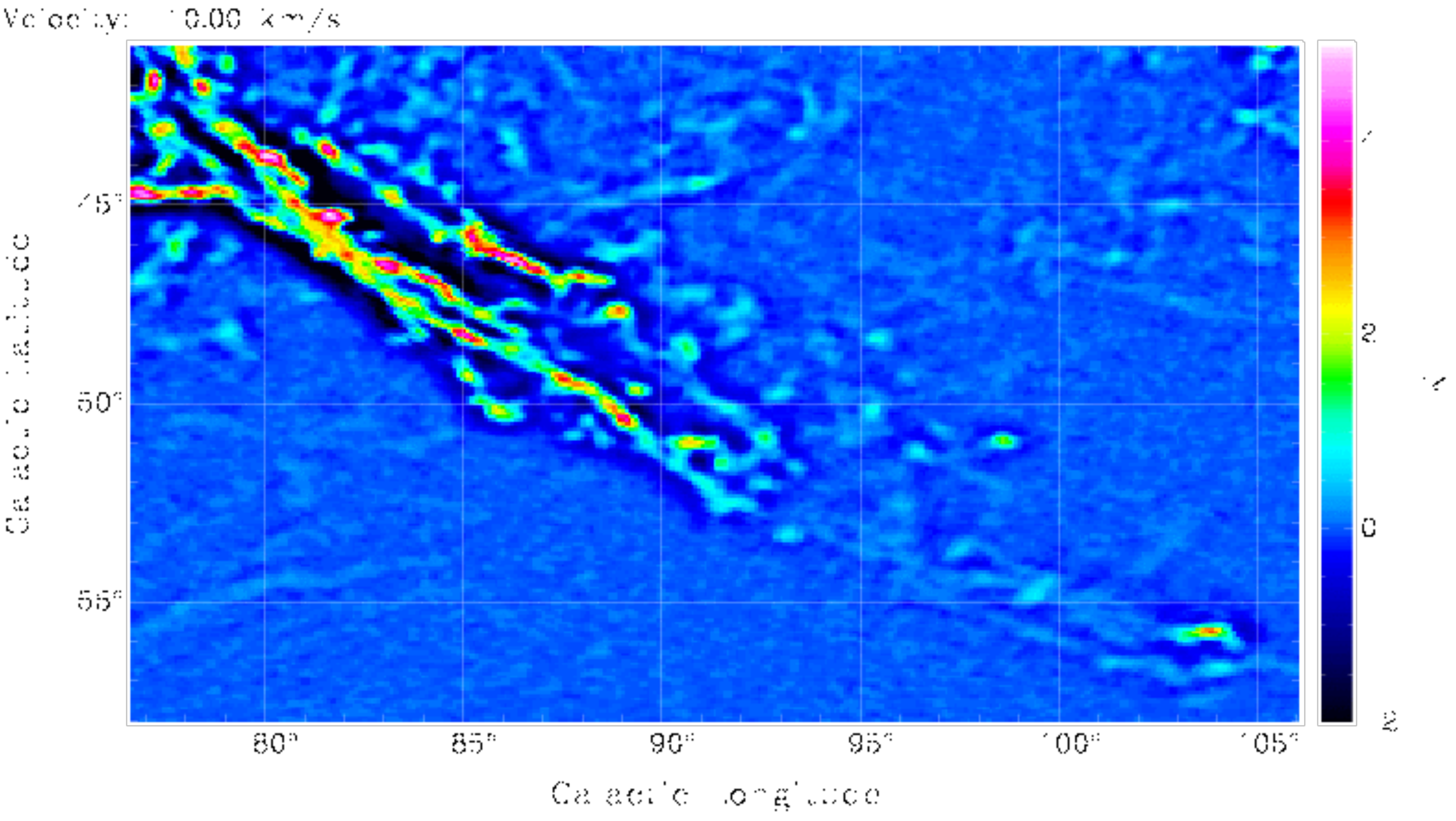}
}
   \centerline{
   \includegraphics[scale=0.3, angle=270]{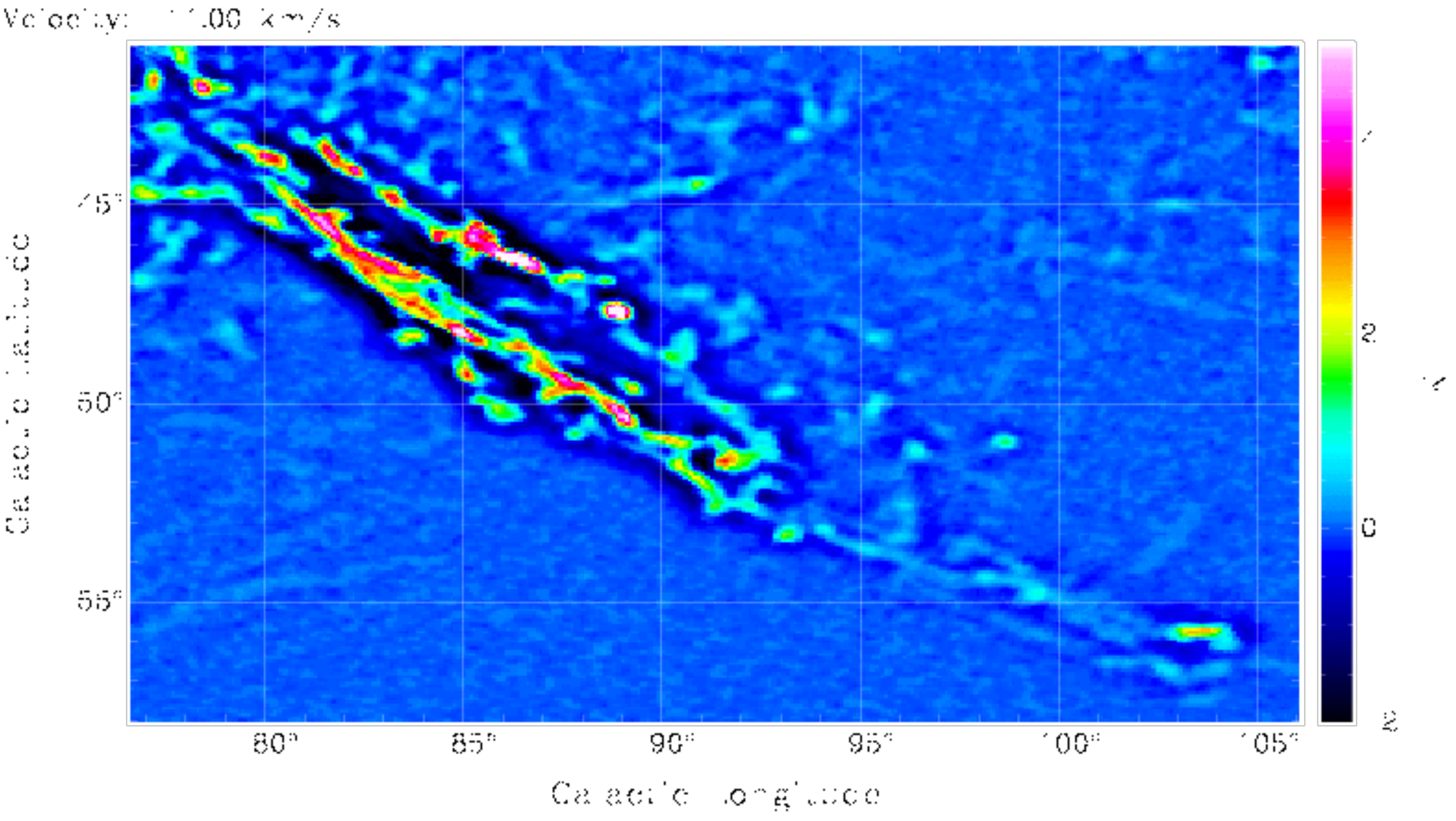}
   \includegraphics[scale=0.3, angle=270]{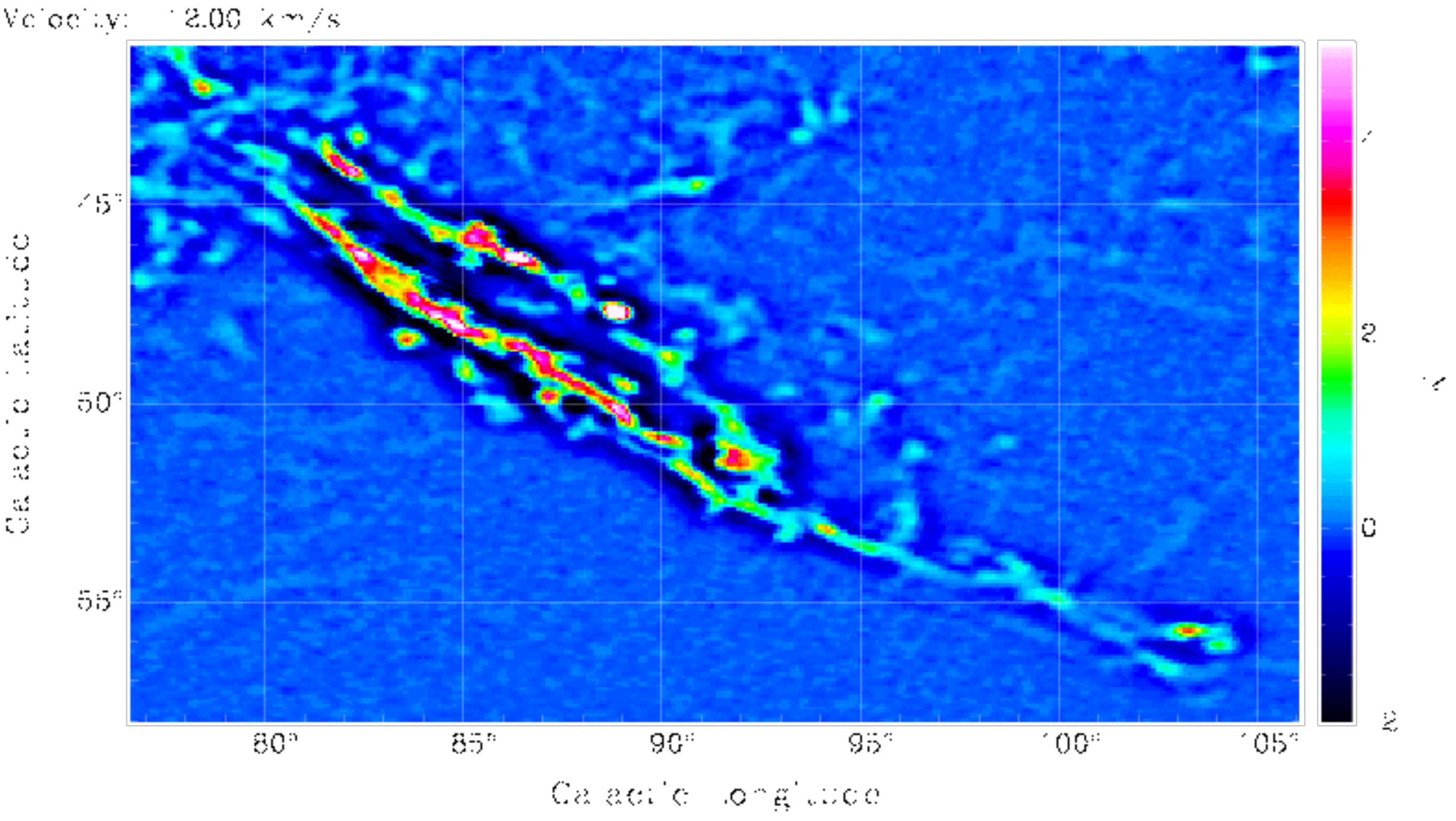}
}
   \centerline{
   \includegraphics[scale=0.3, angle=270]{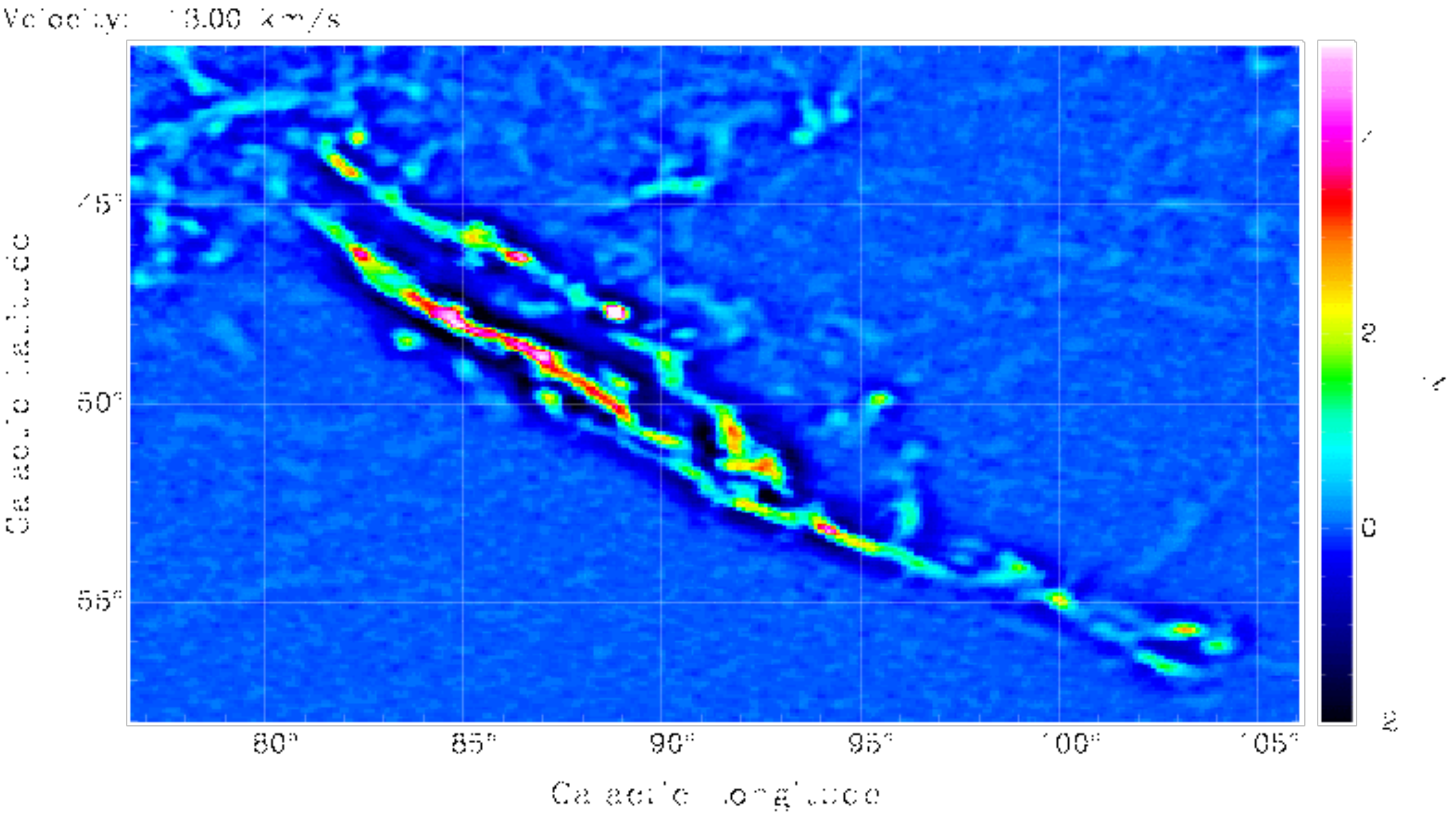}
   \includegraphics[scale=0.3, angle=270]{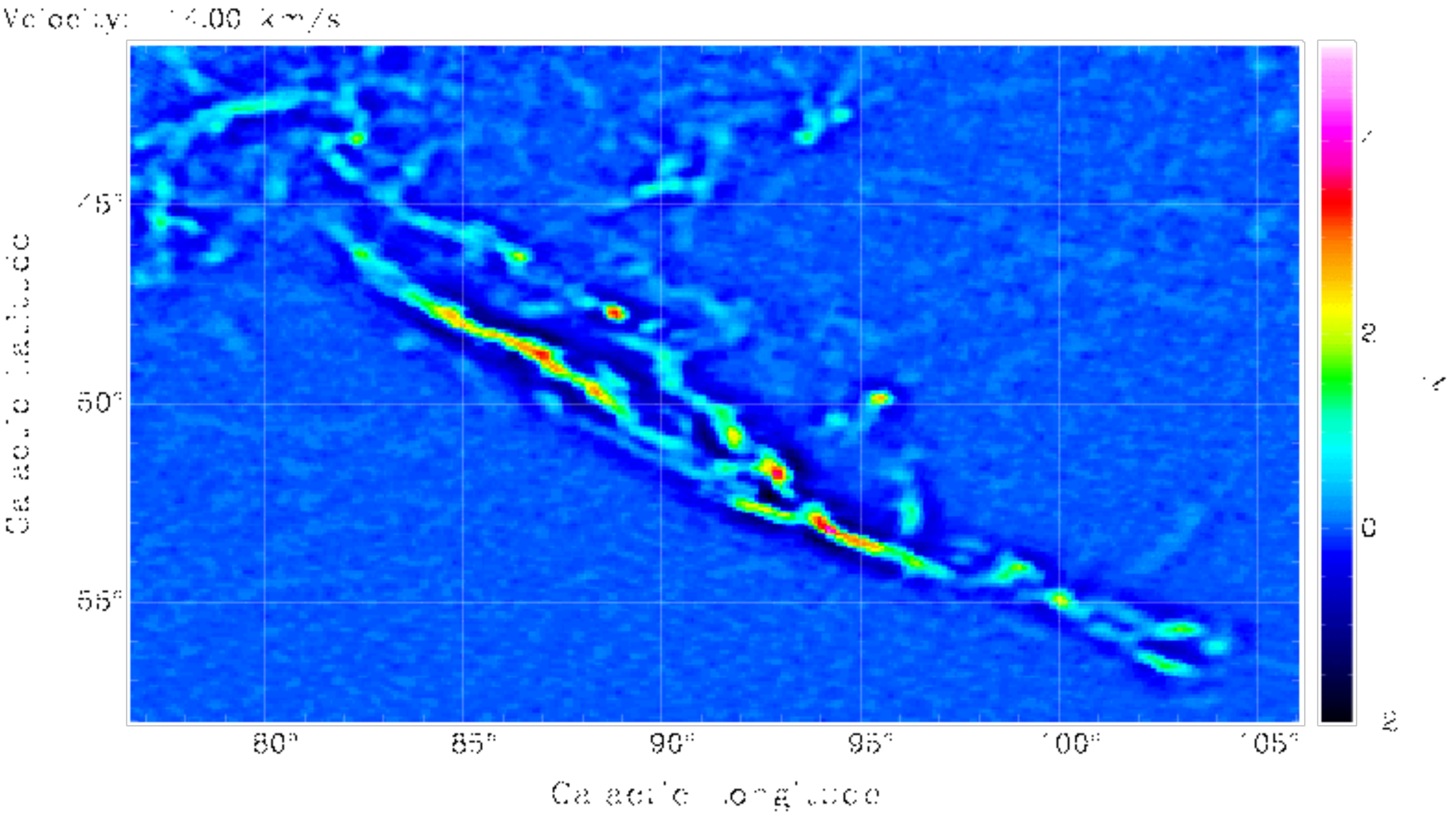}
}
   \centerline{
   \includegraphics[scale=0.3, angle=270]{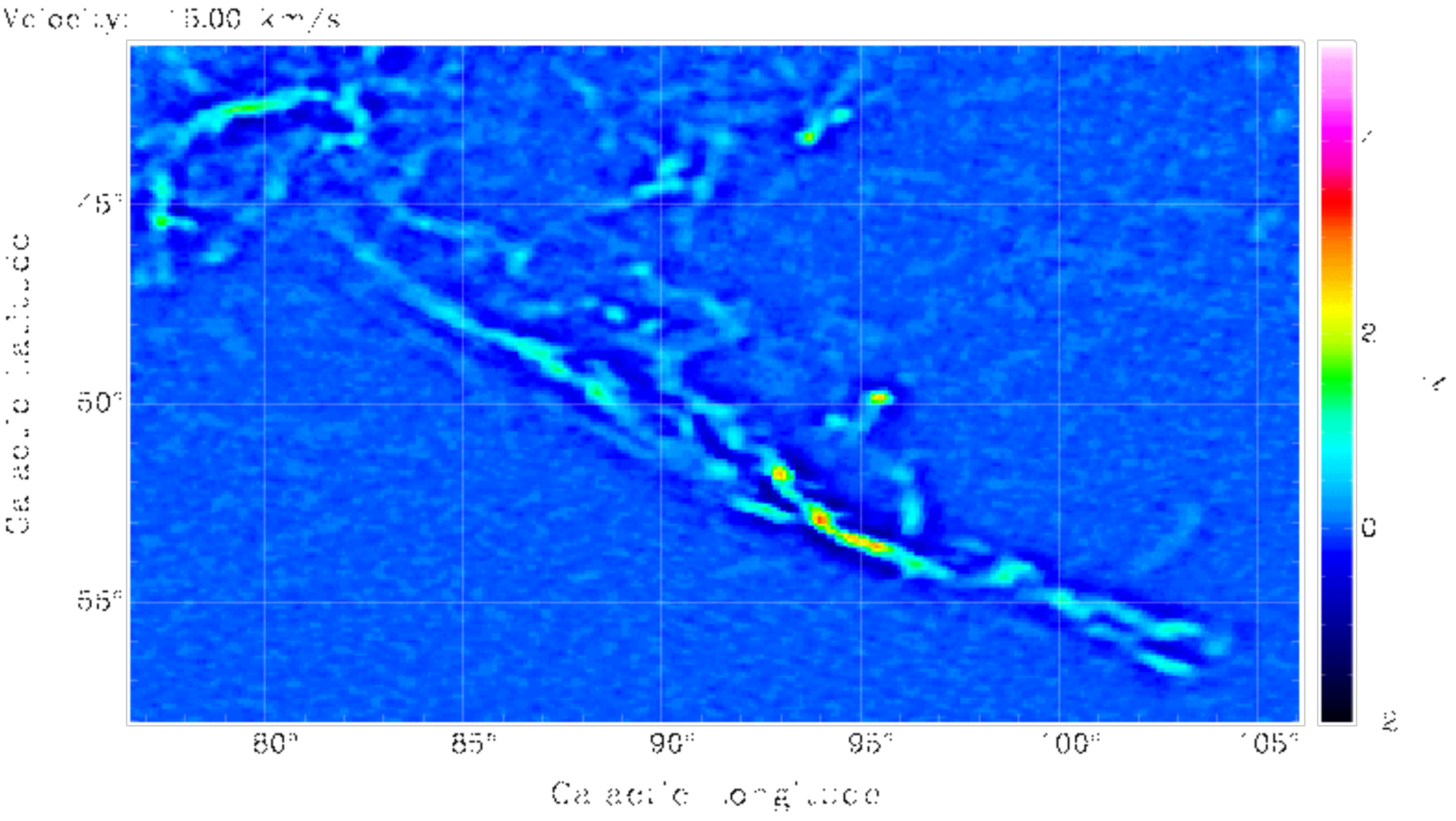}
   \includegraphics[scale=0.3, angle=270]{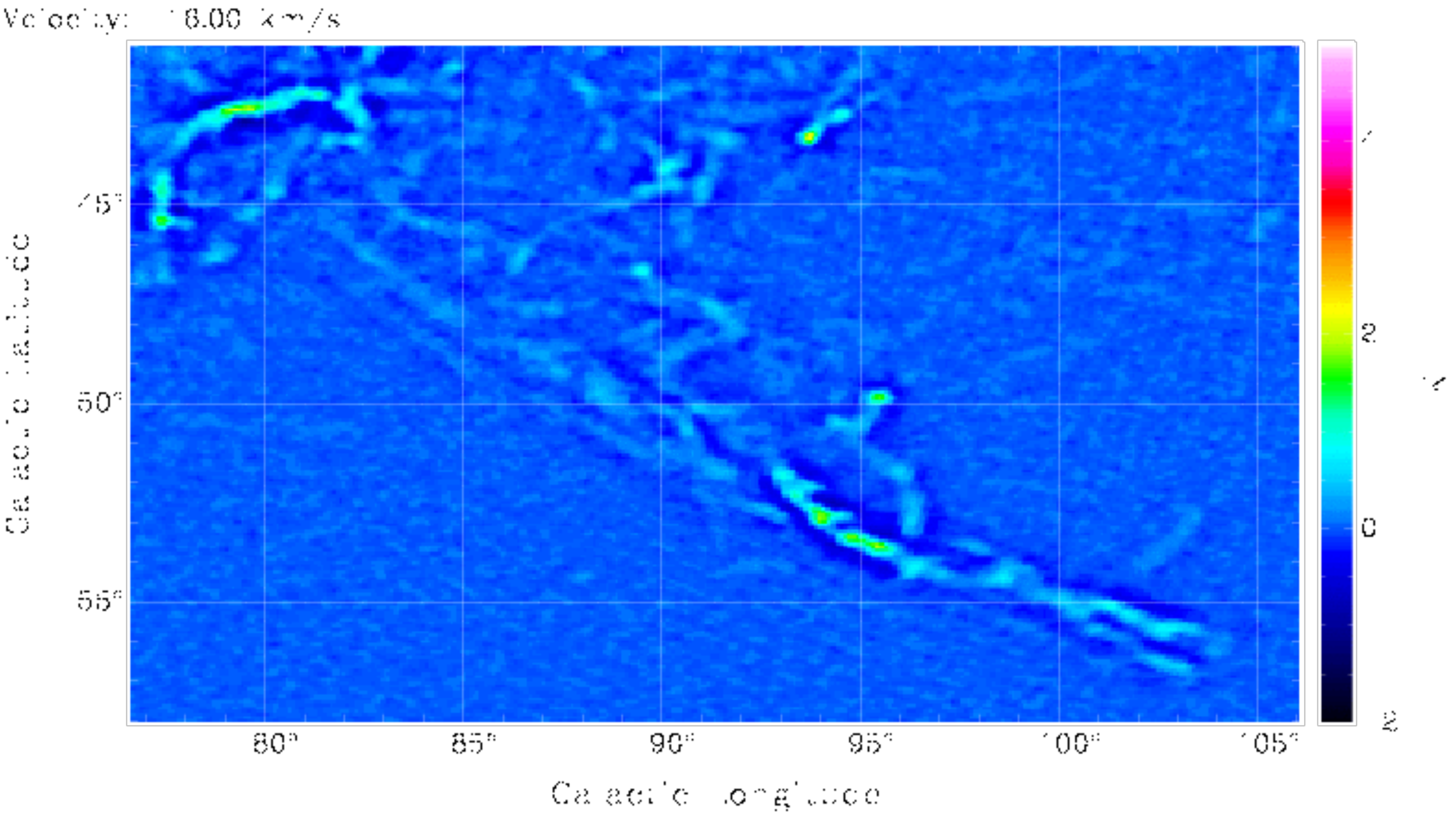}
}
   \caption{USM channel maps for filaments at $l \sim 270\degr, b \sim
     -50\degr$.  }
   \label{FIG_Bfilament}
\end{figure*}

In Fig. \ref{FIG_Bfilament} we present a filamentary structure that
shows for velocity changes obvious gradients in the \hi distribution
along the filaments.

%\end{appendix}


\begin{thebibliography}{}

\bibitem[Andr{\'e} et al.(2014)]{Andre2014} Andr{\'e}, P., Di 
Francesco, J., Ward-Thompson, D., et al.\ 2014, Protostars and Planets VI, 
27 

\bibitem[Bajaja et al.(2005)]{Bajaja2005} Bajaja, E., Arnal, E.~M., Larrarte, J.~J., et al.\ 2005, \aap, 440, 767 

\bibitem[Boulanger et 
al.(1996)]{Boulanger1996} Boulanger, F., Abergel, A., Bernard, J.-P., et al.\ 1996, \aap, 312, 256 

\bibitem[Burton(1971)]{Burton1971} Burton, W.~B.\ 1971, \aap, 10, 76 

\bibitem[Clark(1964)]{Clark1964} Clark, B.~G.\ 1964, Ph.D.~Thesis,
 California Institute of Technology

\bibitem[Clark et al.(2015)]{Clark2015} Clark, S.~E., Hill, 
J.~C., Peek, J.~E.~G., Putman, M.~E., 
\& Babler, B.~L.\ 2015, Physical Review Letters, 115, 241302 

\bibitem[Clark et al.(2014)]{Clark2014} Clark, S.~E., Peek, 
J.~E.~G., \& Putman, M.~E.\ 2014, \apj, 789, 82 


%\bibitem[de Bruyn et al.(2006)]{deBruyn2006} de Bruyn, A.~G., Katgert,   P., Haverkorn, M., \& Schnitzeler, D.~H.~F.~M.\ 2006, Astronomische Nachrichten, 327, 487


\bibitem[Elmegreen 
\& Scalo(2004)]{Elmegreen2004} Elmegreen, B.~G., \& Scalo, J.\ 2004,
\araa, 42, 211

\bibitem[Federrath(2016)]{Federrath2016} Federrath, C.\ 2016, \mnras, 
457, 375 

%\bibitem[Federrath(2015)]{Federrath2015} Federrath, C.\ 2015, arXiv:1510.05654 

\bibitem[Field(1958)]{Field1958} Field, G.~B.\ 1958, Proceedings 
of the IRE, 46, 240 

\bibitem[Field(1959)]{Field1959} Field, G.~B.\ 1959, \apj, 129, 
536 

\bibitem[Field et al.(1969)]{Field1969} Field, G.~B., Goldsmith, 
D.~W., \& Habing, H.~J.\ 1969, \apjl, 155, L149 


\bibitem[Gibson et al.(2000)]{Gibson2000} Gibson, S.~J., Taylor, 
A.~R., Higgs, L.~A., \& Dewdney, P.~E.\ 2000, \apj, 540, 851 

\bibitem[G{\'o}rski et al.(2005)]{Gorski2005} G{\'o}rski, K.~M., 
Hivon, E., Banday, A.~J., et al.\ 2005, \apj, 622, 759 

\bibitem[Hartmann \& Burton(1997)]{Atlas1997} Hartmann, D.~\& Burton, W.~B.\
  1997, Atlas of Galactic Neutral Hydrogen (Cambridge: Cambridge University
  Press)

\bibitem[Haud(2000)]{Haud2000} Haud, U.\ 2000, \aap, 364, 83 

\bibitem[Haud \& Kalberla(2007)]{Haud2007} Haud, U., \& Kalberla, P.~M.~W.\ 2007, \aap, 466, 555 

%\bibitem[Haud(2010)]{Haud2010} Haud, U.\ 2010, \aap, 514, A27 

\bibitem[Haud(2013)]{Haud2013} Haud, U.\ 2013, \aap, 552, A108 

\bibitem[Heiles(1967)]{Heiles1967ApJS} Heiles, C.\ 1967, \apjs, 15, 97 

\bibitem[Heiles 
\& Troland(2003a)]{Heiles2003a} Heiles, C., \& Troland, T.~H.\ 2003a, \apjs, 145, 329 

\bibitem[Heiles 
\& Troland(2003b)]{Heiles2003b} Heiles, C., \& Troland, T.~H.\ 2003b,
\apj, 586, 1067 

\bibitem[Heiles 
\& Crutcher(2005)]{HeilesCrutcher2005} Heiles, C., \& Crutcher, R.\ 2005, Cosmic Magnetic Fields, 664, 137 

\bibitem[Heiles \& Troland(2005)]{Heiles2005} Heiles, C., \& Troland,
  T.~H.\ 2005, \apj, 624, 773

\bibitem[Hennebelle(2013)]{Hennebelle2013} Hennebelle, P.\ 2013, \aap,
  556, A153


\bibitem[Jenkins \& Tripp(2011)]{Jenkins2011} Jenkins, E.~B., \& Tripp,
  T.~M.\ 2011, \apj, 734, 65

\bibitem[Kalberla et al.(2005)]{Kalberla2005} Kalberla, P.~M.~W., Burton,
  W.~B., Hartmann, D.\ et al. 2005, \aap, 440, 775

\bibitem[Kalberla \& Kerp(2009)]{ARAA2009} Kalberla, P.~M.~W., \& Kerp, J.\ 2009, \araa, 47, 27 

\bibitem[Kalberla et 
al.(2010)]{Kalberla2010} Kalberla, P.~M.~W., McClure-Griffiths, N.~M., Pisano, D.~J., et al.\ 2010, \aap, 521, A17 

\bibitem[Kalberla(2011)]{Kalberla2011} Kalberla, P.~M.~W.\ 2011, 
arXiv:1102.4949 

\bibitem[Kalberla 
\& Haud(2015)]{Kalberla2015} Kalberla, P.~M.~W., \& Haud, U.\ 2015,
\aap, 578, A78 

\bibitem[Kalberla, Mebold \& Reich(1980)]{Kalberla1980} Kalberla, P.M.W., Mebold, U., \& Reich, W. 1980, \aap, 82, 275

\bibitem[Kerr(1969)]{Kerr1969} Kerr, F.~J.\ 1969, \araa, 7, 39 

\bibitem[Lallement et 
al.(2014)]{Lallement2014} Lallement, R., Vergely, J.-L., Valette, B., et al.\ 2014, \aap, 561, A91 

%\bibitem[Larson(1979)]{Larson1979} Larson, R.~B.\ 1979, \mnras, 186, 479 
 
\bibitem[Liszt(2001)]{Liszt2001} Liszt, H.\ 2001, \aap, 371, 698 

\bibitem[Malin(1978)]{Malin1978} Malin, D.~F.\ 1978, \nat, 276, 
591 

\bibitem[Martin et al.(2015)]{Martin2015} Martin, P.~G., Blagrave, 
K.~P.~M., Lockman, F.~J., et al.\ 2015, \apj, 809, 153 

\bibitem[McClure-Griffiths et al.(2009)]{Naomi2009} McClure-Griffiths, N.~M.,
  Pisano, D.~J., Calabretta, M.~R., et al.\ 2009, \apjs, 181, 398 

\bibitem[McKee 
\& Ostriker(1977)]{McKe1977} McKee, C.~F., \& Ostriker, J.~P.\ 1977, \apj, 218, 148 

\bibitem[Payne et al.(1980)]{Payne1980} Payne, H.~E., Terzian, 
Y., \& Salpeter, E.~E.\ 1980, \apj, 240, 499 


\bibitem[Planck Collaboration et 
al.(2011)]{Planck2011} Planck Collaboration, Abergel, A., Ade, P.~A.~R., et al.\ 2011, \aap, 536, A24 


\bibitem[Planck Collaboration et 
al.(2014)]{Planck2014} Planck Collaboration, Abergel, A., Ade, P.~A.~R., et al.\ 2014, \aap, 566, A55 


\bibitem[Planck Collaboration et al.(2015a)]{PlanckXIXa} Planck
  Collaboration, Ade, P.~A.~R., Aghanim, N., et al.\ 2015a, \aap, 576,
  A104

\bibitem[Planck Collaboration et 
al.(2015b)]{PlanckXX2015b} Planck Collaboration, Ade, P.~A.~R., Aghanim, N., et al.\ 2015b, \aap, 576, A105 

\bibitem[Planck Collaboration et 
al.(2016)]{Planck2016} Planck Collaboration, Adam, R., Ade, P.~A.~R., et al.\ 2016, \aap, 586, A135 

%\bibitem[Planck Collaboration et al.(2014)]{2014arXiv1409.6728P} Planck
%Collaboration, Adam, R., Ade, P.~A.~R., et al.\ 2014, arXiv:1409.6728 

\bibitem[Riegel \& Crutcher(1972)]{Riegel1972} Riegel, K.~W., \&
  Crutcher, R.~M.\ 1972, \aap, 18, 55

\bibitem[Roy et al.(2013)]{Roy2013} Roy, N., Kanekar, N., 
\& Chengalur, J.~N.\ 2013, \mnras, 436, 2366 

\bibitem[Roy(2015)]{Roy2015} Roy, N.\ 2015, Proceedings of the 
Indian National Science Academy Part A, 81, 583 


\bibitem[Saury et al.(2014)]{Saury2014} Saury, E.,
  Miville-Desch{\^e}nes, M.-A., Hennebelle, P., Audit, E., \& Schmidt,
  W.\ 2014, \aap, 567, A16

\bibitem[Schisano et al.(2014)]{Schisano2014} Schisano, E., Rygl, 
K.~L.~J., Molinari, S., et al.\ 2014, \apj, 791, 27 

\bibitem[Shannon(1949)]{Shannon1949} Shannon, C.~E.\ 1949, IEEE 
Proceedings, 37, 10 

\bibitem[Sofue 
\& Reich(1979)]{Sofue1979} Sofue, Y., \& Reich, W.\ 1979, \aaps, 38, 251 

\bibitem[Spitzer(1968)]{Spitzer1968} Spitzer, L.\ 1968, New York: 
Interscience Publication, 1968,  

\bibitem[Stone et al.(1998)]{Stone1998} Stone, J.~M., Ostriker,  E.~C., \& Gammie, C.~F.\ 1998, \apjl, 508, L99 

\bibitem[Vazquez-Semadeni(1994)]{Vazquez1994} Vazquez-Semadeni, E.\ 
1994, \apj, 423, 681 

\bibitem[Verschuur(1969)]{Verschuur1969} Verschuur, G.~L.\ 1969, 
\apj, 156, 861 

\bibitem[Winkel et al.(2016)]{Winkel2016} Winkel, B., Kerp, J., Fl\"oer,
  L., et al.\ 2015, \aap, 585, A41

\bibitem[Wolfire et al.(2003)]{Wolfire2003} Wolfire, M.~G., McKee, 
C.~F., Hollenbach, D., \& Tielens, A.~G.~G.~M.\ 2003, \apj, 587, 278 


\end{thebibliography}
\end{document}